%% file: main.tex
\ifdefined\reviewversion
  \documentclass[manuscript,screen,review,anonymous]{acmart}
\else
  \documentclass[acmsmall,screen,nonacm]{acmart}
\fi

\usepackage{tikz}
\usepackage{array}
\usepackage{float}
\usepackage{longtable}
\usepackage{placeins}
\usepackage{tcolorbox}
\usetikzlibrary{arrows.meta,calc,fit,positioning}

\newif\ifshowtodos
\ifdefined\reviewversion
  \showtodostrue
\else
  \showtodosfalse
\fi

\newcolumntype{L}[1]{>{\raggedright\arraybackslash}p{#1}}
\newcolumntype{C}[1]{>{\centering\arraybackslash}p{#1}}
\newcommand{\finding}[2][Finding]{
    \begin{center}
    \begin{tcolorbox}[colback=white!15, colframe=black, boxsep=-0.15cm, middle=-0.15cm]
    \textbf{#1:}
    {#2}
    \end{tcolorbox}
    \end{center}
}
\newcommand{\fullcoverage}{\tikz[baseline=-0.55ex]{\fill (0,0) circle (0.52ex);}}
\newcommand{\partialcoverage}{\tikz[baseline=-0.55ex]{%
  \begin{scope}
    \clip (-0.60ex,-0.60ex) rectangle (0,0.60ex);
    \fill (0,0) circle (0.52ex);
  \end{scope}
  \draw[line width=0.35pt] (0,0) circle (0.52ex);}}
\newcommand{\nocoverage}{\tikz[baseline=-0.55ex]{\draw[line width=0.35pt] (0,0) circle (0.52ex);}}

\AtBeginDocument{%
  }

\ifdefined\reviewversion
  \setcopyright{none}
  \copyrightyear{2026}
  \acmYear{2026}
  \acmJournal{CSUR}
\fi

\renewcommand\footnotetextcopyrightpermission[1]{}

\title{Software Engineering for LLM-Based Agents: A Lifecycle-Oriented Survey}
\ifdefined\reviewversion
  \author{Anonymous Authors}
  \affiliation{%
    \institution{Anonymous Institution}
    \city{Anonymous City}
    \country{Anonymous Country}}
  \email{anonymous@example.com}
\else
  \input{authors_public}
  \renewcommand{\shortauthors}{Ge et al.}
\fi

\begin{abstract}
\input{sections/abstract}
\end{abstract}



\begin{document}

\maketitle
\raggedbottom

\input{sections/introduction}
\input{sections/background}
\input{sections/methodology}

\input{sections/landscape}
\input{sections/lifecycle}
\input{sections/discussion}
\input{sections/threats}
\input{sections/conclusion}

\appendix
\input{sections/appendix}

\bibliographystyle{ACM-Reference-Format-citation-order}
\bibliography{references}

\end{document}

%% file: authors_public.tex
\author{Yifei Ge}
\affiliation{%
  \institution{Nanjing University}
  \city{Nanjing}
  \country{China}}
\email{gyf991213@126.com}

\author{Yalin Feng}
\affiliation{%
  \institution{Nanjing University}
  \city{Nanjing}
  \country{China}}
\email{yalinfeng@smail.nju.edu.cn}

\author{Weisong Sun}
\authornote{Weisong Sun and Chunrong Fang are co-corresponding authors.}
\affiliation{%
  \institution{Nanyang Technological University}
  \city{Singapore}
  \country{Singapore}}
\email{weisong.sun@ntu.edu.sg}

\author{Junjie Shi}
\affiliation{%
  \institution{Nanyang Technological University}
  \city{Singapore}
  \country{Singapore}}
\email{CHUNKIT001@e.ntu.edu.sg}

\author{Zhenpeng Chen}
\affiliation{%
  \institution{Tsinghua University}
  \city{Beijing}
  \country{China}}
\email{zpchen@tsinghua.edu.cn}

\author{Chunrong Fang}
\authornotemark[1]
\affiliation{%
  \institution{Nanjing University}
  \city{Nanjing}
  \country{China}}
\email{fangchunrong@nju.edu.cn}

\author{Zhihong Zhao}
\affiliation{%
  \institution{Nanjing University}
  \city{Nanjing}
  \country{China}}
\email{zhaozhih@nju.edu.cn}

\author{Zhenyu Chen}
\affiliation{%
  \institution{Nanjing University}
  \city{Nanjing}
  \country{China}}
\email{zychen@nju.edu.cn}

\author{Yang Liu}
\affiliation{%
  \institution{Nanyang Technological University}
  \city{Singapore}
  \country{Singapore}}
\email{yangliu@ntu.edu.sg}

\author{Zhi Jin}
\affiliation{%
  \institution{Peking University}
  \city{Beijing}
  \country{China}}
\affiliation{%
  \institution{Wuhan University}
  \city{Wuhan}
  \country{China}}
\email{zhijin@pku.edu.cn}

%% file: sections/abstract.tex
Large language model (LLM)-based Agents are used in applications such as software development and information retrieval, where they perform tasks through tool use and interaction with external environments.
Public engineering guidance from companies such as Anthropic describes a task-oriented approach to their development: start from a target task, select a suitable workflow or autonomous Agent design, and refine prompts, tools, and execution strategies through evaluation of task outcomes, latency, and cost.
However, when these capabilities must be delivered and sustained in practice, developers need to establish stakeholder requirements, action boundaries, and ways to evaluate and maintain Agent behavior.
Model-directed execution paths can vary, failures can span model decisions and tools, and changes to prompts, models, or interfaces can affect previously accepted behavior.
These issues call for examining the Agent's integrated implementation: interacting programs, configurations, and tool interfaces determine its behavior.
We therefore view the Agent as a software system requiring ongoing development and maintenance.
Developing such a system involves translating requirements into design and implementation, checking whether its behavior meets those requirements, and revisiting development decisions as the system and its environment change.
We use the software engineering lifecycle as a framework for examining these interdependent development and evolution activities.
Research on Agent development offers methods for individual concerns such as coordination, tool integration, testing, and operational diagnosis.
Surveys centered on Agent capabilities, applications, or particular engineering concerns provide complementary accounts of these contributions.
A lifecycle-oriented synthesis is needed to relate this dispersed knowledge to the full range of software development activities and compare engineering methods across application domains.
Treating the Agent system itself as the software under development, this survey reviews a main collection of 355 papers across five stages: Requirements Engineering, Architecture and Design, Implementation and Integration, Testing and Assurance, and Maintenance and Evolution.
Within each stage, we compare the engineering problems addressed, explain how the methods and practices work, and examine their reported evaluations and limitations.
We synthesize how existing research supports the corresponding development activities and identify open problems and research directions.
The survey provides researchers and practitioners with a lifecycle-oriented reference for understanding, selecting, and further developing methods for engineering Agent software.

%% file: sections/introduction.tex
\section{Introduction}

Large language model (LLM)-based Agents combine model-generated decisions, tool use, and environmental feedback to perform tasks~\cite{wang2024autonomousagentsurvey}.
Research methods such as ReAct~\cite{yao2022react}, Reflexion~\cite{shinn2023reflexion}, and AutoGen~\cite{wu2023autogen} investigate reasoning--action interleaving, reflective feedback, and programmable multi-Agent conversations as ways to construct and improve these systems.
In software products and services, these capabilities are realized through Agent systems with executable control logic, tool interfaces, and dependencies on their operating environments~\cite{wu2023autogen,hadfield2025research}.
This survey studies the engineering of such systems across their development lifecycle, focusing on the methods and practices used to specify, design, implement, integrate, test, and maintain them.

Developers building Agents for real applications face practical choices about how much autonomy a task needs and how to organize its execution.
Anthropic, the developer of Claude, distilled experience from its own implementations and work with customer teams across industries into its 2024 guide, ``Building Effective Agents''~\cite{anthropic2024agents}.
The guide recommends first testing whether a single LLM call, supported by retrieval and examples, can adequately perform the task.
When additional coordination is needed, known subtasks can follow predefined workflows using chaining, routing, or parallel execution; when tasks unfold through intermediate findings, an orchestrator can determine subtasks dynamically, or an autonomous Agent can select its next action from tool and environmental feedback.
Developers evaluate these choices against task outcomes, latency, and cost, inspect executions, and refine prompts and tool interfaces, with tool-interface testing, sandbox evaluation, human checkpoints, and stopping conditions supporting controlled execution~\cite{anthropic2024agents}.
Delivering the resulting task capability as dependable software requires translating stakeholder needs and task goals into requirements for intended behavior, operating constraints, and quality, and obtaining evidence that the system continues to satisfy those requirements as it changes.
For model-directed Agents, execution paths can vary across runs, failures can span model decisions and tool interactions, and changes to prompts, tools, or execution logic can alter system behavior~\cite{anthropic2024agents,hadfield2025research}.
Developers therefore need to specify and test behavioral boundaries across varying execution paths, diagnose and recover from failures across dependencies, and assess the effects of change.
These needs motivate our software-engineering perspective: how developers establish requirements, satisfy them through design and implementation, and gather evidence that the Agent system's behavior and quality remain acceptable throughout its lifetime.

The development of Claude Research, documented in Anthropic's engineering blog, provides a concrete case of these needs~\cite{hadfield2025research}.
The system uses a lead Agent to delegate searches to parallel subagents and synthesize their findings to answer open-ended questions.
In production, users reported that the system missed readily available information, but the team could not determine whether poor queries, source selection, or tool failures caused the problem.
Adding production tracing enabled the team to locate failures and address their causes~\cite{hadfield2025research}.
In software-engineering terms, the answer-quality problem exposed a diagnostic requirement: developers needed execution evidence to determine which part of the system to change.
The system's long-running, stateful executions also complicated releases, because a software update could encounter Agents midway through their work.
The team kept old and new versions running concurrently and gradually shifted traffic to preserve active executions~\cite{hadfield2025research}.
Maintaining the service during updates therefore involved managing version coexistence and ongoing execution state.
In this system, improving and sustaining task capability depended on software design, diagnostic evidence, and change management.

\begin{figure*}[t]
  \centering
  \includegraphics[width=\textwidth]{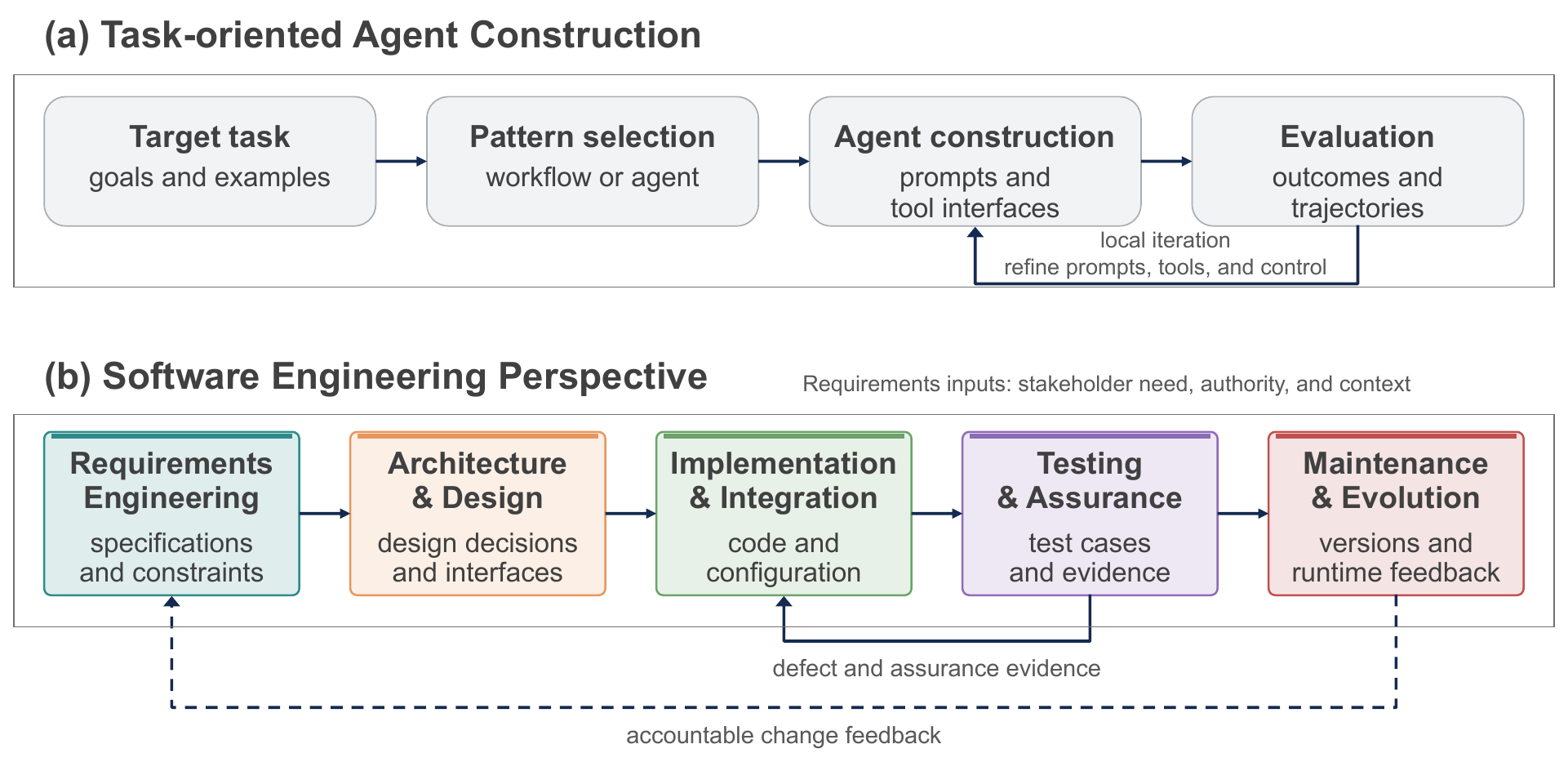}
  \caption{Two perspectives on developing LLM-based Agents. (a) Task-oriented construction follows task definition, pattern selection, implementation, and outcome-driven refinement. (b) The software-engineering perspective organizes development responsibilities across requirements, architecture, implementation, testing, and maintenance. The separate lanes offer two ways to examine development; feedback links schematically illustrate iteration within each view.}
  \Description{Two separate horizontal lanes contrast Agent development perspectives. The upper lane connects target task, pattern selection, Agent construction, and evaluation, with local iteration from evaluation back to construction. The lower lane connects Requirements Engineering, Architecture and Design, Implementation and Integration, Testing and Assurance, and Maintenance and Evolution. Its internal feedback links return from Testing to Implementation and from Maintenance to Requirements. A note identifies stakeholder need, authority, and context as Requirements inputs. No arrows connect the two lanes.}
  \label{fig:agent-engineering-framework}
\end{figure*}

Software engineering organizes such responsibilities through a lifecycle that connects stakeholder needs with the development, operation, and evolution of a software system; its activities can be iterative, concurrent, and incremental~\cite{iso2026softwarelifecycle}.
These responsibilities span development stages: diagnosing a deployed Agent depends on design and implementation choices about execution evidence, while assessing a repair depends on the requirements against which its behavior is checked.
We therefore organize the review into five groups of development activities.
\emph{Requirements engineering} establishes intended behavior, constraints, and acceptance conditions.
\emph{Architecture and design} allocates responsibilities and defines interfaces and quality-related decisions.
\emph{Implementation and integration} realizes those decisions in executable software and connects its dependencies.
\emph{Testing and assurance} investigates whether the system satisfies its requirements and how it fails.
\emph{Maintenance and evolution} diagnoses operational problems and manages changes to deployed software~\cite{washizaki2026swebok}.
Figure~\ref{fig:agent-engineering-framework} contrasts these organizing perspectives: the upper lane follows task definition, construction choices, and outcome-driven refinement; the lower lane follows development responsibilities and the artifacts and evidence they produce.
This lifecycle organization places engineering problems within the activities that establish requirements, implement solutions, evaluate their effects, and manage change, providing a basis for comparing studies across Agent systems.
The survey question is therefore which engineering problems arise at each stage, how existing research addresses them, and what evidence supports the resulting methods.

Existing surveys organize Agent research around different objects and questions.
General surveys synthesize components, capabilities, and applications~\cite{wang2024autonomousagentsurvey}, while Agent-for-SE surveys examine software-development tasks performed \emph{by} Agents~\cite{liu2026llmagentsse,guo2025agenticsesurvey}.
Engineering-oriented surveys address production challenges of foundation-model-based software~\cite{rajbahadur2026productionfmware}, GUI Agents~\cite{yu2026guiengineering}, architecture~\cite{zhou2024agentarchitectureoptions}, validation~\cite{mirto2026agentvalidation}, infrastructure~\cite{li2026harnessengineering}, and operations~\cite{wang2026agentoperations}.
Together they provide foundations for comparing Agent development across application domains and engineering activities.
Our survey brings these concerns into a lifecycle-organized comparison of primary studies, relating each study's development problem to its method, software artifacts, and supporting evidence.
Section~\ref{sec:background} examines the overlaps and distinctions with related surveys.

We study original research in which developing, designing, implementing, testing, or maintaining an LLM-based Agent system is a central concern, across application domains.
For studies that use Agents to write code, elicit requirements, or perform other external tasks, inclusion requires a substantive contribution to engineering the Agent system itself.
Our main analysis examines 355 papers across the five development stages.
Within each stage, we compare engineering problems, methods and practices, software artifacts, evaluation approaches, and reported limitations.
We also analyze explicitly reported connections between development activities.
The aim is to make existing engineering knowledge comparable and identify questions that need further investigation.
Section~\ref{sec:methodology} describes collection and analysis; Section~\ref{sec:collection-results} and Section~\ref{sec:threats} qualify the corpus and evidence.

The resulting synthesis documents methods and evidence across all five stages, but with uneven coverage.
Testing contributes the largest share of the coded engineering contributions, while Requirements contributes the smallest.
Controlled evaluations are common, whereas evidence from authentic engineering work or sustained operation is more limited.
These findings motivate stage-specific research questions and closer examination of the conditions under which engineering methods have been evaluated.

\textbf{Contributions.}
This article makes the following contributions:
\begin{itemize}
  \item A cross-domain map of research on engineering LLM-based Agents, organized across five software-development stages.
  \item A stage-specific synthesis of engineering problems, methods and practices, software artifacts, evaluation approaches, and reported limitations.
  \item An analysis of research coverage, supporting evidence, and reported cross-stage connections that identifies priorities for further research on Agent development.
\end{itemize}

The article proceeds from scope and review methodology, including the collection results, to stage-specific methods, cross-lifecycle findings, and evidence-derived research directions.

%% file: sections/background.tex
\section{Background and Scope}
\label{sec:background}

\subsection{Agents as Software Systems}

An LLM-based Agent is a computational entity that uses a language model as a central decision-making component to select and carry out actions toward a goal~\cite{wang2024autonomousagentsurvey}.
It interacts with an environment, using observations and action outcomes to inform subsequent decisions~\cite{yao2022react,anthropic2024agents}.
In software engineering, a system consists of interacting elements organized to achieve defined purposes~\cite{washizaki2026swebok}; a software system is one in which software is central to meeting stakeholders' requirements~\cite{iso2026softwarelifecycle}.
An implemented Agent brings together such elements: model inference is coupled with executable logic that manages its actions, retains execution state, and communicates with tools or external services~\cite{wu2023autogen,hadfield2025research}.
Its task goals give the system a purpose, its programs and configurations realize its behavior, and its interfaces connect that behavior to the environment.
We therefore treat the integrated Agent implementation as a software system whose behavior emerges from the interaction of these elements.
Model-directed action selection introduces variability into execution, while choices about interfaces, state handling, and control logic affect system behavior and quality~\cite{hadfield2025research}.
This system-level view makes the specification, construction, evaluation, and evolution of the Agent itself the subject of software engineering.

To study Agents from this perspective, we take the integrated implementation of one or more Agents as the engineering object and examine the activities through which developers establish and sustain its behavior and quality.
The development lifecycle organizes our analysis around the system's requirements, design decisions, implementation and integration, testing and assurance, and maintenance and evolution.
For each primary study, we identify the author-stated development problem, the proposed method or result, the software artifacts involved, and the evidence used to assess the contribution.
We place the contribution in the development activity it addresses and compare how studies approach that activity, what they produce, and what their evaluations establish.
The unit of synthesis is thus a source-grounded \emph{contribution to Agent development and its evidence}.
Such a contribution may be a method, system, benchmark, empirical study, experience report, or other original result.
Inclusion follows the scope criteria in Section~\ref{sec:methodology}; reusability is recorded during subsequent content analysis.

\subsection{Software Engineering for Agents}

Building on this view of Agents as software systems, we use \emph{software engineering for Agents} to describe research on the methods, practices, and evidence for developing and sustaining LLM-based Agent software.
Our concern is how developers specify the Agent software's intended behavior and quality, realize and evaluate these in an implementation, and sustain them as the system changes.
The Agent's development lifecycle provides the organizing framework for these activities, spanning requirements, architecture and design, implementation and integration, testing and assurance, and maintenance and evolution.
We focus on studies in which a problem within these development activities is central to the research contribution, and compare the approaches taken and the evidence supporting their findings.
This scope applies across application domains, including Agents used in software development, scientific research, and other services.
It is worth noting that many studies on \emph{Agents for software engineering} investigate how Agents perform SE tasks, such as code generation~\cite{hong2023metagpt}, repository-level issue resolution~\cite{yang2024sweagent}, and intent-aware code summarization~\cite{ge2026draftprecision}.
These studies focus on Agent-assisted SE tasks and the software artifacts they produce.
That topic is outside the scope of this survey, which examines the development and maintenance of the Agent software itself.

\subsection{Agent Development Lifecycle}

An Agent development lifecycle organizes how an Agent is conceived, constructed, evaluated, deployed, and maintained over time.
Our examination of public engineering guidance from Anthropic, Microsoft, and LangChain identifies a recurring starting point: choose a use case, construct an Agent that can address it, and improve its behavior through evaluation and feedback.
Anthropic's ``Building Effective Agents'' recommends selecting simple construction patterns according to task needs and adding complexity when it improves outcomes~\cite{anthropic2024agents}.
Microsoft organizes development around discovery, experimentation, build, deployment, and ongoing operation~\cite{microsoft2026adlc}; LangChain connects building, testing, deployment, and monitoring in a feedback loop~\cite{chase2026adlc}.
We characterize this shared emphasis as task- and delivery-oriented: the Agent is approached through the functionality it provides and the process of making that functionality work in practice.
At the level of lifecycle organization, requirements formulation, architectural design, and implementation and integration are largely embedded within discovery or build, while testing and monitoring support readiness assessment and subsequent improvement.
This organization leaves the engineering progression from a task-solving implementation to a maintainable software system less explicit.
In particular, it leaves open how task goals are developed into stakeholder requirements and quality constraints; how these requirements guide architectural choices and implementation and integration decisions; how testing provides system-level assurance against those requirements across components and interfaces; and how operation, fault resolution, and changes to models, tools, or requirements are managed as maintenance and evolution activities.
These distinctions matter because the requirements, design rationale, implementation, and quality evidence must remain consistent as the Agent and its environment change.
To make this progression and its associated engineering obligations explicit, we adopt an SE lifecycle perspective: we treat the Agent as the software system being developed and examine how its requirements are established, realized, assessed, and sustained throughout its life.

In software engineering, the software lifecycle describes how a system evolves from its initial conception through development, operation, maintenance, and eventual retirement~\cite{iso2026softwarelifecycle,washizaki2026swebok}.
It provides a framework for coordinating the work needed to turn stakeholders' requirements into usable software and to sustain its quality as those requirements and its operating environment change.
Within this lifecycle, requirements engineering establishes what the software should do and the constraints it must satisfy; architecture and design determine how its responsibilities, components, and interfaces are organized; implementation and integration realize these decisions in a working system; testing and assurance examine whether the system meets its requirements and provide evidence about its quality; and maintenance and evolution address faults, changing needs, and changes in the environment~\cite{washizaki2026swebok}.
A lifecycle model organizes these activities, their work products, and their relationships.
Iterative and incremental models revisit activities as understanding improves and software functionality develops; lifecycle processes can also proceed concurrently~\cite{iso2026softwarelifecycle,washizaki2026swebok}.
This organization helps developers coordinate decisions, assess progress and quality, and manage changes throughout the system's life.

We apply this lifecycle framework to the integrated Agent system as the software being developed.
This entails specifying its intended behavior and autonomy constraints, designing its control and interfaces, implementing and integrating its models and tools, assessing its quality and failure behavior, and managing changes to the system and its operating environment.
These activities provide the basis for our five groups: requirements engineering (RE), architecture and design (A\&D), implementation and integration (I\&I), testing and assurance (T\&A), and maintenance and evolution (M\&E).
We place each study's contribution in the activity whose engineering problem it addresses, then synthesize the proposed methods, the artifacts produced or consumed, and the evaluation evidence.
We also examine reported connections between activities, including how evaluation and operational feedback inform subsequent revisions.
This application of the lifecycle provides a common basis for comparing how Agent software is developed and sustained across different architectures and application domains.

\subsection{Positioning Against Existing Surveys}

To establish how this survey relates to existing syntheses, we consider publicly available surveys and evidence syntheses relevant to LLM-based Agent research and the engineering of AI-based software.
A work is eligible when it provides a substantive synthesis of multiple prior studies, methods, or engineering projects; addresses LLM-based Agents or software-engineering knowledge relevant to their development; and provides sufficient scope and synthesis detail to compare its research object and analytical perspective.
We apply these criteria to all works identified within the search scope and cutoff.
Both formally published works and public preprints are considered.
Thematic groupings and development-stage coverage are established through subsequent analysis.
Table~\ref{tab:related-survey-landscape} reports a subset of eligible works with detailed full-text comparisons.

We compare these works along three dimensions in Table~\ref{tab:related-survey-landscape}: \emph{research object}, \emph{synthesis focus}, and \emph{development-stage coverage}.
The first identifies the systems studied and whether the Agent software itself is an engineering target; the second identifies what the review organizes and compares, such as capabilities, architectural decisions, or validation methods; the third examines substantive treatment of RE, A\&D, I\&I, T\&A, and M\&E.
We assess coverage from each work's stated scope, taxonomy, and synthesis sections, looking for analysis of methods, decisions, artifacts, or evaluation evidence relevant to the activity.
Dedicated synthesis of an activity within the review's Agent-system scope receives a filled circle; substantive but embedded treatment, or treatment through a broader software class or subsystem, receives a half-filled circle; an empty circle indicates no substantive treatment identified.
Coverage is grounded in the synthesis of existing work; component names, runtime adaptation, and future research proposals alone do not establish development-stage coverage.
For Agent-for-SE work, a dash marks stages organized around tasks performed by Agents.
An architecture or operations review can fully satisfy the engineering-object criterion while concentrating on a single development activity.
The symbols describe the scope and focus of the synthesis, independently of the maturity of the underlying research.

\begin{table*}[t]
  \centering
  \caption{Full-text comparison of related surveys and evidence syntheses by research object, synthesis focus, and coverage of Agent software development activities.}
  \label{tab:related-survey-landscape}
  \scriptsize
  \setlength{\tabcolsep}{2.5pt}
  \renewcommand{\arraystretch}{1.16}
  \begin{tabular}{L{0.145\textwidth}L{0.145\textwidth}L{0.205\textwidth}C{0.065\textwidth}C{0.043\textwidth}C{0.043\textwidth}C{0.043\textwidth}C{0.043\textwidth}C{0.043\textwidth}}
    \toprule
    Study & Scope / object & Main synthesis target & \shortstack{Agents\\as software} & RE & A\&D & I\&I & T\&A & M\&E \\
    \midrule
    Wang et al.~\cite{wang2024autonomousagentsurvey} & General LLM Agents & Components, capabilities, applications & \nocoverage & \nocoverage & \partialcoverage & \nocoverage & \partialcoverage & \nocoverage \\
    Kang and Li~\cite{kang2026lifecycleagents} & General LLM Agents & Operational capability lifecycle & \nocoverage & \nocoverage & \partialcoverage & \partialcoverage & \partialcoverage & \nocoverage \\
    Liu et al.~\cite{liu2026llmagentsse} & Agents used for SE tasks & SE tasks and Agent mechanisms & \nocoverage & -- & -- & -- & -- & -- \\
    Mart\'inez-Fern\'andez et al.~\cite{martinezfernandez2022se4ai} & AI-based systems & SE knowledge areas and methods & \nocoverage & \partialcoverage & \partialcoverage & \partialcoverage & \partialcoverage & \partialcoverage \\
    Rajbahadur et al.~\cite{rajbahadur2026productionfmware} & Prompt- and Agent-based software & Demo-to-production challenges & \partialcoverage & \nocoverage & \partialcoverage & \partialcoverage & \partialcoverage & \partialcoverage \\
    Yu et al.~\cite{yu2026guiengineering} & GUI Agents & Architecture, evaluation, and SE concerns & \fullcoverage & \partialcoverage & \fullcoverage & \partialcoverage & \fullcoverage & \fullcoverage \\
    Li et al.~\cite{li2026harnessengineering} & Agent harness & Harness layers and mechanisms & \partialcoverage & \nocoverage & \partialcoverage & \partialcoverage & \partialcoverage & \partialcoverage \\
    Wang et al.~\cite{wang2026agentoperations} & Agent operations & Observability, failures, and governance & \fullcoverage & \nocoverage & \nocoverage & \partialcoverage & \partialcoverage & \fullcoverage \\
    Zhou et al.~\cite{zhou2024agentarchitectureoptions} & LLM-Agent architecture & Architectural decisions and options & \fullcoverage & \nocoverage & \fullcoverage & \partialcoverage & \nocoverage & \nocoverage \\
    Mirto et al.~\cite{mirto2026agentvalidation} & Agentic systems & Validation methods and dimensions & \fullcoverage & \nocoverage & \nocoverage & \nocoverage & \fullcoverage & \partialcoverage \\
    
    Dantas et al.~\cite{dantas2026safeagents} & Agent safety & Specification, verification, enforcement & \fullcoverage & \partialcoverage & \partialcoverage & \partialcoverage & \fullcoverage & \nocoverage \\
    
    Yehudai et al.~\cite{yehudai2026agentevaluation} & Agent evaluation & Benchmarks and developer evaluation & \fullcoverage & \nocoverage & \nocoverage & \nocoverage & \fullcoverage & \partialcoverage \\
    
    Gan et al.~\cite{gan2024agentrisks}\textsuperscript{a} & Agent security and privacy & Threats, defenses, and system cases & \fullcoverage & \nocoverage & \partialcoverage & \partialcoverage & \fullcoverage & \nocoverage \\

    Miranda et al.~\cite{miranda2026mdeagents}\textsuperscript{c} & Multi-Agent frameworks & MDE4AI, agent frameworks and metamodel & \fullcoverage & \nocoverage & \fullcoverage & \partialcoverage & \partialcoverage & \nocoverage \\
    
    He et al.~\cite{he2026harnessruntime}\textsuperscript{d} & Agent harness & Control, agency, runtime, harness engineering & \partialcoverage & \partialcoverage & \partialcoverage & \partialcoverage & \partialcoverage & \partialcoverage \\
    \midrule
    \textbf{This survey} & \textbf{Cross-domain Agent systems} & \textbf{Development methods and evidence} & \fullcoverage & \fullcoverage & \fullcoverage & \fullcoverage & \fullcoverage & \fullcoverage \\
    \bottomrule
  \end{tabular}
  \vspace{2pt}

  \begin{minipage}{0.98\textwidth}
    \footnotesize
    \emph{Object:} \fullcoverage\ Agent system is the engineering target; \partialcoverage\ broader software class or surrounding subsystem; \nocoverage\ not the engineering target.\\
    \emph{Stages:} \fullcoverage\ dedicated synthesis within the study's Agent-system scope; \partialcoverage\ embedded, broader-object, or subsystem treatment; \nocoverage\ no substantive treatment identified; -- opposite engineering direction.\\
    \textsuperscript{a}Coverage assessed from arXiv v1; alignment with the TOSEM article~\cite{gan2026agentrisks} is pending.
    \textsuperscript{b}SoK technical report synthesizing research reviews and practitioner evidence.
    \textsuperscript{c}Framework review and original metamodel.
    \textsuperscript{d}Selective evidence synthesis and reporting proposal.
  \end{minipage}
\end{table*}

In requirements engineering, Table~\ref{tab:related-survey-landscape} identifies coverage at different levels of specificity.
The SE4AI survey compares requirements processes, specification approaches, and non-functional requirements for AI-based systems~\cite{martinezfernandez2022se4ai}; its method-level synthesis concerns a broader software class and predates contemporary LLM-based Agents.
The GUI-Agent survey analyzes capability and autonomy requirement signals and identifies the operating boundaries needed to make them testable~\cite{yu2026guiengineering}.
Its analysis characterizes requirements expressed in GUI-Agent research and diagnoses specification weaknesses, while offering limited comparison of methods for eliciting, negotiating, and revising those requirements.
Thus, the compared works provide either broader SE methods or Agent-specific requirement characterizations; a cross-domain synthesis of how developers obtain, formalize, and manage requirements for LLM-based Agent software remains missing.

Architecture and design receive more explicit engineering treatment.
Zhou et al. compare architectural options and quality tradeoffs through a decision model~\cite{zhou2024agentarchitectureoptions}; Miranda et al. synthesize construction and communication concepts into a framework metamodel~\cite{miranda2026mdeagents}; and Yu et al. analyze responsibilities, interfaces, and recovery structures for GUI Agents~\cite{yu2026guiengineering}.
These contributions reach design choices and structural representations, with scopes centered on architectural options, framework abstractions, and a particular class of interactive systems, respectively.
Less systematically compared are the development methods that derive, assess, or revise architectural artifacts: their required inputs, the decisions they support, and the evidence for the resulting system qualities.
This leaves a need to organize Agent architecture research around the work of designing software, alongside its catalogues of possible structures.

Implementation and integration are addressed through framework construction mechanisms~\cite{miranda2026mdeagents}, harness execution environments and tool interfaces~\cite{li2026harnessengineering}, and FMware grounding, orchestration, and production engineering~\cite{rajbahadur2026productionfmware}.
Their depth differs: the framework review chiefly reconciles development concepts; harness engineering examines concrete mechanisms within the model's execution infrastructure; and FMware engineering analyzes production challenges across both prompt-based and Agent-based software.
Together they explain important building blocks and implementation concerns, but the compared works do not provide a dedicated, cross-domain synthesis of Agent software construction and integration methods.
In particular, methods for translating specifications into executable configurations, adapting interfaces, and diagnosing integration defects remain distributed across these different organizing scopes.
A common engineering comparison needs to establish what each method constructs or changes, its environment assumptions, and the evidence for correctness, portability, or integration effort.

Testing and assurance have comparatively detailed, dedicated syntheses.
Yehudai et al. compare benchmarks, evaluation dimensions, and developer frameworks, including stepwise and trajectory assessment~\cite{yehudai2026agentevaluation}; Mirto et al. organize validation methods and evidence around behavioral, safety, temporal, regulatory, and multi-Agent properties~\cite{mirto2026agentvalidation}.
The GUI-Agent survey adds domain-specific analysis of test conditions, action failures, and recovery~\cite{yu2026guiengineering}.
Their coverage extends to assessment mechanisms, supported properties, and limitations of the resulting evidence.
The remaining lifecycle-level synthesis problem is to place this depth of assurance analysis alongside equally explicit accounts of requirements, design, and implementation methods.
The evaluation and validation reviews organize the system primarily as an object to be assessed; they leave the engineering methods that produce its requirements, architecture, and implementation outside their main synthesis.

Maintenance and evolution are synthesized most directly through the operations survey's monitoring, anomaly detection, root-cause localization, and resolution methods~\cite{wang2026agentoperations}.
Its operational depth includes failure taxonomies, diagnostic benchmarks, rollback, and persistent prompt or policy revision, with method comparisons organized around failure handling.
Other reviews examine interface evolution and maintainability within GUI systems~\cite{yu2026guiengineering}, regression and deployment feedback within harnesses~\cite{li2026harnessengineering}, and evidence renewal within validation~\cite{mirto2026agentvalidation}.
FMware engineering covers deployment feedback and updates within the broader production pipeline~\cite{rajbahadur2026productionfmware}.
Coverage therefore concentrates on operational reliability or on changes to particular system layers and evidence artifacts.
What remains insufficiently consolidated is a cross-domain comparison of Agent software change methods that distinguishes fault repair, adaptation to changed requirements or interfaces, and version governance, and examines both the durable artifacts modified and the evidence that a change preserves required behavior.

This comparison exposes two connected limitations in the existing survey literature.
First, development-stage coverage and engineering depth are uneven: dedicated syntheses of architecture, assurance, and operations coexist with embedded or broader-object treatment of requirements and implementation, leaving parts of Agent development without a systematic comparison of methods and supporting evidence.
Second, the question of how to build Agents is organized around different analytical units: components and capabilities in general Agent surveys~\cite{wang2024autonomousagentsurvey,kang2026lifecycleagents}, or particular frameworks, infrastructure layers, and engineering concerns in more focused reviews.
The SE4AI and GUI-Agent surveys demonstrate lifecycle-oriented SE analysis within broader AI software and GUI-specific boundaries, respectively~\cite{martinezfernandez2022se4ai,yu2026guiengineering}.
Across the compared works, a cross-domain account of LLM-based Agent development that combines coverage of all five stages with method-level SE analysis remains missing.
We address this gap by treating the Agent as the software under development and organizing primary studies by the engineering problem, method or practice, software artifacts, and evaluation evidence within each stage.
This yields a lifecycle-wide account of how Agent software is specified, designed, implemented, assured, and maintained, while identifying where existing methods or their evidence remain limited.

%% file: sections/methodology.tex
\section{Survey Methodology}
\label{sec:methodology}

We review research on the engineering of LLM-based Agent systems across the five development stages introduced in Section~\ref{sec:background}.
The review connects three activities: retrieving candidate studies, selecting studies within this scope, and extracting evidence for lifecycle-oriented comparison.
The collection protocol follows established software-engineering review and snowballing practices~\cite{kitchenham2007guidelines,wohlin2014snowballing,chen2025codelmsecurity}.
This section presents the research questions, the procedures used to build and analyze the corpus, and the resulting collection.

\subsection{Research Questions}

The analysis is organized around five research questions, one for each development stage:
\begin{itemize}
  \item \textbf{RQ1 (Requirements engineering):} Which problems and approaches are investigated in requirements engineering for Agent software?
  \item \textbf{RQ2 (Architecture and design):} Which problems and approaches are investigated in the architecture and design of Agent software?
  \item \textbf{RQ3 (Implementation and integration):} Which problems and approaches are investigated in the implementation and integration of Agent software?
  \item \textbf{RQ4 (Testing and assurance):} Which problems and approaches are investigated in the testing and assurance of Agent software?
  \item \textbf{RQ5 (Maintenance and evolution):} Which problems and approaches are investigated in the maintenance and evolution of Agent software?
\end{itemize}
Sections~\ref{sec:req-methods}--\ref{sec:maint-methods} answer RQ1--RQ5 in order.
For each question, we compare the engineering problems studied, the methods and practices used to address them, and the reported evaluation results, then synthesize their support for the development activity and the remaining research needs.
These questions guide the extraction and synthesis described below.

\subsection{Literature Search}

To retrieve research on Agent software development across the five stages, the search combines Agent-related expressions with lifecycle-specific engineering terms.
The vocabulary was developed iteratively from prior reading of Agent research and lifecycle-focused pilot studies.
To account for different ways of describing Agents and Agent-based systems, the Agent block includes \emph{LLM agent}, \emph{language-model agent}, \emph{foundation-model agent}, \emph{AI agent}, \emph{agent system}, \emph{agentic system}, \emph{agent-powered}, and \emph{agent-enabled}.
The lifecycle blocks cover Requirements, Architecture and Design, Implementation and Integration, Testing and Assurance, and Maintenance and Evolution.
Table~\ref{tab:lifecycle-search-concepts} lists representative concepts used to broaden these blocks.
During search development, we checked whether the keyword queries could retrieve papers already identified through prior reading and pilot searches, using missing matches to examine gaps in retrieval coverage.
Appendix~\ref{sec:known-paper-coverage} identifies the reference sets used for these checks, distinguishes their uses from citation-expansion seeds, and reports the corresponding query versions and coverage results.

\begin{table}[H]
  \centering
  \caption{Lifecycle concepts used to construct search queries and broaden retrieval coverage.}
  \label{tab:lifecycle-search-concepts}
  \small
  \renewcommand{\arraystretch}{1.12}
  \begin{tabular}{L{0.20\textwidth}L{0.70\textwidth}}
    \toprule
    Entry point & Representative search concepts \\
    \midrule
    Requirements & elicitation, modeling, analysis, negotiation, specification, goals, capabilities, constraints, permissions, human-control boundaries, acceptance criteria \\
    Architecture and design & architectural concerns, decomposition, responsibilities, interfaces, contracts, interaction topology, reference architectures, design patterns \\
    Implementation and integration & construction, configuration, debugging, frameworks, SDKs, harnesses, protocols, orchestration, tool and environment integration \\
    Testing and assurance & test models, generation, oracles, coverage, benchmarks, verification, validation, reliability, safety, failure diagnosis, regression \\
    Maintenance and evolution & deployment, observability, monitoring, incident analysis, repair, rollback, impact analysis, version management, controlled adaptation, re-validation \\
    \bottomrule
  \end{tabular}
\end{table}

The retrieval sources comprise OpenAlex, arXiv, DBLP, and Semantic Scholar exports.
The search window starts in 2022 and ends at the corpus cutoff.
We use 2022 as the starting year to cover early LLM-centered systems that connect language models with external tools and environments.
Examples first released that year include MRKL's modular integration of language models and external capabilities~\cite{karpas2022mrkl} and ReAct's interleaving of reasoning and actions~\cite{yao2022react}.
To complement keyword retrieval, backward searches inspect the references of included studies and directly relevant secondary sources, while forward searches identify subsequent papers citing them.
Collection stopped at a dated snapshot after diminishing marginal yield.
The screened cohort contains selected citation-expanded candidates and reconciled prior-pool records; it does not cover every retrieved citation candidate.
A fully adjudicated zero-addition citation round has not been established.
The resulting coverage boundary and the unresolved access cases are retained when interpreting the corpus.
Source-specific query strings, retrieval dates, pagination logs, and source--candidate citation links are retained with the replication materials in Appendix~\ref{sec:replication-snapshot}.

\subsection{Study Selection}

Study selection determines which retrieved papers enter the review by applying the scope defined in Section~\ref{sec:background}.
An eligible primary study satisfies five conditions:
\begin{enumerate}
  \item Its research object is an LLM- or foundation-model-based Agent or Agent system with an observation or feedback--decision--action loop.
  \item The engineering activity acts on the Agent system itself.
  \item Its central research question concerns at least one of the five software-development stages.
  \item The lifecycle concern makes a substantive contribution to the paper's central research problem.
  \item The stage can be identified from the authors' stated problem and contribution.
\end{enumerate}
A counterfactual question helps resolve boundary cases: after removing the external task, application domain, or target cognitive capability, do the main problem and conclusions still concern how Agent software is developed, designed, implemented, tested, assured, maintained, or evolved?
This distinguishes Agent engineering contributions from papers whose primary contribution is using an Agent for an external task or improving a runtime capability.
Component diagrams and routine experimental evaluations are considered in relation to the central contribution.

The selection unit is a version family, represented by its latest eligible version after cross-source and version reconciliation.
Records are merged using matching DOI, arXiv, or source identifiers, or compatible normalized titles, authors, and years; conflicting title matches remain separate for review.
Preprocessing also records year, publication type, publication evidence, and full-text availability.
Surveys, mappings, position papers, standards, and engineering reports can support framing and discovery as secondary or contextual sources.
Records without accessible full text are excluded on access grounds and counted separately from topical exclusions.

Screening comprised an initial relevance assessment followed by stricter scope confirmation.
The first two authors independently reviewed the candidates under the same eligibility criteria, without consulting one another's initial decisions.
They examined titles, abstracts, introductions, and bibliographic metadata, consulted the full text as needed, and recorded decisions with supporting passages and page or section locators.
The corresponding author resolved disagreements and rechecked high-risk decisions.
Uncertain cases, vertical-domain studies, and papers with an unclear central engineering contribution received closer examination of the relevant method, system, or conclusion sections.
The stricter pass examined whether the paper's central problem and contribution satisfied the scope conditions, using the counterfactual above to resolve boundary cases.
The authors confirmed the resulting inclusion and exclusion decisions against the recorded source evidence.

\begin{figure*}[t]
  \centering
  \includegraphics[width=\textwidth]{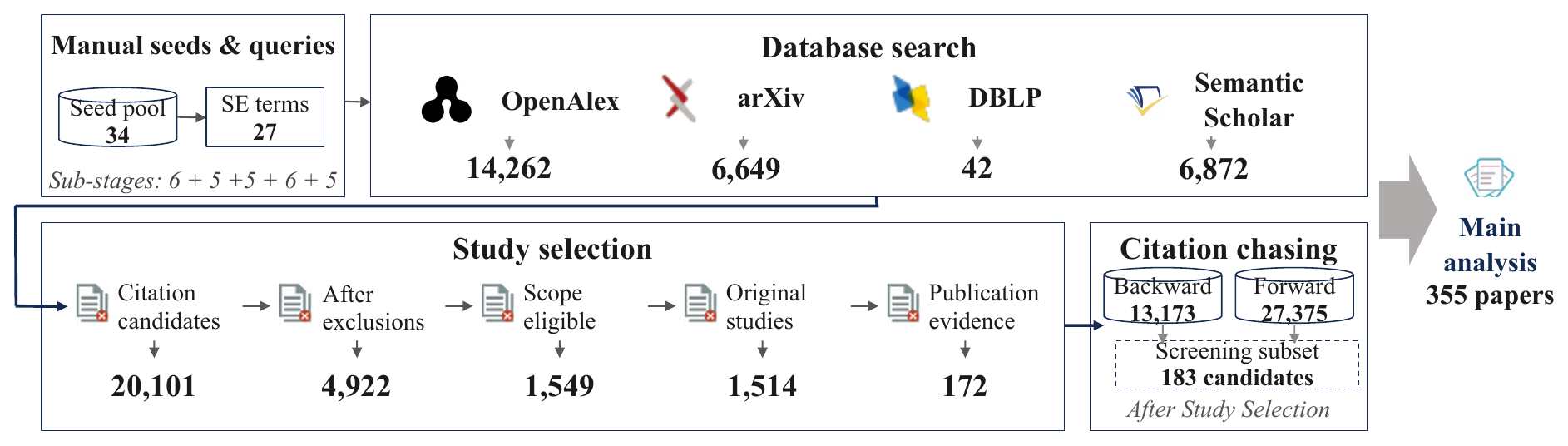}
  \caption{Working overview of search and corpus selection. Database values count within-source records; backward and forward values count citation links. The 20,101 citation candidates are net additions after removing overlap with saved prior pools. Screening covers a selected cohort: the 4,922, 1,549, and 1,514 values describe retained subsets of 5,354 records. The final 355 P-tier papers comprise 172 already in prior pools and 183 citation-only additions. Retrieval, screening, and final-paper provenance use these distinct populations.}
  \Description{The author-maintained diagram shows manual seeds and query terms, four database sources, a study-selection band, a citation-chasing block, and 355 main-analysis papers. The database counts are 14,262, 6,649, 42, and 6,872. The lower band shows 20,101 citation candidates, 4,922 after exclusions, 1,549 scope-eligible records, 1,514 original studies, and 172 prior-pool papers. Citation chasing shows 13,173 backward links, 27,375 forward links, and 183 citation-only papers. Units and overlapping populations are qualified in the caption.}
  \label{fig:search-screening-flow}
\end{figure*}

Figure~\ref{fig:search-screening-flow} summarizes the search and study-selection process, distinguishing retrieved candidates, the screened cohort, and the final analysis set.
The reconciled screening cohort contains 5,354 records; excluding 432 records without accessible full text leaves 4,922 readable records.
Scope screening excludes 3,373 records and retains 1,549, comprising 35 Secondary sources and 1,514 scope-eligible Primary records.
Publication evidence separates the Primary records into 355 P records with captured publication or public acceptance evidence, 1,043 E records with a preprint but no qualifying publication evidence captured at the cutoff, and 116 U records with unresolved publication status.
The main synthesis uses the fixed 355-paper P selection; a separate 150-paper supplement probes coverage outside that selection.
The 355 main-analysis papers comprise 172 from prior pools and 183 identified only through citation expansion.
P includes both index/journal-reference evidence and public acceptance claims, and its publication coverage is incomplete.
Section~\ref{sec:collection-results} summarizes the resulting collection, while Appendix~\ref{sec:corpus-amendments} preserves the preceding screening counts and subsequent identity and scope amendments.

\subsection{Data Extraction and Synthesis}
\label{sec:evidence-coding}

After study selection establishes the corpus, full-text analysis examines how the included studies contribute to the development of Agent software.
We use a common extraction form to connect each paper's research problem, engineering approach, and reported findings, following structured extraction and qualitative comparison in software-engineering reviews~\cite{martinezfernandez2022se4ai,chen2025codelmsecurity}.
For each study, we record the engineering problem and the part of the Agent system it concerns; the proposed method, practice, or infrastructure; and the inputs, outputs, and software artifacts involved.
We also record how the approach is evaluated, the setting and outcomes of that evaluation, and the assumptions and limitations affecting its interpretation.
Bibliographic information supports the description of the research landscape, while explicit uses of an artifact, decision, or feedback item in another development activity support the analysis of cross-stage connections.

The five lifecycle stages provide the top-level organization for synthesizing these observations.
Within each stage, we compare contributions addressing related engineering problems, examining similarities and differences in their methods, the artifacts they produce or consume, and the evidence supporting their reported effects.
A paper can contribute to several stages; we map each distinct engineering contribution from its stated purpose and supporting text so that these different roles remain visible in the comparison.
We combine this qualitative analysis with paper-level distributions to describe which development activities have attracted research attention.
Evaluation comparisons examine the conditions under which approaches have been studied, while cross-stage comparisons trace how outputs from one activity inform another.
These comparisons organize the stage-by-stage discussion of approaches, reported limitations, and research needs.

The first two authors independently checked the extracted information and its interpretation against the full texts, including stage assignments, engineering contributions, evaluation evidence, and cross-stage connections.
The corresponding author reconciled disagreements by returning to the relevant source passages.
Supporting quotations and page or section locators link each comparison to its source evidence.
This review covers both the 355-paper main analysis and the separately reported 150-paper supplement.
The replication materials retain the extraction records; Appendix~\ref{sec:analysis-definitions} defines the evaluation and relationship categories used in the statistics.

%% file: sections/landscape.tex
\subsection{Collection Results}
\label{sec:collection-results}

The following results describe the publication profile and lifecycle coverage of the 355-paper main collection.
The 150-paper supplement is analyzed separately in Appendix~\ref{sec:supplement-stage-sensitivity}.

\noindent\textit{Publication Profile.}
The collection is concentrated in recent work: 326 of the 355 papers (91.8\%) are dated 2025 or 2026 in the bibliographic metadata.
The recorded years comprise one paper in 2023, 28 in 2024, 127 in 2025, and 199 in 2026.
These dates follow the retrieved bibliographic records, which can precede formal publication; the 2026 count covers the period up to the search cutoff.
Of the 355 papers, 201 have index or journal-reference evidence of publication, and 154 have public acceptance claims.
This fixed selection reflects the publication evidence captured during collection; incomplete publication metadata and the separately examined supplement are documented in Appendix~\ref{sec:identity-amendments}.

\begin{figure*}[t]
  \centering
  \includegraphics[width=\textwidth]{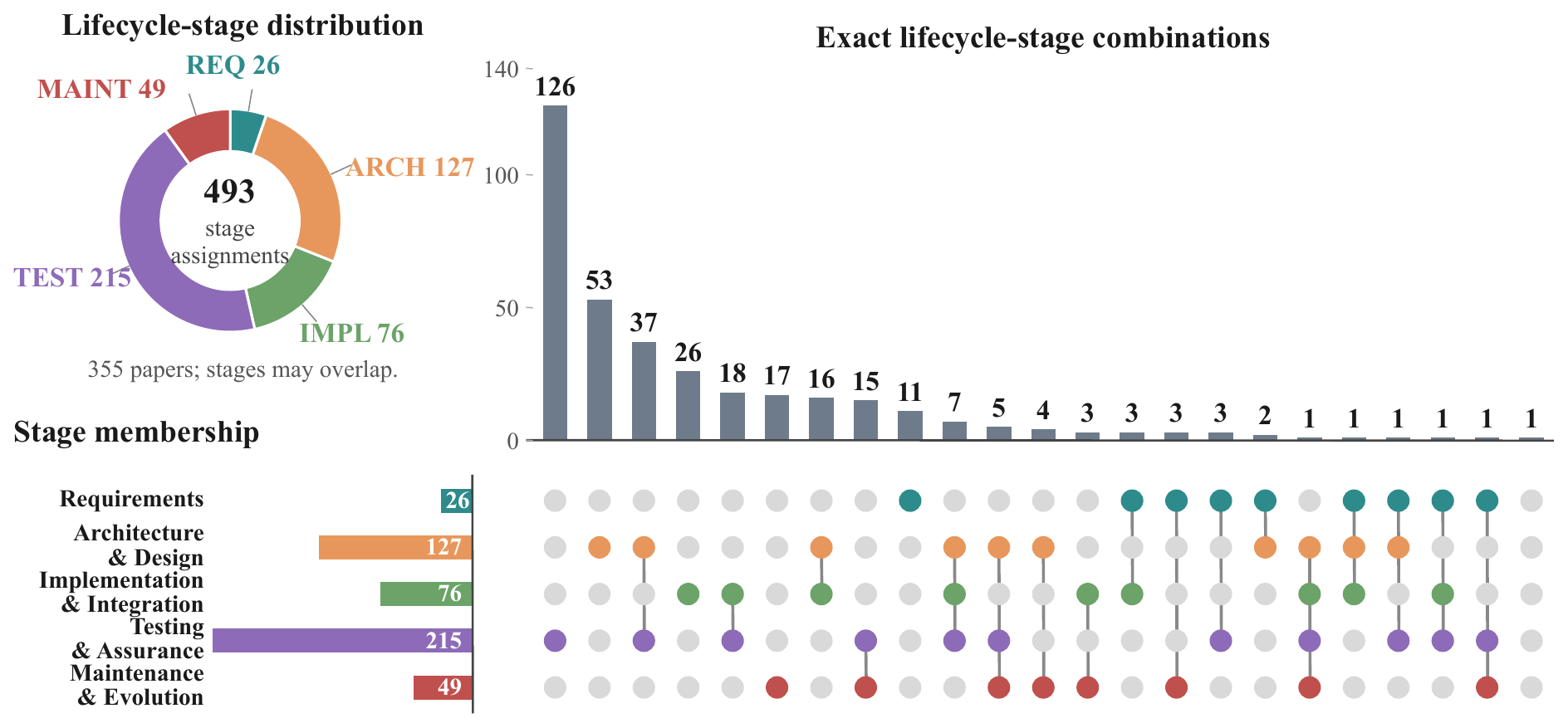}
  \caption{Lifecycle coverage of the 355-paper main collection. The ring and horizontal bars summarize 493 non-exclusive paper--stage memberships. The vertical bars and dot matrix show exact stage combinations, counting each paper once. The all-empty combination denotes one audit-only paper with no five-stage assignment.}
  \Description{The ring and horizontal bars show 26 Requirements, 127 Architecture, 76 Implementation, 215 Testing, and 49 Maintenance paper memberships, totaling 493. Twenty-three columns show exact stage combinations among 355 papers. There are 233 single-stage papers, 121 multi-stage papers, and one audit-only paper. Testing-only is the largest combination with 126 papers, followed by Architecture-only with 53. One paper covers four stages and none covers all five.}
  \label{fig:stable-stage-distribution}
  \label{fig:stable-stage-coverage}
\end{figure*}

\noindent\textit{Lifecycle Coverage.}
Figure~\ref{fig:stable-stage-distribution} summarizes the stages addressed by the included papers.
Testing and Assurance appears in 215 papers, followed by Architecture and Design in 127, Implementation and Integration in 76, Maintenance and Evolution in 49, and Requirements Engineering in 26.
A paper belongs to a stage when at least one of its identified engineering contributions addresses that stage.
As papers can address multiple stages, these counts total 493 paper--stage memberships across the 355-paper collection.
The distribution places testing and architectural questions at the center of much of the collected research, with fewer papers addressing requirements and maintenance.

The stage combinations in Figure~\ref{fig:stable-stage-coverage} show that 233 papers (65.6\%) contribute to exactly one stage, 121 (34.1\%) span at least two stages, and one (0.3\%) provides only cross-cutting evidence.
The largest single-stage patterns are Testing (126 papers), Architecture (53), Implementation (26), Maintenance (17), and Requirements (11).
One paper covers four stages, and none covers all five.
These combinations describe the focus of the papers' engineering contributions; routine development activities can extend beyond that focus.
Section~\ref{sec:lifecycle} examines the engineering problems, methods, and evidence within each stage.

%% file: sections/lifecycle.tex
\FloatBarrier
\section{Engineering Agents Across the Software Lifecycle}
\label{sec:lifecycle}

This section answers RQ1--RQ5 through five stage-specific reviews of Agent software development: Requirements Engineering, Architecture and Design, Implementation and Integration, Testing and Assurance, and Maintenance and Evolution, respectively.
For each stage, we identify the engineering problems studied, compare the methods and practices used to address them, and examine their artifacts and evaluation evidence.
The stage discussions explain the development responsibility, why it matters for Agent software, and how the reviewed approaches address it.
All nine comparison tables use the same problem--approach--key idea--representative study structure; years identify the cited bibliographic records.
The main analysis draws on 570 lifecycle contributions from the 355-paper collection described in Section~\ref{sec:collection-results}; one additional cross-cutting contribution is retained for audit only.
Separately identified supplemental studies extend the source comparisons while retaining their own denominator.

The five stages are parallel analytical perspectives on development responsibilities.
Their presentation follows a familiar development sequence, while the reviewed methods can be used iteratively and a paper can contribute to several stages.
Each stage concludes with a synthesis of its methods, supporting evidence, and open problems; the final subsection brings these observations together to inform the implications and research directions in Section~\ref{sec:discussion}.

\FloatBarrier
\subsection{Requirements Engineering}
\label{sec:req-methods}

Requirements Engineering establishes the needs a software system should serve and the behavior and constraints required to meet them.
Through elicitation, analysis, specification, validation, and change management, it provides an agreed basis for development and for assessing whether the resulting system meets stakeholder needs~\cite{nuseibeh2000requirements}.
For Agent software, this means clarifying the intended service, the authority delegated to the Agent, and the conditions for accepting its behavior.
Making these requirements explicit allows stakeholders and developers to examine decisions that would otherwise be left to the interpretation of a task request during implementation.
Table~\ref{tab:req-method-comparison} organizes the reviewed approaches around three connected problems: identifying needs, specifying acceptable and authorized behavior, and expressing delegation and governance requirements.

\subsubsection{Acquiring and Revising Requirements}
Requirements acquisition, commonly termed elicitation, identifies stakeholders, investigates their work and needs, and clarifies the problem and boundaries of the proposed software system~\cite{nuseibeh2000requirements}.
Needs can be tacit, ambiguous, or conflicting; interviews, observation, workshops, and prototypes help make them explicit and open to agreement.
As needs change, revision involves reassessing requirements and their consequences, with version control and traceability supporting consistent updates~\cite{nuseibeh2000requirements}.
In Agent development, these activities establish whose expectations the Agent should satisfy and how those expectations should evolve.
For an illustrative code-summary Agent, asking for a concise explanation leaves open whether the intended reader needs implementation details, behavioral contracts, or maintenance guidance.
Clarifying the audience and purpose with stakeholders determines what the Agent should explain and how its summaries will be assessed; a subsequent change of audience calls for revisiting both.

The first group in Table~\ref{tab:req-method-comparison} distinguishes stakeholder elicitation, interactive clarification, and empirical analysis of existing specifications.
Axetorn et al. investigate trust requirements for an enterprise HR chatbot through a two-cycle design-science study~\cite{axetorn2026trustrequirements}.
They interview six employees with differing roles and AI experience, asking about trust, prior use, and HR-specific needs.
The interview transcripts are independently coded and consolidated into themes; literature findings and a two-day workshop with five AI experts inform their refinement and candidate designs.
The authors then express the identified concerns as explicit requirements and map them to architectural decisions.
For example, concerns about unreliable answers become requirements for verifiable citations and abstention when supporting information is unavailable, motivating retrieval filtering and answer checking.
Prototype tests with synthetic HR data and follow-up interviews assess the implemented response to those requirements.
This case demonstrates a traceable route from stakeholder accounts to specifications and design; comparative evidence about the elicitation process's completeness or effectiveness remains open.

Yuma implements interactive clarification as a human-supervised construction workflow~\cite{neves2025yuma}.
A requirements-engineer Agent asks follow-up questions about the requested system's goals, functions, technologies, and constraints, then produces a detailed textual specification.
In its literature-review-assistant example, the dialogue clarifies relevance criteria, access to publication databases, and whether screening should be automated or assisted.
The resulting specification guides an architecture proposal, which the user inspects and refines before workflow generation and export to Dify.
The operational contribution is an explicit clarification and approval step between a natural-language request and system construction.
The reported demonstration establishes a working tool flow; the paper identifies ambiguous requirements as a limitation and leaves user studies and performance and scalability evaluation to future work.
Where Axetorn et al. synthesize expectations across stakeholders, Yuma refines a requesting user's specification; their evaluations leave comparison of the resulting requirements' coverage and agreement unresolved.

Configuration studies reveal how expectations are recorded after developers have begun specifying Agent behavior.
Galster et al. review the documentation of five coding tools, identify eight configuration mechanisms, and derive file- and directory-matching rules to examine their presence in 2,926 repositories~\cite{galster2026configuring}.
Their analysis characterizes the use and relationships of context files, skills, and subagents; file presence is a proxy for adoption, with active use unobserved.
Wen examines 878 cybersecurity skill specifications using rule-based indicators of required inputs, expected outputs, scope boundaries, and concrete examples~\cite{wen2026skillspec}.
Only 19.0\% contain example-demonstration cues; inspection of a six-skill subset shows how missing specification details have to be recovered from helper code to construct an initial check.
The coding measures textual cues, with manual validation of the rules and user-comprehension studies still outstanding.
Together, these studies expose the forms and omissions of existing specifications.
They provide empirical evidence about requirements representation, while stakeholder identification, elicitation, and agreement require additional investigation.

Revision is visible in practitioners' work, although the reported practices are less structured than the initial elicitation procedures above.
Through interviews with 17 experienced developers, Dhanorkar et al. document maintenance of repository instructions, adjustment of prohibited actions, and difficulties in deciding how explicitly to state goals and constraints~\cite{dhanorkar2026oversight}.
These accounts identify recurring work and practical difficulties; they supply observational evidence for requirements maintenance rather than an evaluated change-management procedure.
Practitioner elicitation also serves a different object in DiLLS and XAgen: interviews with eight and twelve practitioners, respectively, inform requirements for Agent diagnosis and explanation tools, such as organizing logs and locating failures~\cite{sheng2026dills,wang2026xagen}.
Their contribution here is to characterize developers' support needs; their tool evaluations concern diagnosis and debugging.
Across these examples, the unresolved requirements-management questions are how stakeholders agree on revisions, how conflicting requests are prioritized, and how changes are checked against previously accepted behavior.

\subsubsection{Specifying Safety and Authority}
Safety requirements constrain system behavior to prevent unacceptable outcomes, while authorization requirements state which actors may perform operations or use resources.
Making these constraints explicit allows developers to examine whether the specified service can be delivered safely and within the permissions granted.
Goal-oriented requirements methods refine safety objectives into behavioral conditions and analyze circumstances that could violate them~\cite{vanlamsweerde2001goals}; security requirements models represent ownership, permission, and delegation to check whether intended actions are authorized~\cite{giorgini2005security}.
These methods connect high-level expectations to constraints that can guide design and verification.

For Agent software, these constraints specify when the Agent may act, which outcomes it must avoid, and what constitutes compliance.
For example, a code-summary Agent needs requirements governing access to private repositories alongside requirements about the summaries themselves.
The second group in Table~\ref{tab:req-method-comparison} compares three different starting points: operation contracts, stakeholder losses, and stated behavioral obligations.
Informed abstention makes an external action conditional on sufficient information, verifiable state, and appropriate authorization~\cite{ojewale2026doubt}.
Required fields come from the operation schema; actions requiring approval are designated in deployment policy.
For example, a bonus-processing request must identify the employee, amount, and justification before it forms a valid operation.
A missing field triggers targeted clarification, an unverified state triggers a bounded check, and an unresolved authorization issue requires explicit approval or human handoff.
The implemented tool wrapper tests these conditions before invocation and records the reason for pausing.
Experiments compare runtime checkpoints with prompt-only and baseline conditions in simulated HR and DevOps scenarios.
They evaluate enforcement and unnecessary refusal under authored preconditions; deciding the policy and its approval requirements remains a deployment responsibility.

Doshi et al. address the preceding question of which constraints developers should specify through an adaptation of STPA~\cite{doshi2026safetooluse}.
The process identifies direct and indirect stakeholders and their values, translates those values into potential losses, and analyzes behaviors that could cause the losses.
Developers select the losses to address and derive safety requirements, then refine them into information-flow and tool-ordering constraints.
In the paper's calendar-Agent example, privacy and timely communication lead to requirements to withhold unrelated private information and notify attendees after an event changes.
These become restrictions on data sent through email and on the ordering of update and notification calls.
An Alloy model checks bounded execution traces for violations and illustrates feasible safe traces.
The process therefore supplies an explicit derivation rationale, while a deployed policy engine, developer studies, and evaluation across real workflows remain future work.

Alamdari et al. begin with user-specified behavioral obligations and give them precise temporal semantics~\cite{alamdari2026formal}.
An obligation such as obtaining authorization before a high-value transaction is expressed in linear temporal logic (LTL).
A labeling function maps observed interactions to propositions such as authorization having been granted; the TRAC monitor updates the formula as execution progresses to determine compliance or violation.
This separates interpreting individual events from checking an obligation over a sequence.
Experiments evaluate violation detection on Agent trajectories and compare temporal monitoring with LLM-based judges.
The evidence concerns compliance with supplied properties and depends on the quality of event labeling.
Identifying an adequate set of obligations and validating their correspondence to stakeholder intent remain upstream specification activities.

The distinction is consequently between deriving a requirement and enforcing a requirement that has already been supplied.
The compared studies provide detailed mechanisms for the latter, with STPA offering an explicit but preliminarily evaluated route for the former.
Axetorn et al.'s case further illustrates a possible break between identification and implementation: security concerns enter the requirements, while organization-specific safeguards are deferred from the prototype~\cite{axetorn2026trustrequirements}.
The open issue in these examples is how developers establish, approve, and revise safety and authority requirements, including their coverage of affected stakeholders and interacting tools.
Runtime compliance measurements alone leave these specification decisions unevaluated.

\subsubsection{Operationalizing Delegation and Governance}
In Requirements Engineering, operationalization translates high-level goals into responsibilities and conditions on the operations needed to fulfill them~\cite{vanlamsweerde2001goals}.
When work is delegated across people and software, developers need to distinguish who is expected to perform it from who is permitted to do so, and specify how fulfillment of these commitments will be monitored~\cite{giorgini2005security}.
Otherwise, a stated organizational goal can leave responsibility for its realization unresolved.
Goal refinement, responsibility assignment, and explicit delegation and monitoring relationships provide a way to express these commitments as inspectable requirements~\cite{vanlamsweerde2001goals,giorgini2005security}.

Applied to Agent development, this requirements work specifies which responsibilities the Agent may assume, which evidence it must retain, and when a person should intervene.
It translates professional or organizational expectations into conditions governing delegated execution and its oversight.
The final group in Table~\ref{tab:req-method-comparison} addresses this problem at two scales: professional delegation and organizational control assessment.
Hagar and Diakopoulos propose a three-step translation: identify the commitments relevant to delegated work, turn them into operational instructions, and preserve those instructions in inspectable configuration artifacts~\cite{hagar2026runtimeconfiguration}.
For example, provenance becomes requirements to preserve source records, retain transformation scripts, and distinguish observations from inferences; human judgment becomes an obligation to pause at consequential ambiguities.
Their journalism case compares an unconfigured Agent with researcher-authored configurations emphasizing defensibility or efficiency.
The former asks for documented transformations and approval at ambiguous decisions; the latter permits more autonomous progress with narrower escalation.
Comparing transcripts and produced files exposes differences in provenance, reviewability, and escalation across three runs of one task.
The case illustrates how commitments can be made operational; institutional agreement on those commitments and reliable effects across repeated deployments are not established by that experiment.

AI-ZTMM derives organizational control requirements from public threat taxonomies, mitigation guidance, and research literature~\cite{mok2026actioncentric}.
Recurring control concerns are consolidated into implementation-independent objectives; controls requiring changes for Agent autonomy, tool use, persistent memory, or delegation are retained.
The resulting 31 requirements are mapped to security functions and maturity-assessment criteria, with links back to supporting sources and threats.
Eleven experts review successive versions, prompting changes to requirement grouping and assessment wording; incident mapping examines threat coverage.
These checks support the model's structure and coverage, while the effectiveness of implemented controls remains untested.
Together, the two approaches show how professional commitments or threat knowledge can become inspectable requirements.
Their remaining challenge is to establish how organizations resolve competing commitments, authorize the resulting specification, and assess its adequacy as Agent behavior and operating conditions change.

\begin{table*}[t]
  \centering
  \caption{Engineering problems and approaches in Requirements Engineering. Rows are grouped by the problem addressed; each approach is summarized by its key idea and linked to representative studies discussed in the text.}
  \label{tab:req-method-comparison}
  \small
  \setlength{\tabcolsep}{4pt}
  \renewcommand{\arraystretch}{1.16}
  \begin{tabular}{@{}L{0.18\textwidth}L{0.20\textwidth}L{0.36\textwidth}L{0.20\textwidth}@{}}
    \toprule
    \textbf{Engineering problem} & \textbf{Approach} & \textbf{Key idea} & \textbf{Representative studies} \\
    \midrule
    Identifying needs and making them explicit & Stakeholder elicitation & Code interview accounts into trust themes, refine them with experts, and map explicit requirements to design. & Axetorn et al.~\cite{axetorn2026trustrequirements} (2026) \\
    & Interactive goal refinement & Clarify goals and constraints through dialogue, then use the textual specification to propose a design for user approval. & Yuma~\cite{neves2025yuma} (2025) \\
    & Configuration analysis (observational) & Characterize existing configuration mechanisms and specification cues; identify what artifacts expose or omit. & Galster et al.~\cite{galster2026configuring} (2026); Wen~\cite{wen2026skillspec} (2026) \\
    \midrule
    Specifying acceptable and authorized behavior & Action preconditions & State required information and authority, with clarification or escalation when conditions are unmet. & Informed abstention~\cite{ojewale2026doubt} (2026) \\
    & Hazard-based derivation & Trace stakeholder values through potential losses and unsafe behavior to safety requirements. & Doshi et al.~\cite{doshi2026safetooluse} (2026) \\
    & Temporal specification & Encode supplied obligations in LTL and map observed events to propositions for compliance checking. & Alamdari et al.~\cite{alamdari2026formal} (2026) \\
    \midrule
    Expressing delegation and governance requirements & Situated configuration & Encode professional commitments as persistent instructions, evidence requirements, and escalation rules. & Hagar and Diakopoulos~\cite{hagar2026runtimeconfiguration} (2026) \\
    & Threat-to-control mapping & Translate threats into control objectives, security functions, and organizational assessment criteria. & AI-ZTMM~\cite{mok2026actioncentric} (2026) \\
    \bottomrule
  \end{tabular}
\end{table*}

\paragraph{Synthesis.}
Table~\ref{tab:req-method-comparison} shows several routes from needs or risks to specifications: stakeholder accounts can be coded into requirements, dialogue can clarify a request, hazards can justify constraints, and professional commitments can be translated into delegation rules.
Observational configuration studies reveal what developers record, while contract checks and temporal monitors assess behavior against specified obligations.
These contributions answer different parts of the requirements problem.
A working construction pipeline, successful compliance check, or well-covered control taxonomy establishes evidence about that particular step; elicitation coverage, stakeholder agreement, and the adequacy of a revised specification require their own evaluation.
Twenty of 26 contributions are directly evaluated, including four E3 contributions, but none reaches E4; only 11 report an available or partly available persistent artifact.
These counts include evaluations of the reported contributions, whose objects range from configurations and engineering-support tools to behavior under specified constraints.
The industrial elicitation case and practitioner accounts ground requirements work in concrete settings, while systematic comparison of elicitation procedures and evidence about maintaining agreed requirements through successive versions remain limited in the studies examined here~\cite{dhanorkar2026oversight,axetorn2026trustrequirements}.

The supplemental literature extends the range of requirements methods considered here.
Source-checked supplemental studies derive explainability requirements through personas and scenarios~\cite{zheng2026persona}, or specify authorization through finite effect-trace contracts~\cite{david2026alignmentcontracts}.
These examples broaden the observed ways to derive and represent requirements; their integration into sustained requirements management remains to be evaluated.
AgentBuild further links scientist-authored acceptance criteria to an explicit agent-construction gate~\cite{shin2026agentbuild}.
Its single-domain construction evidence shows why a domain-specific study can contribute to Agent software engineering, while its failed workflow-scope check separates numerical task quality from acceptance against an authored contract.
This case connects authored acceptance criteria to a construction decision, with reserved SNR cases supplying evaluation evidence.
Norm elicitation supplies another link to testing: contextual-integrity work collects necessity and relevance judgements, aggregates disagreement, and supplies labels for evaluating form-filling assistants~\cite{ghalebikesabi2024contextual}.
In the inspected preprint, identified norms can override elicited labels, and the benchmark scores an ask-the-user decision as correct for unsure labels; the aggregation and override rules are therefore part of the evaluation oracle.
The elicited requirements thus shape evaluation, while the population represented and the rules for resolving disagreement remain important conditions on that evaluation.

\finding[Findings]{For requirements acquisition, interviews and interactive clarification translate stakeholder expectations into specifications that guide Agent design, with evidence mainly from prototype demonstrations and follow-up feedback~\cite{axetorn2026trustrequirements,neves2025yuma}. Safety and authority studies make constraints explicit through risk-based derivation, behavioral contracts, and temporal specifications, and examine their enforcement through model checking and execution experiments~\cite{ojewale2026doubt,doshi2026safetooluse,alamdari2026formal}. Delegation and governance studies express organizational expectations as configurations and control requirements, supported by configuration comparisons, expert review, and threat-coverage analysis~\cite{hagar2026runtimeconfiguration,mok2026actioncentric}. These contributions support requirements in guiding design and behavior checks, but methods and validation remain limited for adequately covering stakeholder needs, reconciling conflicts, and maintaining agreement as systems change. Further work should evaluate existing elicitation and specification methods more deeply and develop negotiation and change-management methods, assessing requirements coverage, agreement, and consistency across multiple stakeholders and successive versions.}

\subsection{Architecture and Design}
\label{sec:arch-methods}

Software architecture identifies the system structures and significant decisions needed to satisfy requirements; design develops those decisions into responsibilities, interfaces, and interactions that guide construction~\cite{washizaki2026swebok}.
These activities make quality trade-offs inspectable before they become embedded in an implementation: distributing work can increase concurrency while introducing communication costs and additional failure paths.
In Agent development, the corresponding questions include which decisions remain under program control, how Agents exchange results, where state resides, and which boundaries constrain tool effects.
Similar capabilities can be exposed through different interfaces and yield different consistency and efficiency, as the controlled coding-Agent comparison by Xu et al. demonstrates~\cite{xu2026toolarchitecture}.
Developers consequently need a basis for choosing an arrangement and understanding its assumptions.
Table~\ref{tab:arch-structure-comparison} groups approaches to responsibilities, coordination, service connections, and state or resources; Table~\ref{tab:arch-control-comparison} groups approaches to disclosure, authority, supervision, and recovery.
Together, they organize how the reviewed studies turn engineering concerns into structural decisions and control boundaries.

\subsubsection{Describing and Realizing Coordination}
Coordination design allocates responsibilities and specifies the ordering, communication, and synchronization needed for components to work together~\cite{washizaki2026swebok}.
Describing these relationships exposes dependencies that would otherwise remain implicit in individual implementations.
For Agent systems, developers must decide both who performs a subtask and how model-selected actions interact with program-controlled scheduling and shared resources.
The responsibility and coordination groups in Table~\ref{tab:arch-structure-comparison} show three ways to support those decisions: describe or assess a structure, select its participants, and specify or generate their interactions.
AIOS separates application logic from shared resource management through an SDK that translates application requests into categorized system calls~\cite{mei2025aios}.
A scheduler dispatches those calls to LLM, memory, storage, and tool services; a context manager snapshots and restores interrupted model contexts.
Native applications use the SDK directly, while adapters connect existing Agent frameworks to the same resource-management layer.
The evaluation checks task performance alongside throughput, latency, and concurrent scaling across framework-based Agents.
It therefore examines whether centralizing resource management preserves useful application behavior while improving execution efficiency.

Architecture descriptions and assessment procedures help developers inspect a design before or alongside its implementation.
Rausch and Wittek adapt C4 views and UML-style notation to describe Agent tasks, tools, exchanged artifacts, and quality gates~\cite{rausch2026agenticc4}.
Their industrial examples expose a practical documentation trade-off: deployment views retain protocols but omit some artifact exchanges that reappear at finer levels.
AgentArcEval begins by clarifying Agent goals with stakeholders, reviewing governance obligations, and identifying quality requirements and guardrails~\cite{lu2026agentarceval}.
Evaluators then review the architecture, construct and prioritize context scenarios, and examine design documents, logs, or evaluation results for risks and trade-offs.
The resulting recommendations identify which design decisions or safeguards should change; operational observations can trigger scenario reprioritization and another review.
The tax-copilot case illustrates the procedure and resulting recommendations, while comparative evaluation of its effectiveness across systems remains future work.
These outputs serve distinct purposes: communicating structure, justifying design changes, and exercising the implemented architecture.

Constrained configuration adds a selection problem, with different sources of information about candidates.
BAMAS provisions a pool of model instances using integer linear programming under a budget~\cite{yang2026bamas}.
Its lexicographic objective prioritizes performance tiers derived from a leaderboard, using estimated token costs; optimality for that formulation is not a guarantee of optimal task quality or realized execution cost.
The online Knapsack Composer instead tests candidate tools or Agents on generated skill queries and accepts components using a value-to-cost threshold~\cite{yuan2025composer}.
Its reported task evaluation fixes the selected configuration after composition, so candidate testing supplies design-search evidence within that fixed evaluation.
The comparison exposes two design dependencies: the fidelity of a supplied ranking and cost model, and the representativeness of local component tests.
Composition choices consequently depend on the task, available inventory, and budget.
BAMAS also separates selecting the model pool from choosing how its members collaborate, allowing developers to distinguish a budget-allocation decision from an interaction-design decision.

Coordination design makes responsibilities, control flow, and state or result transfer explicit.
An explicit coordination contract can support centralized or distributed control and can be authored before or during execution.
For example, the OpenHands platform separates an Agent that consumes event history and emits an action from a runtime that returns observations, with an event stream recording their interaction~\cite{wang2025openhandsplatform}.
This arrangement coordinates an Agent with its environment and user interface through the recorded action--observation exchange.

Workflow-based methods make a different artifact explicit: the permitted interactions and the programs that realize them.
ZipperGen starts from a global message-sequence specification recording communication and ownership of control decisions, then projects it into local Agent programs~\cite{bollig2026coordination}.
The specification can be handwritten or generated at runtime by an LLM planner: generated workflows undergo typing and dataflow checks before projection and execution.
Global specification and projection therefore support both authored and runtime-generated workflows with local execution.
The correctness argument relates terminating executions and establishes a semantic no-stuck property under the formal model.
Its arithmetic-planner check also exposes the remaining semantic obligation: structurally accepted workflows can execute the wrong arithmetic convention.
Developers therefore still need task-level checks, together with assumptions about action termination and scheduling, when using the generated programs.

DynTaskMAS constructs and amends a dependency graph during execution and schedules ready subtasks asynchronously~\cite{yu2026dyntaskmas}.
Its graph vertices denote subtasks, with status labels and dependency edges consumed by a ready queue; semantic context routing and a metric-driven controller support execution.
The implemented queue is centralized on one node, with distributed queues left to future work.
Latency and utilization comparisons against an internal serial baseline, together with component ablations, evaluate the benefits of asynchronous scheduling under the tested serving configuration.
The dependency graph remains model-inferred, making missing or spurious edges a separate correctness concern; distributed scheduling and controlled evaluation of failure-driven adaptation remain open extensions.

Caldeira and Banerjee investigate a different coordination choice, comparing final, per-step, and hybrid validation in a question-answering pipeline~\cite{caldeira2025control}.
Their 50-question experiment reports model-judged output measures and runtime, not a distributed-fault or adversarial-security test.
These studies inform specification semantics, scheduling efficiency, and validation placement, respectively; selecting among them requires an explicit target property.

Task-conditioned coordination can select participants and interaction patterns without generating a new topology.
BAMAS selects from four predefined collaboration templates and instantiates the selected model pool~\cite{yang2026bamas}.
ANTAP makes a narrower choice: it dispatches a query to an Agent using numerical operators calibrated from observed behavior, rather than interpreting Agent self-descriptions at routing time~\cite{alsheich2026linguistic}.
Its output is an execution assignment, not a new graph; its security boundary still depends on trusted offline calibration and the embedding model.
In its sleeper-agent experiment, calibration includes queries containing the adversarial trigger tokens; the reported suppression therefore does not establish protection against triggers absent from calibration.
Thus, a learned selector can change who performs work without synthesizing the structure in which that work occurs.

AgentNet makes coordination decisions locally through a router and executor in each Agent~\cite{yang2025agentnet}.
Its network vertices are Agents rather than DynTaskMAS's subtasks; connection weights are updated from interaction success between tasks and weak edges are pruned.
The routing decisions and local trajectory memories therefore evolve together, rather than producing only an offline workflow artifact.
Its task benchmarks, routing ablations, and small-team scaling study support this adaptive coordination design, but do not directly evaluate cross-organization privacy or fault tolerance under injected node failures.
The contrast with DynTaskMAS concerns the represented object, decision authority, and execution interface, not whether either method is dynamic.

Other methods derive design artifacts from process evidence or comparative cases.
PM4AA maps repository contributors' behavior to a prescribed role vocabulary, partitions event logs by role, and derives process specifications used to generate Agents~\cite{bala2026pm4aa}.
The example consolidates eight process roles into five Agents and uses three routing smoke tests to check basic execution.
This supplies a trace-to-design procedure whose evidence currently establishes construction feasibility; evaluating the adequacy of the derived responsibilities requires broader behavioral tests.
Eger et al. instead derive guidance for choosing single- or multi-agent structures from two prototypes and ten constructed healthcare cases~\cite{eger2026architecture}.
The output is guidance for architects grounded in those prototypes and cases, with transferability depending on the new design context.
These studies expose different dependencies: adequate historical process traces, and the transferability of design advice beyond the cases from which it was derived.

Structure synthesis expands the decision space beyond selecting a participant or a predefined template.
EvoMAS searches over structured configurations of roles, models, prompts, tools, and communication links, using execution feedback to mutate and recombine candidates~\cite{hu2026evomas}.
Its crossover inherits one parent's topology while combining Agent attributes; mutation can change the topology itself.
AutoRAS generates primitive sequences that compile into communication and behavioral structures, conditioned on the query and execution-derived robustness diagnostics~\cite{yue2026autoras}.
Its output is a generated structure rather than a provisioned pool of components.
The studies assess generated systems through execution validity and task outcomes, with robustness diagnostics additionally guiding AutoRAS's search.
These results help select a working configuration; maintainability and compatibility after subsequent changes require evaluation across versions.
Across the compared approaches, authored structure, generation, and adaptation can coexist: DynTaskMAS generates and executes a graph, while ZipperGen accepts both handwritten and generated specifications.
The useful design question is which structural choices a method exposes and how it checks them.

For a developer, the comparison in Table~\ref{tab:arch-structure-comparison} therefore begins with the unresolved design question.
Architecture assessment supports a reasoned choice; component selection allocates a budget; workflow specification controls exchanges; scheduling and routing decide how the structure executes.
Combining them requires checking that the selected participants satisfy the interfaces and assumptions of the coordination mechanism.

\subsubsection{Discovering Services and Selecting Protocols}
Service-oriented design separates finding a component that supplies a needed function from agreeing on the interface and protocol through which it can be used~\cite{washizaki2026swebok}.
For distributed Agents, advertised capabilities help select a participant, while message formats, session behavior, reconnection, and error handling determine whether collaboration can proceed reliably.
The service-connection group in Table~\ref{tab:arch-structure-comparison} distinguishes these two design decisions.
DAWN distinguishes a Principal Agent that composes work from Gateway Agents that register, discover, and execute distributed resources~\cite{aminiranjbar2025dawn}.
This separation exposes responsibility boundaries for resource metadata, orchestration, and communication.
Its experiment evaluates planning and composition, including task-to-Agent matching; the broader security, compliance, and global-scale properties of the reference architecture remain architectural proposals.

ProtocolRouter addresses a narrower decision within an already distributed system: which protocol should connect each module under explicit requirements~\cite{du2026protocolbench}.
It produces a per-module protocol plan and uses adapters and stateless bridges across protocol boundaries.
ProtocolBench keeps workflows and other non-protocol factors fixed while preserving native streaming, retry, and reconnection behavior, so task outcomes, latency, overhead, and recovery can be compared under shared conditions.
The resulting trade-offs are scenario-dependent: the router improves some measures, while fixed protocols remain preferable on others.
Together, these studies distinguish discovering a suitable service from choosing how services interact; a discovery result alone does not determine an appropriate protocol or establish its end-to-end guarantees.

\begin{table*}[t]
  \centering
  \caption{Engineering problems and approaches in structural design. The groups distinguish responsibility allocation, coordination, service connections, and state or resource organization.}
  \label{tab:arch-structure-comparison}
  \small
  \setlength{\tabcolsep}{4pt}
  \renewcommand{\arraystretch}{1.16}
  \begin{tabular}{@{}L{0.18\textwidth}L{0.20\textwidth}L{0.36\textwidth}L{0.20\textwidth}@{}}
    \toprule
    \textbf{Engineering problem} & \textbf{Approach} & \textbf{Key idea} & \textbf{Representative studies} \\
    \midrule
    Assigning and assessing responsibilities & Shared runtime services & Separate Agent applications from kernel-managed resources through explicit interfaces. & AIOS~\cite{mei2025aios} (2025) \\
    & Scenario-based assessment & Relate stakeholder goals and quality scenarios to design risks and recommendations. & AgentArcEval~\cite{lu2026agentarceval} (2026) \\
    & Controlled interface comparison & Vary interface abstractions while keeping capabilities similar to expose design trade-offs. & Xu et al.~\cite{xu2026toolarchitecture} (2026) \\
    \midrule
    Selecting and coordinating participants & Budgeted composition & Select model or tool inventories using supplied rankings or local component tests. & BAMAS~\cite{yang2026bamas} (2026); Knapsack Composer~\cite{yuan2025composer} (2025) \\
    & Specification and projection & Check a global communication specification and project it into local Agent programs. & ZipperGen~\cite{bollig2026coordination} (2026) \\
    & Dependency-aware scheduling & Maintain a subtask graph and dispatch ready work as dependencies change. & DynTaskMAS~\cite{yu2026dyntaskmas} (2026) \\
    & Template or executor selection & Choose a predefined collaboration pattern or a behavior-calibrated executor for a query. & BAMAS~\cite{yang2026bamas} (2026); ANTAP~\cite{alsheich2026linguistic} (2026) \\
    & Decentralized coordination & Route work locally and update collaboration links from interaction outcomes. & AgentNet~\cite{yang2025agentnet} (2025) \\
    & Process-derived design & Map historical process traces to prescribed roles and role-specific process specifications. & PM4AA~\cite{bala2026pm4aa} (2026) \\
    & Configuration search & Mutate and recombine roles, models, prompts, tools, and communication links using execution feedback. & EvoMAS~\cite{hu2026evomas} (2026) \\
    \midrule
    Connecting distributed services & Discovery and composition & Separate orchestration from resource registration, discovery, and execution. & DAWN~\cite{aminiranjbar2025dawn} (2025) \\
    & Requirement-guided protocol choice & Select per-module protocols and bridge their interfaces under explicit requirements. & ProtocolRouter~\cite{du2026protocolbench} (2026) \\
    \midrule
    Organizing state and resources & State separation and scoping & Separate specifications from recoverable execution state, or align memory access with ownership. & OpenHands SDK~\cite{wang2026openhands} (2026); HLTM~\cite{xu2026hierarchicalmemory} (2026) \\
    & Resource-path design & Split cache-loading paths and schedule transfers to manage serving bottlenecks. & DualPath~\cite{wu2026dualpath} (2026) \\
    \bottomrule
  \end{tabular}
\end{table*}

\FloatBarrier
\subsubsection{Locating Control and Recovery Boundaries}
Control and recovery design assigns responsibility for enforcing restrictions, containing failures, and restoring acceptable operation~\cite{washizaki2026swebok}.
The placement of these mechanisms matters because a check can govern only the interactions or effects visible at its boundary.
In an Agent system, a tool call that follows the workflow may still disclose private data or commit an irreversible change, requiring explicit decisions about authorization, supervision, and recovery.
Table~\ref{tab:arch-control-comparison} compares four responses to that problem, moving from disclosure and authority boundaries to supervision and consistency mechanisms.
Disclosure control limits the information made available across a trust boundary.
AirGapAgent first minimizes user data using a trusted task context, then lets a separate conversational Agent interact with an untrusted third party using only that subset~\cite{bagdasarian2024airgapagent}.
Third-party messages cannot flow back into the minimizer.
The architecture limits what data the conversational Agent can disclose; it still depends on the quality of minimization and does not prevent inference from retained information.
When legitimate interaction needs omitted data, the design routes escalation back to the user, making privacy and task completion an explicit authorization trade-off.
Its synthetic profiles and model-generated privacy labels also leave alignment with actual users' privacy expectations unestablished.

Disclosure and information-flow controls can rely on different trust assumptions.
MINIM sanitizes structured UI observations locally, using predicted necessity and sensitivity to retain, abstract, or remove elements before remote disclosure~\cite{yu2026minim}.
It assumes an honest-but-curious remote Agent and excludes active prompt injection; its node-preservation and disclosure measures are not end-to-end task-success guarantees.
In contrast, tacit treats Agent-generated code as untrusted and compiles it in a capability-safe Scala subset before execution~\cite{odersky2026tracked}.
Classified values and tracked effects constrain information flow under trusted compiler, library, and runtime assumptions.
External commands fall outside that static boundary and require appropriately restricted permissions or additional containment.
These methods constrain different information paths: the observation supplied to the model and the effects permitted in generated code. Any proposed combination must preserve both trust boundaries.

Execution isolation addresses propagation across application boundaries.
IsolateGPT places applications in distinct spokes and mediates collaboration through a trusted hub, non-LLM operators, and user permissions~\cite{wu2024isolategpt}.
Where AirGapAgent restricts a downstream Agent's data, IsolateGPT restricts cross-application interaction.
Its target is propagation to other applications or the system; attacks confined to the application's own scope remain possible.
Isolation therefore bounds the extent of a compromise, while within-application behavior requires additional controls.

Externalized policy and human-review coordination separate authority or review from task planning, with enforcement timing and decision procedures varying across mechanisms.
AgenticRei extracts a typed invocation into a subject--action--resource triple, evaluates deontic policy outside the LLM, and returns a decision with attached obligations~\cite{joshi2026deontic}.
Its design distinguishes permission to act from duties triggered by acting.
The prototype illustrates policy composition and decision latency, but credential verification is simulated through trusted-issuer matching and production-runtime integration remains ongoing.
The decision log makes these outcomes inspectable; tamper-evident logging would require additional integrity mechanisms.

CUGA illustrates why a policy layer need not have a single enforcement mode~\cite{shlomov2026governance}.
Its typed policies combine intent blocking and tool approval with playbook injection, tool-description guidance, and output formatting.
Some triggers use semantic retrieval, and natural-language policy conflicts can invoke an LLM to select a policy.
Externalizing policy therefore makes the decision and intervention points inspectable without making every decision deterministic.
Demonstrations and benchmark comparisons with policy ablations examine how these intervention modes affect task completion.
Coverage against adversarial bypass remains a separate requirement for using the layer as a security boundary.

Human review raises two further design questions: when to request it and what the reviewer can change.
Zhou et al. schedule confirmation points by balancing confirmation, diagnosis, correction, and re-execution time~\cite{zhou2026confirmation}.
Their 48-participant controlled simulation evaluates oversight timing under modeled errors, not mandatory authorization before every external effect.
The current objective optimizes time; irreversible harms and mandatory high-stakes checkpoints require additional constraints.
AgentClick supplies an artifact-editing interface instead: a skill-guided Agent submits proposals to a browser-backed review service and consumes approval or correction~\cite{zhuang2026agentclick}.
Its walkthroughs demonstrate the interaction path, not controlled improvements in reviewer performance or non-bypassable interception of all actions.
A review interface, a checkpoint schedule, and an authorization policy therefore address different parts of governed execution.

Execution-state supervision assesses an action, observation, trajectory, or interaction to recognize risk or ineffective progress and support a response.
Timing is not the dividing line: a risk critic can inspect a proposed action before execution, just as a review schedule can ask a user to inspect actions already completed.
MCIP combines structured information-flow trajectories with a safety-aware Guardian that identifies and categorizes risks~\cite{jing2025mcip}.
Unlike a deterministic policy decision over an explicit invocation, this supervision depends on learned risk recognition.
Its experiments evaluate that recognition and training intervention; the authors do not enumerate the full diversity of concrete adaptive attacks.
Reported classification improvements should therefore not be interpreted as complete mediation of unsafe actions.

Tang et al.'s risk verifier critiques proposed actions using the execution trajectory, with an abstractor that makes the underlying risk easier to recognize~\cite{tang2025riskknowledge}.
The verifier and acting Agent use the same base LM in separate assessment and action roles.
The proposal--critique loop stops when the verifier accepts an action or the iteration limit is reached.
SupervisorAgent takes a selective approach to supervision cost: an LLM-free, heuristic filter flags explicit errors, repetitive behavior, or excessive observation length; the supervisor then selects context-dependent guidance, correction, or verification~\cite{lin2026supervisoragent}.
These methods supply risk recognition and runtime process control under their respective supervision conditions.
Evidence for an intended control point must also be separated from evidence of its deployment: AASL tests its proposed interception logic retrospectively on recorded failures, not as a deployed runtime layer~\cite{yu2026operationalhallucination}.

Execution contracts specify admissible calls, returns, messages, state transitions, or external effects, with consistency obligations scoped to the controlled interface.
LLM-as-Code places the call graph under program control and exposes child-Agent results as typed returns that a parent can accept, retry, or replace~\cite{qi2026llmascode}.
This makes control and return obligations explicit without requiring a fixed workflow or a global transaction protocol.
G2CP instead validates messages and authorization before executing graph operations and logging their results~\cite{benkhaled2026g2cp}.
Deterministic operation semantics are relative to the supplied graph and operation; they do not make LLM-based query interpretation or fuzzy entity linking correct.
SEMAP combines behavioral contracts, structured messages, and a task-state machine with verification~\cite{mao2025semap}.
Although evaluated on coding and vulnerability-detection tasks, its engineering intervention changes the multi-Agent protocol rather than the external code artifact.
Its preliminary task experiments support a protocol-level comparison, while component ablations and formal protocol verification remain future work.
These approaches constrain different interfaces; a typed return, a valid graph operation, and a checked task-state transition are not interchangeable evidence of task completion.

SagaLLM preserves goals and dependencies, validates outputs before commitment, and associates workflow actions with compensation agents~\cite{chang2025sagallm}.
This addresses what must happen after a partially executed workflow fails, a different concern from deciding whether its first action is permitted.
Its experiments exercise selected sequential and reactive planning cases; formal verification of generated compensation code remains future work.
Compensation design should consequently not be equated with proven rollback correctness for arbitrary external effects.
ACRFence addresses the related mismatch between restored Agent state and effects already committed to external tools~\cite{zheng2026acrfence}.
Its proposed effect log and semantic comparison distinguish replay from an explicit fork when an Agent regenerates a call after restoration.
The paper evaluates the underlying attacks; the effect-aware defense is proposed and has not yet been implemented or evaluated.

Coordination specifies how execution proceeds; a contract additionally specifies what constitutes an admissible exchange or effect and what happens on violation.
The two can coexist, as can authorization and runtime supervision. Their implementation status, assumptions, and evaluated properties must be recorded separately.

\begin{table*}[t]
  \centering
  \caption{Engineering problems and approaches in architectural control. The comparison separates disclosure and isolation, authorization, supervision, and consistency or recovery.}
  \label{tab:arch-control-comparison}
  \small
  \setlength{\tabcolsep}{4pt}
  \renewcommand{\arraystretch}{1.16}
  \begin{tabular}{@{}L{0.18\textwidth}L{0.20\textwidth}L{0.36\textwidth}L{0.20\textwidth}@{}}
    \toprule
    \textbf{Engineering problem} & \textbf{Approach} & \textbf{Key idea} & \textbf{Representative studies} \\
    \midrule
    Bounding disclosure and propagation & Data minimization & Restrict third-party interaction to a task-relevant data subset behind a one-way boundary. & AirGapAgent~\cite{bagdasarian2024airgapagent} (2024) \\
    & Mediated isolation & Separate applications and mediate their exchanges through trusted operators and permissions. & IsolateGPT~\cite{wu2024isolategpt} (2025) \\
    \midrule
    Governing authority & External policy decisions & Evaluate typed invocations against explicit permissions and obligations outside the acting model. & AgenticRei~\cite{joshi2026deontic} (2026) \\
    & Policy-triggered checkpoints & Place blocking, approval, and advisory policies at explicit interaction points. & CUGA~\cite{shlomov2026governance} (2026) \\
    \midrule
    Recognizing risky execution & Information-flow supervision & Structure execution trajectories so a trained Guardian can identify and categorize risk. & MCIP~\cite{jing2025mcip} (2025) \\
    & Proposal--critique control & Abstract a proposed action and its context, critique risk, and request revision. & Risk verifier~\cite{tang2025riskknowledge} (2025) \\
    & Selective supervision & Use lightweight triggers to request context-dependent correction or verification. & SupervisorAgent~\cite{lin2026supervisoragent} (2026) \\
    \midrule
    Preserving execution consistency & Protocol contracts & Combine typed messages, behavioral contracts, and checked task-state transitions. & SEMAP~\cite{mao2025semap} (2025) \\
    & Compensation design & Preserve dependencies, validate before commitment, and associate effects with recovery actions. & SagaLLM~\cite{chang2025sagallm} (2025) \\
    & Effect-aware restoration & Use a proposed effect log and semantic comparison to distinguish replay from an explicit fork. & ACRFence~\cite{zheng2026acrfence} (2026) \\
    \bottomrule
  \end{tabular}
\end{table*}

\subsubsection{Designing State and Resource Substrates}
State and resource design determines data ownership, persistence, access, and the allocation of execution capacity~\cite{washizaki2026swebok}.
These decisions support continuity of service and predictable resource use when components share infrastructure.
Agent systems bring conversation histories, tool effects, tenant-specific knowledge, and model-serving resources into this design problem: a resumed run needs a consistent history, retrieval must respect ownership, and concurrent model calls must fit resource limits.
The final group in Table~\ref{tab:arch-structure-comparison} compares state separation and scoping with resource-path design because they address different obligations of the execution substrate.
The OpenHands SDK separates immutable component specifications from mutable conversation state and an append-only event log; typed actions separate validation from tool execution~\cite{wang2026openhands}.
Configured persistence supports conversation-state replay and crash recovery, which the paper measures separately from task performance; these measurements do not establish recovery of external tool effects.
HLTM instead aligns LinkedIn's memory hierarchy with enterprise ownership, restricts retrieval to the caller's subtree, and localizes incremental updates~\cite{xu2026hierarchicalmemory}.
Its evaluation includes retrieval leakage, incremental indexing, and an operational use case.
These examples make distinct state obligations explicit: recovering an execution history and enforcing tenant-scoped retrieval each require a dedicated mechanism and evaluation.

Resource scheduling addresses a different substrate-level obligation.
DualPath addresses the cache-transfer bottleneck of multi-turn Agentic inference by combining storage-to-prefill and storage-to-decode loading paths with traffic isolation and adaptive scheduling~\cite{wu2026dualpath}.
Its cluster experiments and ablations measure serving performance under concrete hardware, network, and workload conditions.
These measurements characterize the Agent serving substrate under the evaluated configurations and workloads.

Controlled architectural comparisons can inform interface decisions without proposing a new runtime.
Xu et al. compare six coding-Agent tool architectures while keeping capabilities similar and varying interface abstraction or lightweight reasoning scaffolds~\cite{xu2026toolarchitecture}.
Similar overall task resolve rates coexist with differences in consistency, exploration, and efficiency; some effects depend on the actor model.
The comparison illustrates why task success alone is an incomplete architecture criterion.
Its domain is coding, but the manipulated object is the Agent's interface design, not the external code artifact it produces.

\paragraph{Synthesis.}
Architecture studies supply descriptions, assessment procedures, executable structures, and control contracts for distinct design decisions.
Of 135 contributions, 121 are directly evaluated, eleven are coded E3/E4, and 91 report available or partly available persistent artifacts.
The evaluated properties include coordination semantics, isolation, protocol performance, state recovery, and resource efficiency.
Selecting a design therefore requires relating the desired property to the method's assumptions and evaluated conditions.

\finding[Findings]{Architecture research provides scenario-based assessment, explicit coordination specifications, and shared runtime structures, supported by case applications, formal arguments, and controlled execution comparisons~\cite{lu2026agentarceval,bollig2026coordination,mei2025aios}. Isolation, protocol selection, and scoped state mechanisms further make privacy, communication, and recovery decisions inspectable and testable~\cite{wu2024isolategpt,du2026protocolbench,wang2026openhands,xu2026hierarchicalmemory}. These methods support the SE role of connecting structural choices to required system properties. Their evidence is strongest for particular mechanisms under stated assumptions; it is less complete for composing controls, comparing architectural alternatives under common requirements, and preserving those properties through dependency or interface changes. Further work should evaluate architectural trade-offs and combinations across configurations and versions, linking each decision to the quality scenario it is intended to satisfy.}

\subsection{Implementation and Integration}
\label{sec:impl-methods}

Software implementation realizes a design in executable code and configuration; integration assembles components and checks the interfaces through which they cooperate~\cite{washizaki2026swebok}.
The engineering objective is a working system whose behavior remains consistent with its design as separately developed parts are connected.
For Agents, this includes translating workflow or skill descriptions into runtime behavior, connecting models and tools, adapting data formats, and implementing constraints around generated calls.
The MCP issue study makes this practical burden visible in incomplete tool descriptions and integration changes~\cite{oyelayo2026mcpchallenges}.
The 91 contributions address how to carry out that work: Table~\ref{tab:impl-construction-comparison} covers executable realization, configuration authoring, and data compatibility; Table~\ref{tab:impl-execution-comparison} covers execution cost, call conformance, authority, and recovery.
The following comparison connects these implementation choices to the problems they solve and the behavior that must survive integration.

\subsubsection{From Specifications to Executable Integrations}
Executable realization turns a specification into code, build artifacts, and runtime configuration, using translation rules and interface checks to preserve the intended behavior~\cite{washizaki2026swebok}.
In Agent development, the specification may describe a workflow, a tool action, or a skill; its realization must bind those descriptions to operations supported by the target environment.
The first two groups in Table~\ref{tab:impl-construction-comparison} compare constructing runtime services, separating execution logic from configurable choices, and translating authored specifications into deployable configurations.
OpenHands separates an Agent's action selection from action execution: a session-specific container exposes a REST API for shell, notebook, and browser operations, and a build process augments a developer-supplied base image with the runtime client~\cite{wang2025openhandsplatform}.
This gives developers a concrete construction path from a base image to an action-execution service.
Task benchmarks exercise the resulting environment, while security depends on the permissions and tools included in the constructed image.

EnCompass and Agent Spec separate different implementation decisions.
EnCompass compiles a Python workflow with programmer-selected branchpoints into a search space; the search algorithm and budget are supplied separately at execution time~\cite{li2025encompass}.
Its three case studies compare search outcomes and the code changes needed to switch strategies, not controlled measurements of developer time.
The abstraction does not remove the developer's responsibility to choose branchpoints and intermediate scores.
Agent Spec instead serializes Agent definitions and control/data-flow configurations, then uses runtime adapters to instantiate them in different frameworks~\cite{amini2025agentspec}.
Cross-runtime experiments expose the limits of this portability: outcomes remain runtime dependent, and the evaluated AutoGen and CrewAI adapters lacked support for the tested non-agentic flows.
Thus, separating workflow logic from search policy and separating configuration from framework implementation create different validation obligations.

Construction can also preserve human decisions in inspectable artifacts.
Botender links community proposals, generated interaction cases, and deployment of event-triggered bot behavior~\cite{kuo2026botender}.
Its field study demonstrates collaborative construction in Discord communities; the evaluated version supports single-turn conversational bots, not unrestricted long-horizon automation.
AgentBound instead generates capability manifests from MCP-server code and documentation; developers review and refine the declared filesystem, environment, and network access before those permissions are enforced outside the model~\cite{buehler2026agentbound}.
These construction artifacts make desired behavior and authorized resources available for human decisions.

SkCC addresses repeated adaptation of a skill to different frameworks through a compiler pipeline~\cite{ouyang2026skcc}.
It parses Markdown and metadata into an abstract syntax tree, then normalizes procedures, permissions, schemas, and constraints into a typed intermediate representation.
Shared structural, dependency, and permission checks operate on that representation before target-specific emitters produce deployable artifacts.
The evaluation reports compilation outcomes and triggered checks, establishing the pipeline's ability to accept or reject skills under its implemented rules.
Those rules cover structural and declared-resource problems; prompt injection and some dependency threats remain outside the assessed coverage.
Thus, portability of the authored representation and security of the deployed behavior require distinct checks.

The supplemental MCP-description study examines tool descriptions as an integration artifact~\cite{hasan2026mcpsmells}.
Its augmentor derives examples and limitations from tool-execution traces; a configurable client then supplies original, augmented, or component-selected descriptions without changing the server.
This operation supports authoring and experimentation rather than tool-permission enforcement.
Description-quality scores and downstream success remain different outcomes: the study reports regressions for some model--domain combinations despite improved description scores.
Its fully automated trace orchestration was added after the evaluation, and three models' comparison baselines come from the earlier benchmark study. The automated toolchain and the reported performance comparison therefore have different validation boundaries.

\begin{table*}[t]
  \centering
  \caption{Engineering problems and approaches in construction and integration. The groups connect executable realization, configuration authoring, and state or data compatibility to representative solutions.}
  \label{tab:impl-construction-comparison}
  \small
  \setlength{\tabcolsep}{4pt}
  \renewcommand{\arraystretch}{1.16}
  \begin{tabular}{@{}L{0.18\textwidth}L{0.20\textwidth}L{0.36\textwidth}L{0.20\textwidth}@{}}
    \toprule
    \textbf{Engineering problem} & \textbf{Approach} & \textbf{Key idea} & \textbf{Representative studies} \\
    \midrule
    Realizing executable behavior & Runtime construction & Build a session environment and expose shell, notebook, and browser actions through a runtime API. & OpenHands~\cite{wang2025openhandsplatform} (2025) \\
    & Programmable search abstraction & Compile workflow branchpoints into a search space while supplying search policy separately. & EnCompass~\cite{li2025encompass} (2025) \\
    & Specification adapters & Instantiate a shared declarative Agent specification through framework-specific adapters. & Agent Spec~\cite{amini2025agentspec} (2026) \\
    \midrule
    Authoring deployable configurations & Participatory construction & Refine community proposals using interaction cases and deploy the resulting bot behavior. & Botender~\cite{kuo2026botender} (2026) \\
    & Typed skill compilation & Parse skill text into an intermediate representation, check it, and emit target-specific artifacts. & SkCC~\cite{ouyang2026skcc} (2026) \\
    \midrule
    Integrating state and data & Managed persistence and retrieval & Reconcile versioned facts and connect storage, retrieval, and context assembly to Agent execution. & Routray et al.~\cite{routray2026tieredchatbot} (2026) \\
    & Schema-based data adaptation & Normalize source trajectories, convert them for target harnesses, and validate their structure. & ADP~\cite{song2026adp} (2026) \\
    & Empirical integration analysis & Examine reference-server issues and linked fixes to identify recurring interface and implementation problems. & MCP study~\cite{oyelayo2026mcpchallenges} (2026) \\
    \bottomrule
  \end{tabular}
\end{table*}

\subsubsection{State and Data as Integration Artifacts}
Data integration reconciles representations and preserves the meaning and provenance needed by consuming components; persistence additionally requires consistent update and recovery behavior~\cite{washizaki2026swebok}.
For Agents, adding a memory or dataset leaves practical questions about conflicting facts, conversation structure, and framework-specific action formats.
These questions determine whether retrieved or converted information can be used correctly by the running system.
The state and data group in Table~\ref{tab:impl-construction-comparison} compares managed persistence with schema-based conversion; the empirical MCP study complements both with observed integration problems.
Routray et al. describe versioned fact reconciliation, dual persistence, and a staged retrieval pipeline consumed by a stateful Agent~\cite{routray2026tieredchatbot}.
The memory unit couples fact reconciliation with versioned storage, dual-write rollback, and managed background consolidation.
The retrieval unit instead integrates format-specific ingestion, per-session quotas, parent/source metadata, configurable reranking, and an on-demand search tool with the Agent's stateful execution path.
The distinction matters for implementation: updating a stored fact and delivering retrieved context require different consistency and interface checks, alongside the tool-connection manager and state-graph runtime.
Its retrieval-ranking ablation does not isolate the quality or portability of that Agent integration.
The evidence comes from synthetic-profile, retrieval, and simulated-load experiments; cross-deployment behavior remains to be assessed, and public code release is described as future work.

ADP addresses a different boundary: heterogeneous source datasets and target Agent harnesses~\cite{song2026adp}.
A typed trajectory schema represents ordered actions and observations, including API calls, executable code, messages, and textual or web observations.
Source converters map raw records into that schema; a separate converter for each harness maps the common records to its action space, prompts, and training conversation format.
Automated checks then inspect tool-call structure and conversation completion.
Experiments train Agents in three frameworks, demonstrating that the converted data is consumable downstream.
The task gains evaluate the resulting training pipeline; measuring adaptation effort and compatibility independently would establish how much engineering work the shared schema saves.

Training compatibility also depends on what the model is asked to learn, not only on whether examples share a format.
In Workflow Graphs, a conversation can contain responses from nodes with different system prompts; response masking excludes other nodes' outputs from the loss under the current node's prompt~\cite{park2025workflowgraphs}.
MCIP instead converts function-call dialogues into synthetic information-flow records and risk labels for Guardian training~\cite{jing2025mcip}.
These mechanisms define supervision at different boundaries: node provenance, information flow, and harness format.
The Workflow Graphs comparison combines fine-tuning with masking rather than isolating the masking rule; MCIP's synthesized records are not collected deployment logs.
Data-quality and integration-reliability checks must therefore accompany downstream task or risk-classification measurements.

Two supplemental studies make the supervision target itself an integration decision.
AgentGate constructs query--candidate--structured-output examples, including hard negatives, to adapt compact dispatch models~\cite{cheng2026agentgate}.
CommandSans instead labels AI-directed instruction tokens and trains an encoder tagger whose outputs control sanitization of tool responses~\cite{das2025commandsans}.
Their delivered artifacts are a specialized router and an instruction detector, not interchangeable forms of trajectory conversion.
Validation must follow that distinction: AgentGate's fine-tuning probe reports mixed changes across routing metrics, while CommandSans separately compares token tagging, attack outcomes, and utility under differing input distributions.
Some CommandSans attack-success rates are estimates based on whether injection text remains in a tool response, not observed harmful execution; those values should not be pooled with execution-based rates.
Reliability assessment of the surrounding Agent system must additionally examine how errors in these learned components propagate through integration.
These examples show why choosing the supervision target belongs to integration design: a router's misassignment and a sanitizer's missed instruction propagate through different execution paths.

AgentGym illustrates a different construction decision: its gather-and-filter trajectory pipeline supplies data, while AgentSTaR organizes repeated exploration, reward filtering, and retraining~\cite{xi2025agentgym}.
The latter procedure retrains from the initial Agent at each iteration, rather than simply updating the running Agent in place.
Its cross-environment task evaluation concerns the constructed training pipeline, with reward filtering and retraining jointly contributing to the resulting Agent.

\FloatBarrier
\subsubsection{Execution Efficiency and Representation Changes}
Performance optimization changes an implementation to reduce time or resource consumption while checking the behavior that the transformation must preserve~\cite{washizaki2026swebok}.
Profiling, caching, and representation changes are useful when their savings exceed lookup, reconstruction, and validation costs.
For Agents, plans, message encodings, and histories depend on context, so reusing or shortening them can also change subsequent decisions.
The first group in Table~\ref{tab:impl-execution-comparison} therefore distinguishes the representation reused or transformed, rather than treating every reduction in computation as the same optimization.
APC extracts reusable plan templates from successful executions and adapts cache hits to the current context~\cite{zhang2025apc}.
Its hit/miss routing switches between planners; the engineering object is this serving path, evaluated alongside task quality.
Prompt Choreography reuses message encodings in a global KV cache through separate prefill and decode operations~\cite{bai2026choreography}.
Its design assumes relative-position attention, an entirely GPU-resident cache, and calls without long waits for human or tool input.
AgentDiet inserts an externally controlled reflection module that shortens selected trajectory steps within a sliding window~\cite{xiao2026agentdiet}.
Thus, reusable plans, cached encodings, and reduced histories have different invalidation and information-loss risks.
In particular, Prompt Choreography can change output distributions; fine-tuning mitigates that change rather than proving semantic equivalence.
AgentDiet's evaluation includes reflection overhead and cache invalidation in its cost accounting, an important distinction from counting removed tokens alone.

Cache-reuse mechanisms also have materially different preconditions.
KVCOMM estimates context-dependent KV-cache offsets from an online anchor pool for Agents sharing an identical RoPE-based model checkpoint but using different prefixes~\cite{ye2025kvcomm}.
It falls back to dense prefilling when reuse conditions are unmet; its sensitivity experiments expose a reuse--accuracy trade-off rather than semantic equivalence.
LRAgent targets multi-LoRA Agents: it shares a base cache while retaining low-rank adapter contributions, and its shared-A variant also reuses the low-rank cache~\cite{jeon2026lragent}.
Its attention kernel avoids materializing full-dimensional adapter outputs.
Shared activations are similar rather than identical, and the main accuracy experiments use shared-A weights; the controlled efficiency traces isolate computation from changes in Agent trajectories.
Their reported speedups also use different measurement units: KVCOMM measures per-agent TTFT, whereas LRAgent sums prefill latency across Agent steps.

GraphFlow shares operation-level base caches and stores sparse prefix-specific residuals~\cite{li2026graphflow}.
Its method retains residuals for frequent paths and recomputes them on demand for rare paths, trading storage against reconstruction work.
Nova changes a different resource: it streams vision-encoder weights through reusable physical GPU layers, overlapping transfers with computation~\cite{xu2025nova}.
This preserves the supplied weights rather than approximating context, but its benefit depends on transfer bandwidth and execution length.
Agent-X combines reordered, cacheable prompt prefixes with example-derived speculative token drafts~\cite{chung2026agentx}.
Target-model verification and selective autoregressive fallback remain part of decoding; end-to-end experiments evaluate this combination in its macOS/TinyAgent implementation.
Together, these studies require separate accounting of storage, reconstruction, verification, and fallback costs before comparing apparent reuse gains.
Table~\ref{tab:impl-f10-boundaries} extends the comparison to state recovery and execution controls across 13 source papers and 15 coded contributions.

The supplemental cases extend this comparison beyond inference-cache reuse without changing the main frequencies.
AgentGit restores recorded execution state and branches from checkpoints to explore alternatives~\cite{li2025agentgit}.
Its all-output enumeration model quantifies avoided prefix work, whereas its prompt-combination experiment compares the combined framework; neither isolates a general parallel-branching speedup or proves external-state reversibility.
Cortex instead caches remote query results, checking their suitability for a new query before avoiding a live fetch~\cite{ruan2026cortex}.
It separately limits retention through a user-defined time to live; semantic matching and aging address different reasons why a cached response may be unsuitable.
DualSpec selects inexpensive action drafts and uses a semantic progress check to decide whether full reasoning is needed~\cite{zhong2026dualspec}.
The engineering distinction is what must remain valid when work is skipped: the restored state, the retrieved information, or the proposed next action.
Restoration does not establish reversal of external effects, semantic matching alone does not guarantee freshness, and a progress judgment does not enforce an authorization policy.
These source contrasts motivate representation-specific validation and separate accounting of the gains and obligations associated with each transformation.

XGrammar-2 instead turns changing output constraints into efficient decoding operations, using tag-triggered grammar dispatch, shared grammar caches, and just-in-time compilation~\cite{li2026xgrammar2}.
Its engine experiments measure compilation, latency, and throughput under dynamic tool-calling workloads.
Syntactic conformance is valuable at an integration boundary, but a well-formed call can still select an inappropriate tool or supply a semantically incorrect argument.
The engineering assessment must therefore establish which properties survive the representation change alongside any execution-speed improvement.

\begin{table*}[t]
  \centering
  \caption{Engineering problems and approaches in execution implementation. Rows distinguish reducing execution cost, checking calls, enforcing authority, and recovering effects.}
  \label{tab:impl-execution-comparison}
  \small
  \setlength{\tabcolsep}{4pt}
  \renewcommand{\arraystretch}{1.16}
  \begin{tabular}{@{}L{0.18\textwidth}L{0.20\textwidth}L{0.36\textwidth}L{0.20\textwidth}@{}}
    \toprule
    \textbf{Engineering problem} & \textbf{Approach} & \textbf{Key idea} & \textbf{Representative studies} \\
    \midrule
    Reducing repeated computation & Plan reuse & Retrieve successful plan templates and adapt cache hits to the current execution context. & APC~\cite{zhang2025apc} (2025) \\
    & Shared message encodings & Reuse global KV-cache representations through separate prefill and decode operations. & Prompt Choreography~\cite{bai2026choreography} (2026) \\
    & Context-adjusted cache reuse & Estimate prefix-dependent offsets from online anchors, with dense-prefill fallback. & KVCOMM~\cite{ye2025kvcomm} (2025) \\
    & Adapter-aware cache sharing & Share base caches and retain low-rank adapter contributions during attention computation. & LRAgent~\cite{jeon2026lragent} (2026) \\
    & History compression & Shorten selected trajectory steps using an external reflection procedure. & AgentDiet~\cite{xiao2026agentdiet} (2026) \\
    \midrule
    Making tool interactions conform & Grammar-constrained decoding & Dispatch and cache grammars to constrain dynamically selected tool-call formats. & XGrammar-2~\cite{li2026xgrammar2} (2026) \\
    & Invocation middleware & Check calls before execution, process returned information, and detect silent failures. & ALTK~\cite{wright2026altk} (2026) \\
    \midrule
    Enforcing authorized execution & Learned policy interception & Classify policy--trajectory compliance and use the decision to intercept actions. & PolicyGuard~\cite{wen2026policyguard} (2026) \\
    & Resource permissions & Convert reviewed capability manifests into external filesystem, network, and environment restrictions. & AgentBound~\cite{buehler2026agentbound} (2026) \\
    & Dataflow authority tracking & Propagate sink-specific authority with values and restrict composed tool calls. & ChainCaps~\cite{jiang2026chaincaps} (2026) \\
    \midrule
    Recovering partial execution & Effect-aware recovery & Record tool effects and choose retries, alternatives, or compensating operations. & RAC~\cite{perera2026rac} (2026) \\
    \bottomrule
  \end{tabular}
\end{table*}

\subsubsection{Enforcement, Recovery, and Integration Evidence}
Defensive implementation checks inputs and operations at interfaces, handles exceptional conditions, and provides recovery behavior when execution fails~\cite{washizaki2026swebok}.
For an Agent, these checks surround model-generated calls whose format, meaning, and permissions can fail independently.
Recovery also has to account for already completed tool effects, since retrying a call may repeat an external change.
The remaining groups in Table~\ref{tab:impl-execution-comparison} compare call validation, authority enforcement, and effect-aware recovery as separate implementation responsibilities.

Constraint configuration, violation detection, invocation control, and recovery act at different points in an execution.
LCO separates prompt-level constraint construction from candidate selection: Self-Thought synthesizes task-specific constraints, and Guided Evolutionary Exploration searches and selects candidates under those constraints~\cite{wan2026lco}.
Module ablations examine the contributions of constraint construction and candidate search separately.
Both operate through model-mediated generation and assessment: the implementation can reject a candidate judged unsafe, while the quality of the constraint and judgment determines which unsafe candidates remain possible.

Similarly, the dark-pattern study configures graded prompt postscripts, whereas PolicyGuard makes a classification available to an action interceptor~\cite{ersoy2025darkpatterns,wen2026policyguard}.
The supplemental CRAFT study tests policy priority and repeated full or selected policy reminders; textual priority does not reliably enforce that hierarchy under adversarial interaction~\cite{nakash2025craft}.
These examples expose three implementation decisions: how a constraint is represented, how a violation is recognized, and which execution point enforces the decision.
They also locate different failure modes: a model can disregard textual guidance, and an interceptor can act on an incorrect classification.

PolicyGuard learns policy--trajectory compliance and is also exercised as an online interceptor in a closed-loop web-Agent simulation~\cite{wen2026policyguard}.
This evidence distinguishes detection from intervention, while its error analysis still identifies false negatives under cumulative constraints and ambiguous action descriptions.
AgentBound enforces filesystem, environment, and network permissions outside the model, but cannot prevent application-level attacks that remain inside allowed resources~\cite{buehler2026agentbound}.
RAC addresses failures after actions have begun: interceptors record tool events, and a recovery manager retries, seeks alternatives, or compensates prior effects~\cite{perera2026rac}.
Its LangGraph implementation and failure scenarios exercise that control path; compensation still depends on available inverse operations and valid input mappings, sometimes obtained through an LLM.
Correctness assessment must cover both the control mechanism and the behavior of the permitted or recovered execution.

State restoration and durable writeback address other failure boundaries.
AGDebugger checkpoints Agent state before messages and forks sessions for interactive debugging, but cannot undo every external effect~\cite{epperson2025debugging}.
ClawVM instead validates scoped page updates before persistence across compaction or reset~\cite{rafique2026clawvm}.
Its fault tests concern persistence and update handling; validating the truth of stored content is a further obligation.
The bounded live-hook comparison does not trigger compaction or reset, leaving the policies' behavioral differences under those conditions unmeasured.
He et al. propose session identifiers and separated interaction histories, including correlation with model requests; the paper's sandbox attack experiments do not directly validate this session-isolation contribution~\cite{he2025securityagents}.
The relevant distinction is between restoring internal state, compensating external effects, and keeping concurrent states separate.

Execution controls also differ in the decision they enforce.
TraceFix's topology monitor checks declared coordination operations and resource access, while step-order discipline remains in prompts~\cite{xia2026tracefix}.
Its formal verification checks bounded coordination safety, not liveness; the verification scope and the runtime enforcement boundary must therefore be reported separately.
BlindGuard detects anomalous participants and prunes message connections; the Guardian in BAD-ACTS classifies messages and terminates an episode when it detects unsafe content~\cite{miao2026blindguard,noether2026badacts}.
Evaluating these interventions requires tracking both prevented harm and the legitimate work lost through pruning or termination.
Koala II exposes participation thresholds and response-placement settings to conversation participants, addressing interaction control rather than an adversarial authorization boundary~\cite{houde2025participation}.
The cases require separate accounts of coordination conformance, participant isolation, episode termination, and interaction control.

Tool middleware also separates invocation quality from authority over a chain of calls.
ALTK places middleware around the exchange between an Agent and its tools~\cite{wright2026altk}.
Before a call, LLM-based semantic validation can request regeneration; after execution, output extraction and silent-failure checks help determine what the Agent should consume or retry.
The paper evaluates selected components on separate benchmarks, leaving their combined reliability to be established in an integrated execution path.

The supplemental CAI defense combines container isolation, command-tool response filtering, script-write checks, and configurable input/output guards~\cite{mayoral2025caiinjection}.
Tool-path governance depends on these concrete insertion points; enforcement of information-flow security requires controls beyond a natural-language data marker.
The authors observe zero successful attacks in 140 defended attempts from their fixed attack set, providing evidence for that configuration and threat coverage.

ChainCaps instead propagates sink-specific authority with values and restricts composed tool calls~\cite{jiang2026chaincaps}.
Its non-amplification argument assumes correct manifests, proxy-visible data movement, and no declassification; the implementation provides a separate, explicit declassification path.
The attack study is a stress test, not an estimate of deployment risk, and residual shell indirection illustrates the proxy's visibility boundary.
Unlike RAC's compensation after effects occur, these controls act on the acceptability or authority of an invocation; unlike AgentBound's resource restrictions, ChainCaps also tracks authority through composed dataflow.
Combining them requires tests of their interaction under shared execution and failure conditions.

Empirical studies complement constructed systems with evidence about deployment and validation practice.
Oyelayo et al. analyze 136 issues from seven MCP reference servers, including incomplete or overlapping tool descriptions, and relate resolved issues to code churn~\cite{oyelayo2026mcpchallenges}.
This identifies concrete integration work that an interface specification alone does not remove.
The sample is limited to reference servers, and changed lines are not a direct measurement of developer time.

Supplemental studies extend construction to generated harness code and configuration.
AutoHarness searches executable code using environment feedback; its generated verifier subsequently rejects illegal action proposals~\cite{lou2026autoharness}.
MemoHarness instead separates global configuration search from feedback-free, case-specific construction: frozen experience conditions a concrete harness configuration, represented by a policy file and accompanying operating instructions~\cite{huang2026memoharness}.
The latter is a configuration-construction operation, but the reported system comparisons do not isolate its individual effect.
Together, these studies show how construction can produce executable controls or operating instructions, with evaluation still needed to isolate their contribution within the surrounding toolchain.

\paragraph{Synthesis.}
Implementation has the highest direct-evaluation share, with 87 of 91 contributions evaluated.
Eighty-one are coded E2, two E3 and one E4; the last is DLBP's duration-unresolved operational case (Appendix~\ref{sec:evidence-setting-cases}). Persistent artifacts are available or partly available for 70.
The predominance of controlled evaluations provides evidence about particular adapters, transformations, and execution controls; broader integration assessments also need to measure their interaction in the assembled system.

\finding[Findings]{Shared specifications, compilation pipelines, and trajectory adapters provide concrete ways to construct Agent software and connect heterogeneous frameworks, with validation through compilation checks and downstream execution~\cite{amini2025agentspec,ouyang2026skcc,song2026adp}. Cache and history transformations reduce execution costs under measured reuse conditions, while external permissions and recovery managers control resource access and partial failures~\cite{ye2025kvcomm,xiao2026agentdiet,buehler2026agentbound,perera2026rac}. These results support executable construction and specific integration checks, but behavior can still depend on runtime interpretation, model-based judgments, and external side effects. Comparative evidence remains limited on developer effort, portability, dependency upgrades, and the joint behavior of multiple controls. Further evaluation should follow the same integration across frameworks and versions, measuring both engineering effort and preservation of its intended behavior.}

\subsection{Testing and Assurance}
\label{sec:test-methods}

Software testing executes a system under selected conditions and compares observed behavior with an expected outcome; assurance brings testing together with analysis and other evidence about required properties~\cite{washizaki2026swebok}.
The purpose is to expose defects and support decisions about whether software is fit for use and whether changes are acceptable.
For Agents, test conditions include prompts, tool state, policies, and model configuration, while the outcome can involve an entire sequence of actions and external effects.
A task-completion score consequently needs to be related to policy compliance, the validity of the test execution, and behavior across repeated runs or revised versions.
These distinctions appear directly in policy-aware benchmarks and specification-driven test construction~\cite{levy2026stwebagentbench,ahmed2026specops}.
The stage's 264 contributions address how to make such judgments supportable.
Table~\ref{tab:test-design-comparison} groups methods for making requirements testable, challenging adversarial behavior, and controlling test executions.
Table~\ref{tab:test-assurance-comparison} follows with methods for trustworthy assessment, diagnosis, regression checks, and runtime compliance.

\subsubsection{Constructing Test Conditions}
Test design turns requirements and fault hypotheses into inputs, preconditions, execution procedures, and expected outcomes~\cite{washizaki2026swebok}.
These elements make a failure interpretable: the intended behavior must be reachable in the prepared environment, and the check must distinguish acceptable from defective behavior.
Agent tests additionally need to specify what the user, tools, or adversary can change and how natural-language expectations will be checked against actions and state.
The groups in Table~\ref{tab:test-design-comparison} distinguish clarifying test expectations, supplying adverse inputs, and constructing an environment in which the target behavior can actually occur.
DoomArena separates attack configuration from environment-specific gateways, allowing a declared threat model to control where an attack enters an Agent--environment interaction~\cite{boisvert2025doomarena}.
Botender instead generates interaction cases to expose ambiguous intentions and community disagreements; the cases support deliberation about desired behavior, not only pass/fail validation~\cite{kuo2026botender}.
These are complementary ways to construct revealing inputs: one varies an adversary's authority, while the other challenges the specification itself.

Even for indirect prompt-injection tests, the starting state changes what a test establishes.
InjecAgent supplies an already obtained tool response and checks the ensuing tool calls, including a two-step exfiltration case; it does not require the Agent to reach that state during a complete task~\cite{zhan2024injecagent}.
Its attack success rate over valid outputs differs from the rate over all outputs, so parsing failures must remain visible.
AgentDojo instead executes tools over mutable environment state and checks task outcomes against that state~\cite{debenedetti2024agentdojo}.
It separates benign utility, utility under attack without adversarial side effects, and targeted attack success.
When aggregating a collection of attacks, it counts a security case as compromised if any attack in that collection succeeds.
The contrast is between susceptibility at a supplied interaction point and security together with task execution, not two interchangeable robustness scores.

The execution environment and oracle also require engineering.
SpecOps translates feature descriptions into a bundled specification of environment setup, the prompt to the tested Agent, and expected behavior~\cite{ahmed2026specops}.
A test-generation specialist proposes cases and a second specialist checks prompt completeness, setup feasibility, and whether the oracle admits different valid execution paths.
An infrastructure specialist prepares the environment and updates the shared specification when actual constraints require changes; an execution specialist launches the tested Agent and records screen changes.
The validation stage combines these observations with environment evidence to judge the expected outcome.
Experiments on five GUI-based Agents compare planning, execution, and bug detection with two automation baselines; failure analysis separates inability to set up or activate the tested Agent from defects in its behavior.
This makes the test infrastructure itself an evaluated part of the method, since a broken setup can otherwise produce a misleading bug report.

ST-WebAgentBench focuses on defining acceptable behavior within a successfully executed task by associating web tasks with explicit policies and evaluating trajectory actions, sequences, URLs, and content~\cite{levy2026stwebagentbench}.
Completion Under Policy credits only successful, policy-compliant tasks.
The reported pass@3 criterion credits success in any of three attempts, whereas the stricter all-pass@3 requires successful, policy-compliant completion in every run.
User confirmation is simulated by auto-approval and template checks on requests, not assessed with real users.
A small gap between completion and policy-compliant completion can reflect failure to reach policy-relevant states rather than safer behavior.
Together with AgentDojo's attack aggregation, this illustrates why an attempt budget and aggregation rule must accompany a success rate: finding one failure and sustaining compliant success are different assurance targets.

\begin{table*}[t]
  \centering
  \caption{Engineering problems and approaches in test construction. The groups show how studies make requirements testable, challenge adversarial behavior, and control the execution under test.}
  \label{tab:test-design-comparison}
  \small
  \setlength{\tabcolsep}{4pt}
  \renewcommand{\arraystretch}{1.16}
  \begin{tabular}{@{}L{0.18\textwidth}L{0.20\textwidth}L{0.36\textwidth}L{0.20\textwidth}@{}}
    \toprule
    \textbf{Engineering problem} & \textbf{Approach} & \textbf{Key idea} & \textbf{Representative studies} \\
    \midrule
    Making requirements testable & Boundary-challenging cases & Generate interactions that reveal ambiguous intentions and support stakeholder deliberation. & Botender~\cite{kuo2026botender} (2026) \\
    & Policy-grounded scenarios & Couple tasks with explicit policies and check compliant completion across repeated runs. & ST-WebAgentBench~\cite{levy2026stwebagentbench} (2026) \\
    \midrule
    Challenging adversarial behavior & Configurable attack injection & Define attacker authority and inject attacks through environment-specific gateways. & DoomArena~\cite{boisvert2025doomarena} (2025) \\
    & Interaction-point testing & Supply an injected tool response and examine the Agent's subsequent calls. & InjecAgent~\cite{zhan2024injecagent} (2024) \\
    & Stateful security testing & Execute benign and attacked tasks over mutable state to assess utility and adversarial effects. & AgentDojo~\cite{debenedetti2024agentdojo} (2024) \\
    & Misuse testing & Exercise malicious-user requests with synthetic tools, harm rubrics, and refusal checks. & AgentHarm~\cite{andriushchenko2025agentharm} (2025) \\
    & Computer-use red teaming & Inject content into isolated web--OS tasks and distinguish attempted from executed attacks. & RedTeamCUA~\cite{liao2026redteamcua} (2026) \\
    & Backdoor assessment & Poison fine-tuning data and test triggered actions alongside ordinary task behavior. & BadAgent~\cite{wang2024badagent} (2024) \\
    \midrule
    Controlling test executions & Coupled test harness & Coordinate case generation, environment setup, execution, and validation through a specification. & SpecOps~\cite{ahmed2026specops} (2026) \\
    & Tool emulation & Simulate risky tool conditions and validate whether the resulting traces are realizable. & ToolEmu~\cite{ruan2024toolemu} (2024) \\
    & Controlled protocol comparison & Hold workflows and non-protocol settings fixed while exercising native protocol behavior. & ProtocolBench~\cite{du2026protocolbench} (2026) \\
    & Environment fault injection & Inject a cloud-service fault to assess an incident-handling Agent under known conditions. & AIOpsLab design~\cite{shetty2024autonomousclouds} (2024) \\
    \bottomrule
  \end{tabular}
\end{table*}

Evaluation environments also differ in their threat models and measurement procedures.
AgentHarm tests misuse by a malicious user, using side-effect-free synthetic tools, partial-credit rubrics, narrow semantic checks, and a separate refusal judge~\cite{andriushchenko2025agentharm}.
Benign counterparts help distinguish refusal from inability to carry out the tool interactions; the synthetic tasks remain proxies rather than demonstrations of real-world harm.
RedTeamCUA tests indirect injections into an otherwise benign computer-use task in an isolated web--OS environment~\cite{liao2026redteamcua}.
Its decoupled evaluation starts at the injection site, and it distinguishes execution-checked attack success from judge-assessed attempts.
The separate end-to-end evaluation uses a deliberately selected high-risk subset, not an unselected estimate of deployment risk.
Thus, threat authority, exposure, and the distinction between attempted and completed behavior must accompany a benchmark score.

Controlled comparison further requires deciding what to hold constant.
ProtocolBench fixes workflows and non-protocol settings while preserving protocol-native retry, reconnect, and streaming behavior~\cite{du2026protocolbench}.
It therefore compares native implementations under shared conditions, not protocols under every possible workload.
Agent Spec fixes the declarative design and tool interfaces, while leaving prompting, tool calling, and orchestration to each runtime~\cite{amini2025agentspec}.
Its task benchmarks compare these implementation stacks; they do not isolate the causal effect of each differing mechanism.
The AIOpsLab design exercises incident handling through injected cloud faults, but the reported proof of concept concerns a single misconfiguration incident~\cite{shetty2024autonomousclouds}.
The injected fault and restored state belong to the external cloud service; the measured subject is the Agent's incident-handling behavior, not recovery from a defect in the Agent's own implementation.
This supplies evidence about the controlled assessment of incident handling, not general recovery from Agent software defects.
BadAgent changes the intervention point again: it poisons fine-tuning data and tests triggered harmful actions alongside otherwise correct task steps~\cite{wang2024badagent}.
Its tested defenses and model sizes delimit the result; a surviving backdoor is not evidence that every defense will fail.

\subsubsection{Validating Judgments and Execution Evidence}
In software testing, a test oracle determines whether an observed result is acceptable; the evidence supplied to it must also correspond to the execution under assessment~\cite{washizaki2026swebok}.
An incorrect oracle can reject valid behavior or overlook a defect, making oracle validation part of the testing problem.
For Agents, natural-language outputs and long interaction histories motivate learned judges, while delegated execution raises the additional question of whether the reported trace and configuration are authentic.
The assessment group in Table~\ref{tab:test-assurance-comparison} therefore separates human-calibrated judgments, configuration-relative authentication, and verifiable protocol execution.
ToolEmu illustrates why infrastructure and assessment cannot be conflated: an emulated tool failure is useful only if its inputs, outputs, and state evolution could describe a realizable execution~\cite{ruan2024toolemu}.
The study checks this instantiability with human annotations and evaluates helpfulness as well as safety, since an inactive Agent can avoid harm without accomplishing useful work.
AgentAuditor constructs an oracle from prior labeled interactions~\cite{luo2025agentauditor}.
It extracts scenario, risk, and behavior features, clusters their representations to select representative cases, and generates reasoning examples conditioned on those cases' reference labels.
For a new record, content similarity retrieves candidate examples and feature similarity reranks them; the selected examples guide the final LLM judgment.
Evaluation against human-annotated safety and security cases, together with analyses of memory quality, tests whether this procedure improves judgment accuracy.
The evidence concerns binary, English-language assessments, with the quality and relevance of stored cases remaining important conditions on the oracle.

The supplemental Leni study illustrates a further distinction between preserving task utility and validating a verifier's false-positive rate~\cite{dastidar2026reliability}.
Its released rubric scores can be recomputed~\cite{dastidar2026leniartifacts}, but that operation does not independently establish the binary over-rejection labels needed for a calibration claim.
The assurance target therefore requires an explicit event definition and observation unit alongside aggregate task scores.

VET establishes a different property: component proofs and transcript consistency authenticate an output relative to a declared Agent configuration~\cite{grigor2026vet}.
Its trust assumptions concern the underlying cryptographic and external components; authentication does not imply semantic correctness, liveness, or freshness.
Likewise, G2CP makes communication replayable through graph operations and gives protocol properties relative to the graph~\cite{benkhaled2026g2cp}.
The resulting guarantee depends on the supplied graph and its construction.
Its evaluation includes a synthetic industrial knowledge base and a small pilot using real industrial cases; 21 cases remain after four lacking expert consensus are excluded.
These experiments exercise protocol execution under the specified graph-construction and case-selection conditions.
For both methods, verifying a record or operation and assessing its substantive correctness remain separate assurance tasks.

\FloatBarrier
\subsubsection{Diagnosing Defects and Characterizing Practice}
Software diagnosis relates an observed failure to a suspected fault in code, configuration, or interaction, providing a starting point for debugging and repair~\cite{washizaki2026swebok}.
The last visible error may be a consequence of an earlier fault, so diagnostic methods need a way to distinguish symptom from cause.
In Agent systems, the search may span source code, generated actions, tool responses, and messages between Agents, with different methods operating on different parts of that evidence.
The defect-location group in Table~\ref{tab:test-assurance-comparison} compares source analysis, trajectory attribution, and constraint-guided diagnosis; empirical studies establish which problems and practices these tools need to address.
Agentable first filters a project to its code files and builds a code property graph capturing syntax, control flow, and data flow~\cite{ning2026agentdefects}.
Defect-specific queries locate relevant initialization and tool code; an enrichment module retrieves source and syntax-tree context, while LLM-based analysis checks semantic relationships such as agreement between a tool's description and implementation.
The combined results produce a defect report, allowing structural and natural-language interface problems to be investigated together.
Evaluation checks reported defects in collected projects and recall on a separate defect dataset.
Its implementation targets Python, standalone Agents; defects dependent on unavailable external state remain a limitation of source-based inspection.

Trace-based localization asks a different question from source-level defect detection.
Who\&When supplies annotated failed runs and compares full-log, stepwise, and binary-search judgments~\cite{zhang2025whowhen}.
Its target is the earliest decisive Agent--step error, defined through a hypothetical correction that changes the outcome with a corrected continuation.
The empirical reference labels come from expert annotation and consensus, not replayed repair interventions.
Agent-level agreement, exact-step agreement, and agreement within a step tolerance therefore measure distinct localization outcomes, none of which alone validates a repair.

AgentRx adds intermediate evidence before localization: tool schemas, optional policies, and trajectory prefixes generate constraints; executable or semantic checks produce a step-indexed violation log~\cite{barke2026agentrx}.
A judge uses that log and a failure taxonomy to identify the earliest unrecoverable failure.
The log informs rather than dictates the judgment: generated constraints may be wrong, and a genuine violation may be downstream of the critical failure.
This contrasts with stopping at the first detected mistake, as in Who\&When's stepwise baseline.
Both studies contribute datasets for comparing localization against annotated failures.
The next engineering step is to test whether acting on a suggested location yields a repair that restores the expected behavior.

Tangent complements such detectors by mining repository tests and consulting practitioners~\cite{pan2026tangent}.
Encoded tests and reported practice reveal different activities: repository extraction cannot establish that unrecorded testing or operational checks do not occur.
The supplemental TRIAGE study separately examines what evolved harnesses encode and how their benefits transfer, complementing evidence about existing tests with evidence about the controls produced through evolution~\cite{yang2026triage}.

Runtime measurements also need a well-defined denominator.
SWEnergy measures CPU/GPU energy with idle-baseline subtraction and records task outcomes under a fixed hardware configuration~\cite{tripathy2026swenergy}.
Its small-model experiments include configurations that consume little energy but resolve no issues, so lower energy cannot alone establish more efficient successful work.
The operational tap study analyzes repeated review artifacts during development of the collaboration protocol itself~\cite{kim2026tap}.
Its observations are useful longitudinal evidence, but repeated reviews are not independent defects, and model, reviewer-role, environment, and time effects are confounded.
The pairing comparison is observational, with causal model effects unresolved under these confounding conditions.

Maintaining test evidence requires preserving the conditions under which a result was obtained.
ATAG's vulnerability database retains attack procedures, model versions, tool permissions, outcomes, and sources as reusable threat-assessment records~\cite{gandhi2026atag}.
This preserves the setting needed to interpret an attack result across versions.
The database provides an organization for version-specific threat evidence.
Evaluation of the record-management mechanism remains absent, even though the wider paper illustrates applications of the framework.

\subsubsection{Regression Gates and Specification-Guided Repair}
Regression testing re-exercises relevant behavior after a modification to detect unintended changes; specification-guided repair uses violated properties to direct the modification itself~\cite{washizaki2026swebok}.
Together, they connect a proposed fix to evidence that the changed system meets its obligations.
For Agent software, an altered skill or coordination protocol may fix one trajectory while disrupting another, and a verified protocol must still be realized faithfully by the runtime.
The candidate-change group in Table~\ref{tab:test-assurance-comparison} compares two responses: test the actual candidate against an evaluation contract, or use counterexamples to revise a protocol and check its runtime realization.
AEVAL colocates a skill's evaluation contract with its implementation and separates execution from read-only grading~\cite{anand2026aeval}.
The first-attempt rule prevents an executor's self-repair from being credited as an initially passing version.
This exposes the behavior of the candidate version before corrective retries obscure an initial failure.
Reported grader disagreements and differences between backend-specific assertion sets identify remaining obstacles to consistent acceptance decisions.

TraceFix uses bounded TLA+ counterexamples to repair coordination protocols, then compares ways of enforcing them at runtime~\cite{xia2026tracefix}.
Its paired experiment holds the runtime fixed while varying protocol verification, supporting a narrower conclusion than comparing whole architectures alone.
The paper also exposes a handoff obligation: a runtime that enforces topology and mutual exclusion but leaves step ordering to prompts need not preserve all properties of the checked model.
Model-level repair therefore requires an implementation-level conformance argument; bounded protocol checks do not establish unbounded liveness or task semantics.

\begin{table*}[t]
  \centering
  \caption{Engineering problems and approaches in assessment and assurance. The groups connect judgments and execution records to defect diagnosis, release checks, and runtime assurance.}
  \label{tab:test-assurance-comparison}
  \small
  \setlength{\tabcolsep}{4pt}
  \renewcommand{\arraystretch}{1.16}
  \begin{tabular}{@{}L{0.18\textwidth}L{0.20\textwidth}L{0.36\textwidth}L{0.20\textwidth}@{}}
    \toprule
    \textbf{Engineering problem} & \textbf{Approach} & \textbf{Key idea} & \textbf{Representative studies} \\
    \midrule
    Establishing trustworthy assessments & Memory-assisted judgment & Use feature and reasoning memories to assess interactions against human-labeled cases. & AgentAuditor~\cite{luo2025agentauditor} (2025) \\
    & Configuration-relative authentication & Compose component proofs and check transcripts against a declared Agent configuration. & VET~\cite{grigor2026vet} (2026) \\
    & Verifiable protocol execution & Check graph-grounded operations and retain replayable communication records. & G2CP~\cite{benkhaled2026g2cp} (2026) \\
    \midrule
    Locating defects & Source analysis & Combine code-property-graph patterns and semantic analysis to identify Agent-program defects. & Agentable~\cite{ning2026agentdefects} (2026) \\
    & Trace-based attribution & Locate the responsible Agent and decisive error step using annotated failure trajectories. & Who\&When~\cite{zhang2025whowhen} (2025) \\
    & Constraint-guided diagnosis & Check trajectory constraints and use violation evidence to identify a critical failure. & AgentRx~\cite{barke2026agentrx} (2026) \\
    \midrule
    Checking candidate changes & Contract-based regression & Separate execution from grading and assess the first attempt against a colocated evaluation contract. & AEVAL~\cite{anand2026aeval} (2026) \\
    & Counterexample-guided repair & Use bounded protocol counterexamples to revise coordination and compare runtime enforcement. & TraceFix~\cite{xia2026tracefix} (2026) \\
    \midrule
    Assessing runtime compliance & Policy-to-trace checking & Formalize privacy policies, annotate observed practices, and check compliance with automata. & AudAgent~\cite{zheng2026audagent} (2026) \\
    & Activation-based detection & Detect deviations from benign activation prototypes and steer regeneration. & AcMAS~\cite{xu2026acmas} (2026) \\
    & Internal-signal scoring & Classify attention features to estimate dimension-specific violations for downstream controls. & A-Trust~\cite{he2026atrust} (2026) \\
    \bottomrule
  \end{tabular}
\end{table*}

\FloatBarrier
\subsubsection{Runtime Assurance and Its Evidence Boundary}
Runtime assurance supplements pre-release checks by monitoring the deployed execution for violations of specified properties and triggering an appropriate response~\cite{washizaki2026swebok}.
Its usefulness depends on what can be observed, how a violation is recognized, and whether intervention occurs before an unacceptable effect.
For Agents, the monitored evidence can be an external action trace or an internal model signal, with different access requirements and interpretation risks.
The final group in Table~\ref{tab:test-assurance-comparison} compares policy-to-trace checking with learned activation or attention signals; each exposes different properties and requires different access to the running system.
AudAgent translates privacy policies into a structured model, annotates observed data practices, and checks them through ontology alignment and automata~\cite{zheng2026audagent}.
It connects a stated policy to a runtime trace, but simplification of collection and processing conditions can lose policy nuance.
AcMAS instead learns benign activation prototypes, detects deviations, and applies activation-level correction during regeneration~\cite{xu2026acmas}.
A-Trust trains dimension-specific classifiers on attention features; its trust-management system separately consumes the scores through threshold-based message controls~\cite{he2026atrust}.
The scoring mechanism must be evaluated separately from the downstream management operations that consume its estimates.
These learned internal signals can support intervention where the requisite model access is available; their outputs remain fallible estimates, and prototype or threshold calibration is part of the assurance procedure.
AcMAS explicitly identifies domain shift and adaptive attacks as unresolved limits.

\paragraph{Synthesis.}
Of 264 Testing contributions, 242 are directly evaluated and 230 are E2.
Five are coded E3 and five E4; 208 report available or partly available persistent artifacts.
The reviewed methods provide extensive support for constructing conditions, judging trajectories, diagnosing failures, checking protocols, and assessing runtime controls.
Their findings depend on the tested version, environment, property, oracle, observation scope, attempt budget, and aggregation rule.
These conditions determine whether two reported results are comparable and whether a result can support a release or repair decision.

\finding[Findings]{Testing research provides policy-grounded scenarios, controlled attack environments, and specification-driven test execution, making Agent behavior assessable against more explicit expectations~\cite{levy2026stwebagentbench,debenedetti2024agentdojo,ahmed2026specops}. Human-labeled judgment studies, defect datasets, and bounded protocol checks further support oracle validation, diagnosis, and candidate-change assessment~\cite{luo2025agentauditor,barke2026agentrx,xia2026tracefix}. These methods give substantive support to defect detection and checking required properties. However, the acceptance decision still depends on fallible judges, sampled execution conditions, and whether the implemented system preserves the checked specification. The large body of controlled tests is less conclusive about repeated-run reliability and prediction of operational failures. Future studies should validate oracles, evaluate repairs through re-execution and regression tests, and connect version-specific test results to observed behavior in use.}

\subsection{Maintenance and Evolution}
\label{sec:maint-methods}

Software maintenance modifies an existing system to correct faults, adapt to its environment, or improve its behavior while preserving the integrity of the software; evolution concerns its continuing development through such changes~\cite{washizaki2026swebok}.
Operational observation, impact analysis, modification, validation, and version control help ensure that the system continues to meet user needs.
For Agents, change can affect code, prompts, tool dependencies, policies, or persistent knowledge, and a successful intervention in one run may leave the source of a recurring problem unchanged.
Developers therefore need to connect observed symptoms to causes, distinguish an immediate response from a durable revision, and assess the resulting version.
The 54 contributions address parts of this process.
Table~\ref{tab:maint-observation-comparison} groups approaches to collecting operational evidence, inspection, anomaly detection, and fault diagnosis; Table~\ref{tab:maint-change-comparison} groups execution-level responses, durable revision, and version governance.
Their combination makes the maintenance question concrete: what needs to change, on which evidence, and with which acceptance check?

\subsubsection{Making Executions Inspectable and Failures Localizable}

Operational monitoring records events and conditions that help maintainers understand system behavior; diagnosis uses those observations to locate a suspected cause of failure~\cite{washizaki2026swebok}.
Useful instrumentation preserves the context needed to relate a symptom to the component or interaction that produced it.
In Agent systems, this requires connecting model calls, delegated work, and tool effects, since a final answer or a flat log can obscure how an error propagated.
Table~\ref{tab:maint-observation-comparison} follows the work needed to close this information gap, from instrumentation and readable views to anomaly detection and fault localization.
AgentSight illustrates system-boundary instrumentation: eBPF probes collect LLM communication and operating-system events, and a correlation engine joins them into logs for subsequent analysis~\cite{zheng2025agentsight}.
Its temporal and process-based links are heuristic; they provide an investigation record rather than a proof of the Agent's internal reasoning.
The reported overhead experiment covers three developer workflows with three repetitions each, delimiting the operational cost evidence.
Agent Behavior Mining extends an XES-compatible event model with prompts, model calls, tool executions, delegation, and associated costs~\cite{vu2026agentbehaviormining}.
Events share conversation identifiers and record Agent or activity names, timestamps, duration, and token use, enabling process discovery, conformance checking, performance analysis, and comparison of execution variants.
This gives maintainers both an execution representation and views for asking where observed behavior differs from an expected process or consumes disproportionate resources.
An 18-practitioner workshop uses a constructed scenario and traces with induced failures to assess inspectability and perceived utility.
Evidence about incident reduction in production remains absent; session-level correlation and unavailable reasoning traces also constrain the analysis.

Human-facing tools organize this information for diagnosis, but evaluate different parts of the inspection-to-correction process.
Agent Trajectory Explorer normalizes heterogeneous trace formats for visual inspection and positive/negative feedback~\cite{desmond2025trajectoryexplorer}.
The demonstration establishes the inspection and feedback path; how that feedback should produce an Agent revision remains an open design question.
DiLLS presents layered summaries of plans, actions, and operations for centralized multi-Agent systems~\cite{sheng2026dills}.
Its 12-participant comparison with chronological logs evaluates diagnostic performance, confidence, and workload, without establishing the long-term effect on maintenance practice.
XAgen links workflow visualization, human feedback, configuration edits, and LLM-judge explanations~\cite{wang2026xagen}.
Following formative interviews with 12 practitioners, its eight-participant study compares subjective understanding, usability, helpfulness, and satisfaction across two tasks.
Objective correction accuracy and investigation time remain proposed evaluation outcomes, so the current evidence concerns practitioners' perceived assistance.

\begin{table*}[t]
  \centering
  \caption{Engineering problems and approaches in operational observation and diagnosis. The groups distinguish collecting evidence, supporting inspection, detecting anomalies, and locating or characterizing faults.}
  \label{tab:maint-observation-comparison}
  \small
  \setlength{\tabcolsep}{4pt}
  \renewcommand{\arraystretch}{1.16}
  \begin{tabular}{@{}L{0.18\textwidth}L{0.20\textwidth}L{0.36\textwidth}L{0.20\textwidth}@{}}
    \toprule
    \textbf{Engineering problem} & \textbf{Approach} & \textbf{Key idea} & \textbf{Representative studies} \\
    \midrule
    Collecting operational evidence & System-boundary instrumentation & Correlate LLM communication with process and operating-system events. & AgentSight~\cite{zheng2025agentsight} (2025) \\
    & Event-model standardization & Represent calls, delegation, and costs for process and conformance analysis. & Agent Behavior Mining~\cite{vu2026agentbehaviormining} (2026) \\
    & Step-level instrumentation & Persist structured events before and after runtime steps. & NEXUS~\cite{langay2026nexus} (2026) \\
    \midrule
    Supporting maintainer inspection & Hierarchical summarization & Organize plans, actions, and operations into navigable diagnostic summaries. & DiLLS~\cite{sheng2026dills} (2026) \\
    & Trace normalization & Align heterogeneous trajectories for visual inspection and feedback. & Trajectory Explorer~\cite{desmond2025trajectoryexplorer} (2025) \\
    & Workflow-linked explanation & Connect logs and judge rationales to feedback and configuration editing. & XAgen~\cite{wang2026xagen} (2026) \\
    \midrule
    Recognizing anomalous behavior & Repetition analysis & Combine repeated calls and output similarity to identify harmful cycles. & Cycle Detection~\cite{george2025cycledetection} (2025) \\
    & Message-boundary heuristics & Flag candidate unfaithful refusals from tool failures and response wording. & Guardrails as Scapegoats~\cite{singh2026scapegoats} (2026) \\
    \midrule
    Localizing faults & Replay-based attribution & Rank suspicious Agents and actions using replayed and abstracted trajectories. & Famas~\cite{ge2026famas} (2026) \\
    & Dependency-graph analysis & Rank candidate causes using trace dependencies and positional features. & AgentTrace~\cite{wang2026agenttracecausal} (2026) \\
    & Query-based diagnosis & Generate executable trace queries to identify a deficiency and affected component. & Pathfinder~\cite{atreja2026pathfinder} (2026) \\
    \midrule
    Characterizing recurring problems & Issue-to-change analysis & Relate reported integration problems to merged changes and code churn. & MCP developer study~\cite{oyelayo2026mcpchallenges} (2026) \\
    \bottomrule
  \end{tabular}
\end{table*}

Event- and trace-based detectors target different execution conditions and fault hypotheses, each requiring a corresponding reference and validation procedure.
The payload--response heuristic in Guardrails as Scapegoats flags candidate unfaithful safety refusals at the message boundary~\cite{singh2026scapegoats}.
It combines a silent tool-failure signal with policy language in the final response, producing an anomaly candidate rather than an explanation of the underlying defect.
On the inspected baseline, it produces no false positives but misses the sole unfaithful-safety-refusal instance because that response uses capability rather than policy language.
The missed instance reveals a dependence on surface wording that must be addressed before the heuristic can reliably prioritize these failures.

Unsupervised Cycle Detection combines call-sequence repetition with output similarity to classify harmful cycles~\cite{george2025cycledetection}.
The inspected author version evaluates 1,575 trajectories from one stock-market application, including 57 labeled bad cycles; it detects an execution condition rather than locating the defect that caused it.
Famas supplies a different output: it replays executions, abstracts actions, and applies spectrum analysis to rank suspicious Agents and actions~\cite{ge2026famas}.
Evaluation on the annotated Who\&When benchmark tests attribution rather than whether a subsequent repair removes the defect.
AgentTrace reconstructs sequential, communication, and data-dependency edges, then ranks candidate causes with weighted structural and positional features~\cite{wang2026agenttracecausal}.
Its 550-scenario synthetic benchmark uses injected bugs with removal checks to establish reference locations, but this reference construction is not an intervention performed by the diagnostic algorithm on each analyzed system.
The authors report limitations from single-root-cause scenarios and a preference for early positions; production-trace validation remains future work.

Pathfinder instead generates executable queries over trace data to predict a deficiency type and affected component~\cite{atreja2026pathfinder}.
The inspected 23-page author version describes evaluation on executions of four Agent frameworks with injected code defects, using injection type and location as the reference.
This differs from both binary cycle detection and ranking an annotated action: the target, ground truth, and search space change together.
Automated repair is explicitly future work, so the reported scores concern diagnosis rather than repair effectiveness.
The author version also gives inconsistent descriptions of shared versus separate model weights, limiting reproducibility of the learned analyzer.
The fault-localization group in Table~\ref{tab:maint-observation-comparison} thus compares how a suspected cause is obtained: replay, dependency analysis, or generated trace queries.
Their reported accuracy remains relative to different fault targets and reference labels, so selecting a diagnostic approach also requires matching it to the kind of defect a maintainer needs to investigate.

Repository and issue studies supply evidence about maintenance work already performed.
The MCP developer study links reported issues to merged pull requests and measures added and deleted lines~\cite{oyelayo2026mcpchallenges}; SelfHeal derives repair patterns from reported Agent defects~\cite{islam2026selfheal}.
These studies connect problems to changes already made, complementing tools that diagnose one execution.
Their categories describe observable repository and reporting practices; the code-churn measure estimates change size, with elapsed repair time and recurrence requiring further observation.

\subsubsection{From Detection to Bounded Runtime Intervention}

Operational incident handling connects detection to triage, containment, and restoration of service~\cite{washizaki2026swebok}.
The response must be proportionate to the evidence because automatic blocking can interrupt legitimate work, while delayed intervention can allow harmful effects to continue.
For Agent systems, a response may block a pending disclosure, append corrective guidance, or restrict a participant's communication during the current execution.
The execution-response group in Table~\ref{tab:maint-change-comparison} distinguishes triage, preventive blocking, corrective guidance, and participation control according to the response they enable.
ADR spans distinct contributions to sensing, staged detection, and prevention~\cite{li2026adr}.
Its fast triage selectively invokes deeper contextual analysis; a separate pre-prompt hook blocks recognizable credential exposures.
These layers have different operating conditions and error costs.
In particular, the paper distinguishes analyst-labeled production alerts from per-task benchmark outcomes, so their false-positive percentages cannot be compared as if they had a common denominator.
The false-success study likewise positions lightweight, domain-calibrated detectors as triage signals rather than autonomous monitors~\cite{advani2026falsesuccess}.
Its evaluation is offline, and the authors require calibration and stronger environment checks before high-stakes use.

Wink turns detected misbehavior into guidance appended to the Agent's trajectory~\cite{nanda2026wink}.
Its 15-day production A/B test provides evidence about runtime intervention.
The recovery analysis uses an LLM judge given the pre-intervention history, the intervention, and 15 subsequent steps; recovery requires both non-recurrence of the detected misbehavior and clear forward progress within that window.
This establishes local behavioral recovery within the observation window; whole-task outcomes and recurrence in later runs require separate measurements.
Process-Centric Analysis makes intervention depend on the diagnosed process condition~\cite{liu2026processcentric}.
The monitor suppresses problematic steps or adds corrective guidance, updating its trajectory representations accordingly.
Updating the monitored representation keeps subsequent interventions aligned with the revised trajectory.
Evaluation selects 86 reproducibly problematic trajectories from prior SWE-agent executions and requires success in every reproducible run for baseline resolution.
The results concern recovery on that selected failure set; restoring trajectory state also leaves external tool effects to be handled by a separate compensation mechanism.

A-Trust's management system retains timestamped Agent-level trust records to support filtering, isolation, and developer inspection~\cite{he2026atrust}.
Unlike guidance that changes the next step, these controls change which messages or participants can continue; both depend on a fallible assessment of the current execution.

\subsubsection{Revising Durable Artifacts and Refreshing Persistent State}

Corrective and adaptive maintenance change persistent software artifacts to remove a fault or accommodate a changed environment; impact analysis and regression checks assess the consequences of the modification~\cite{washizaki2026swebok}.
In Agent systems, persistent behavior may be governed by code, skill instructions, configuration, or indexed knowledge.
Revising these artifacts addresses problems that would otherwise recur after an immediate execution-level correction has ended.
The durable-artifact group in Table~\ref{tab:maint-change-comparison} compares defect-guided code repair, test-gated replacement, instruction revision, and incremental index refresh.
The critical questions are what persists into the next execution and what evidence permits that change.
LLM-as-Code proposes replacing a generated function only after it passes the caller's tests~\cite{qi2026llmascode}.
This makes the acceptance condition explicit in its self-evolution algorithm, but the reported GUI-task experiment does not evaluate that maintenance loop.
SelfHeal derives repair rules from reported defects and fixes, first summarizing individual cases and then consolidating them by fix pattern~\cite{islam2026selfheal}.
Given buggy code and its intended behavior, a repair Agent retrieves relevant rules and API information to produce a candidate; a separate critic compares code changes, checks API usage and format, and returns acceptance or revision feedback.
The loop runs for at most three iterations and returns the last candidate even after a final rejection.
Because debugging uses inspection rather than code execution, this returned artifact still requires executable validation before release.
The 37-instance AgentDefect evaluation measures repair outcomes on examples from two libraries, providing bounded repair evidence while leaving repository-scale changes and release safety to further study.

SkillForge changes a different artifact: skill instructions and reference documents held in a virtual filesystem~\cite{liu2026skillforge}.
The studied skill format explicitly excludes executable scripts.
Failure analysis informs a diagnostic report, the optimizer edits the documents, and the resulting filesystem state is committed as the next version.
The inspected procedure does not specify a separate test-passing gate before this commit.
Its three revision rounds use chronological historical-ticket splits, reserving a fourth split for evaluation; model-judged consistency with expert responses measures offline revision outcomes rather than live deployment safety.
The comparison in Table~\ref{tab:maint-change-comparison} therefore exposes a consequential difference in acceptance: the procedures produce a test-gated replacement proposal, a critic-guided candidate, and a committed document revision, respectively.
The revision mechanism and the evidence required to admit its output need to be designed together.

The supplemental TRIAGE method extends this comparison to versioned harness files: failure evidence guides persistent edits whose predicted effects are checked in later rounds~\cite{yang2026triage}.
This makes the proposed behavioral effect of a persistent edit available for subsequent checking, an important step between generating a revision and establishing that it solves the observed problem.

Persistent-index refresh changes stored state rather than generating a software patch.
LinkedIn's hierarchical memory system refreshes changed leaves and affected ancestor paths instead of rebuilding the complete index~\cite{xu2026hierarchicalmemory}.
The authors compare incremental and full indexing on production data over two weeks and report more than six months of deployment.
Here the validation object is index content and topology, together with update cost, rather than a patch verdict or recovery message.
The comparison establishes alignment and update costs for the studied indexing procedure and production dataset.

\begin{table*}[t]
  \centering
  \caption{Engineering problems and approaches in intervention and evolution. The groups separate execution-level responses, changes to durable artifacts, and governance of versions and releases.}
  \label{tab:maint-change-comparison}
  \small
  \setlength{\tabcolsep}{4pt}
  \renewcommand{\arraystretch}{1.16}
  \begin{tabular}{@{}L{0.18\textwidth}L{0.20\textwidth}L{0.36\textwidth}L{0.20\textwidth}@{}}
    \toprule
    \textbf{Engineering problem} & \textbf{Approach} & \textbf{Key idea} & \textbf{Representative studies} \\
    \midrule
    Responding to execution problems & Staged detection and prevention & Escalate suspicious sessions for contextual analysis and block recognizable credential exposure. & ADR~\cite{li2026adr} (2026) \\
    & Calibrated failure triage & Prioritize apparently successful executions that may require investigation. & False-success triage~\cite{advani2026falsesuccess} (2026) \\
    & Corrective guidance & Append targeted guidance after detecting misbehavior. & Wink~\cite{nanda2026wink} (2026) \\
    & Process-conditioned intervention & Suppress problematic steps or revise guidance using monitored process conditions. & Process-Centric Analysis~\cite{liu2026processcentric} (2026) \\
    & Trust-based participation control & Use timestamped trust estimates to filter messages or isolate Agents. & A-Trust TMS~\cite{he2026atrust} (2026) \\
    \midrule
    Revising durable artifacts & Defect-guided repair & Generate candidate code changes from defect patterns and API information, with critic feedback. & SelfHeal~\cite{islam2026selfheal} (2026) \\
    & Test-gated replacement & Propose admitting a generated function only after caller-owned tests pass. & LLM-as-Code~\cite{qi2026llmascode} (2026) \\
    & Instruction revision & Diagnose failures, edit skill documents, and commit the revised filesystem state. & SkillForge~\cite{liu2026skillforge} (2026) \\
    & Incremental index refresh & Update changed leaves and affected ancestor paths instead of rebuilding the full index. & LinkedIn HLTM~\cite{xu2026hierarchicalmemory} (2026) \\
    \midrule
    Governing versions and releases & Build identity & Use build-context hashes and version tags to control image reuse or rebuilding. & OpenHands~\cite{wang2025openhandsplatform} (2025) \\
    & Authority-change review & Compare versioned descriptions of permissions, tools, memory access, and approval policies. & AgentRiskBOM~\cite{dutta2026agentriskbom} (2026) \\
    & Version-aware distribution & Package Agents with dependencies and preserve version history and integrity checks. & Cerebrum~\cite{rama2025cerebrum} (2025) \\
    & Evaluation-guided rollout & Combine versioned prompts, offline checks, progressive rollout, and operational feedback. & Nubank framework~\cite{gupta2026customersupport} (2026) \\
    \bottomrule
  \end{tabular}
\end{table*}

\subsubsection{Governing Versions, Distribution, and Operational Feedback}

Configuration and release management identify software versions and dependencies, control changes, and organize their delivery and possible rollback~\cite{washizaki2026swebok}.
This makes a reported result attributable to a specific configuration and gives maintainers a recoverable baseline when a change fails.
For Agents, that baseline can include prompts, runtime images, tools, permissions, and approval policies; governing a release requires both identifying these items and deciding what evidence warrants deployment.
The final group in Table~\ref{tab:maint-change-comparison} distinguishes build identity, authority-change review, version-aware distribution, and evaluation-guided rollout.
OpenHands uses build-context hashes and generic version tags to decide when runtime images can be reused or rebuilt~\cite{wang2025openhandsplatform}.
AgentRiskBOM instead makes changes in permissions, tools, memory access, and approval policies reviewable as a versioned authority description~\cite{dutta2026agentriskbom}.
AgentRiskBOM's commands validate these records, score risk, compare versions, and map controls, giving reviewers a concrete description of how a proposed version changes the system's authority.
The output supports a deployment review; enforcement of the release decision and checking runtime compliance remain separate responsibilities.
By contrast, Ponduru explicitly recommends staged promotion through replay, sandboxing, canary deployment, independent review, and wider release, with rollback criteria~\cite{ponduru2026unifiedagents}.
The procedure is a recommendation whose implementation and effectiveness have yet to be evaluated.
The comparison distinguishes the information provided by a record from the decisions enforced by a release process, while retaining each method's implementation status.
Cerebrum bundles Agents and dependencies for version-aware distribution, yet explicitly identifies the absence of a formal vetting process for uploaded Agents~\cite{rama2025cerebrum}.
Integrity checks establish which package was distributed; approving its behavior requires the missing review process.

The enterprise CRM study separately describes telemetry-driven alerts connected to circuit-breaker suspension and prompt-version records linking cryptographic identifiers, approval, and deployment time~\cite{pothukuchi2026enterprisecrm}.
The alert-response path and version record have different outputs; the presence of both does not establish that each system change was validated.

The Nubank framework gives a more connected example of operational change~\cite{gupta2026customersupport}.
It describes versioned prompt components, an offline evaluation gate, progressive rollout, human escalation, and model-specific baselines before migration.
Its runtime routing is also explicit: situations outside defined routines, failed data retrieval, or low response confidence lead to human transfer.
This connects recognized operating conditions to a concrete change in who handles the task, although the escalation policy's isolated effect and confidence calibration remain unmeasured.
Fast feedback targets components associated with evaluation failures; a slower loop relates production signals to possible routine, interface, or context-allocation changes.
Five reported deployments provide operational evidence for the evaluation-driven workflow within customer support.
Its slower feedback loop specifies how operational evidence can inform architecture, although the reported deployment outcomes do not trace a particular architecture revision through that procedure.

AgentEval provides a narrower link from detected regression to configuration change: its four-month development pilot reports 14 changes motivated by evaluation findings~\cite{guo2026agentevaldag}.
The operational record separates eight genuine regressions from 12 borderline alerts and three false positives.
This supports traceable local revision, while different measurement instruments before and during the pilot prevent attributing a precise investigation-time reduction to the tool alone (Table~\ref{tab:field-evidence-conditions}).

A supplemental counterpoint is PolicyBank: pre-deployment test feedback revises persistent, tool-level policy interpretations that guide subsequent tasks~\cite{choi2026policybank}.
The benchmark evaluates a feedback-to-artifact loop under stipulated feedback, extending the maintenance mechanisms considered alongside the main corpus.
Production requirements-change management remains outside that evaluation; a conflict between the illustrated revision and appendix clarification also leaves semantic convergence unresolved.
Harness Handbook addresses an earlier change-management step: locating the code sites implicated by a behavioral change request and maintaining their description as the harness evolves~\cite{wang2026harnesshandbook}.
Its source-localization and edit-planning evaluation is distinct from validating the resulting edits, adding a maintenance concern not reducible to runtime fault detection or repair generation.
Together, the supplemental cases extend maintenance from reacting to failures to understanding and validating requested changes in persistent behavioral controls.

\paragraph{Synthesis.}
Forty-seven of the 54 Maintenance contributions are directly evaluated; nine are coded E4, and 38 report available or partly available persistent artifacts.
The nine E4 contributions come from seven studies, because ADR contributes three of them.
This is the highest within-stage E3/E4 share; its interpretation depends on shared deployments and study-specific observation periods.
For example, NEXUS records structured events over roughly ten weeks, yet its real-trace pilot contains only 15 production tasks and lacks labels to validate drift detection~\cite{langay2026nexus}.
That observation establishes a functioning collection path, while drift sensitivity requires labeled changes and a larger longitudinal sample.

\finding[Findings]{Monitoring and diagnostic methods turn Agent executions into inspectable events, summaries, and suspected fault locations, with evidence from instrumented workflows, user studies, and labeled failure sets~\cite{zheng2025agentsight,sheng2026dills,ge2026famas}. Runtime interventions show local recovery in operational use, while code or instruction revision and incremental indexing provide mechanisms for changes that persist across runs~\cite{nanda2026wink,islam2026selfheal,liu2026skillforge,xu2026hierarchicalmemory}. Versioned evaluation and rollout practices supply examples of managing these changes in deployed systems~\cite{gupta2026customersupport}. The studies support several maintenance activities, but evidence connecting diagnosis to a validated revision and sustained improvement across releases remains fragmented. Further work should follow repairs and upgrades through regression testing and deployment, measuring recurrence, interacting changes, and maintainer effort alongside immediate recovery.}

\FloatBarrier
\subsection{Cross-Stage Synthesis}

The five stages reveal a range of engineering work on Agent systems.
Requirements methods identify needs and specify acceptable behavior; architecture methods organize responsibilities, interactions, and control; implementation methods realize and integrate those decisions; testing methods assess behavior and diagnose defects; maintenance methods support observation, intervention, and persistent change.
The corresponding artifacts include specifications, architecture descriptions, executable configurations, test assets, diagnostic records, and versioned software or operational state.
Together, these comparisons make the methods available for each development responsibility visible to a developer or researcher.

Methods within a stage often establish different properties, even when they act on similar artifacts.
A policy can express an authorization requirement, supply an implementation constraint, or provide a testing oracle; a trace can support execution replay, defect localization, or maintenance diagnosis.
The engineering purpose and evaluated outcome determine the contribution being compared.
Across stages, the source studies repeatedly require checks on specification fidelity, interface assumptions, test judgments, and the validity of changes.
These concerns explain why a useful method comparison must retain the problem, operation, artifact, and evidence together.

The stage-level syntheses also identify different open problems: sustained requirements management, comparison and composition of architectural choices, integration compatibility and effort, the validity and operational relevance of tests, and reliable evolution across versions.
Section~\ref{sec:discussion} develops practical implications and research directions from these stage-specific findings.

\FloatBarrier

%% file: sections/discussion.tex
\section{Implications and Research Directions}
\label{sec:discussion}

The preceding analysis identifies engineering problems and methods within each development stage and examines their reported results and limitations.
This section considers what those findings mean for developing Agent software.
We first revisit the lifecycle perspective with the benefit of the primary-study synthesis, relate the findings to existing system views, and then derive practical implications and research directions.
The recommendations follow the problems and evidence found in the reviewed studies; evaluating their combined use is part of the research agenda.

\subsection{What the Lifecycle Perspective Reveals}

The task-oriented construction perspective introduced through Anthropic's guide starts with a target task and the choice of a suitable workflow or Agent design~\cite{anthropic2024agents}.
The lifecycle analysis expands the questions asked about the resulting software.
Requirements studies examine whose needs and authority should govern its behavior; architecture studies compare the organization of responsibilities and control; implementation studies investigate executable integrations; testing studies develop procedures for assessing behavior; maintenance studies address diagnosis and continuing change.
The primary literature supplies concrete methods for these activities, including stakeholder elicitation, architectural assessment, framework adaptation, policy-aware testing, and version-aware maintenance.
This gives the lifecycle perspective substance as a way to organize engineering knowledge and compare development choices.

The comparison also changes how evidence is interpreted.
A task score can help assess a constructed system, while a requirements method additionally needs evidence that it captures intended needs, an architecture mechanism needs evidence for its claimed property, and a maintenance method needs evidence about the change it produces.
The source studies make these differences concrete: an executable workflow, an authenticated transcript, a localized failure, and an accepted revision establish different outcomes.
The Claude Research case in the Introduction illustrates how diagnostic and update concerns arise during task delivery; the stage-specific synthesis identifies research addressing those concerns and others across development.
The resulting view supports selecting methods for particular engineering responsibilities and identifying where their evidence remains incomplete.

\subsection{Relationship to Existing Lifecycle and System Views}

Existing whole-system accounts provide useful abstractions for this engineering work.
FMware and GUI-Agent surveys organize production and lifecycle concerns, harness engineering examines system-level coupling and regression feedback, and CARE and LADL describe development phases with outputs and feedback~\cite{rajbahadur2026productionfmware,yu2026guiengineering,li2026harnessengineering,ramachandran2026care,taj2026ladl}.
The present synthesis complements these views with comparisons of primary-study methods, their artifacts, and the evidence supporting particular development activities.
For example, the Requirements analysis separates empirical understanding of stakeholder needs from constructing and validating a specification.
The Testing analysis distinguishes building an execution environment, producing test cases, and validating the judgment procedure.
These distinctions explain what is involved in carrying out activities that a higher-level lifecycle can name.

The empirical support also varies within those activities.
Controlled comparisons and executable demonstrations are common, whereas observations of actual engineering work and continuing operation are less frequent.
A method may be well characterized under its tested conditions while its use across dependencies, organizations, or successive releases remains open.
Table~\ref{tab:headline-claim-boundaries} summarizes these findings and their interpretation.
Together, they motivate stage-specific research as well as investigation of the connections through which methods are combined.

\begin{table}[H]
  \centering
  \caption{Main synthesis findings and their interpretation.}
  \label{tab:headline-claim-boundaries}
  \small
  \renewcommand{\arraystretch}{1.12}
  \begin{tabular}{L{0.21\textwidth}L{0.20\textwidth}L{0.26\textwidth}L{0.23\textwidth}}
    \toprule
    Finding & Evidence & Engineering implication & Interpretation scope \\
    \midrule
    Methods address all five development stages & Source comparisons over 570 lifecycle contributions & Select methods by engineering problem, resulting artifact, and evaluated property & Coverage within the selected literature; effectiveness is method- and context-specific \\
    Contributions have different stage coverage & Stage patterns over 355 papers & Combine stage-specific knowledge when examining a development process & Counts describe research contributions, with routine project activities outside the coding \\
    Evaluation is common and predominantly bounded & 517 directly evaluated contributions; 464 E2 and 37 E3/E4 & Extend mechanism evaluation with studies of actual engineering work and continuing use & Setting, study quality, independence, and duration require separate interpretation \\
    Many contributions expose artifacts and configuration details & 418 with full/partial artifact access; 347 with high configuration completeness & Use accessible artifacts and stated conditions to support inspection and reuse & Availability and reporting are signals; successful reproduction needs execution evidence \\
    Reported connections favor construction and assurance & 172 stage links; 135 implemented/evaluated & Examine how specifications, implementations, tests, and operational changes are used together & Relations reported in papers; industrial prevalence and process benefit need direct study \\
    \bottomrule
  \end{tabular}
\end{table}

\subsection{Implications for Engineering Practice}

The five stages suggest complementary questions for developers selecting and applying the reviewed methods.
These questions concern what should be specified, designed, implemented, assessed, and maintained for the system at hand.

\paragraph{Requirements: make intended behavior and authority inspectable.}
Requirements artifacts can record stakeholder needs, delegated authority, acceptance conditions, and routes for human intervention.
The Requirements comparison shows that elicitation, risk derivation, formal specification, and governance configuration address different parts of that work (Table~\ref{tab:req-method-comparison}).
A temporal obligation can be monitored once specified~\cite{alamdari2026formal}, while the interpretation of domain commitments requires authorized stakeholder judgment~\cite{hagar2026runtimeconfiguration}.
Developers should therefore check both the rationale for a requirement and the fidelity of its representation, then identify how its satisfaction will be assessed.
Changes in intended use or authority provide reasons to revisit those artifacts.

\paragraph{Architecture: relate design choices to explicit properties.}
Coordination, isolation, policy enforcement, human review, and recovery solve different design problems.
Workflow projection addresses coordination semantics, IsolateGPT limits cross-application propagation, and SagaLLM organizes compensation for interrupted workflows~\cite{bollig2026coordination,wu2024isolategpt,chang2025sagallm}.
A design record should identify the property sought, its assumptions, and the interface or control point responsible for it.
For generated workflows, this includes checking structural validity and intended task semantics.
For human review, it includes distinguishing authorization rules, checkpoint timing, and the interface through which a reviewer can act~\cite{joshi2026deontic,zhou2026confirmation,zhuang2026agentclick}.
Evaluation should follow these decisions, including the interactions between mechanisms when they are combined.

\paragraph{Implementation: preserve integration assumptions in executable configurations.}
Framework adapters, data schemas, caches, and middleware make Agent software operational under specific compatibility conditions.
Agent Spec supports cross-framework execution with runtime-dependent behavior; KVCOMM and LRAgent impose different model and cache assumptions~\cite{amini2025agentspec,ye2025kvcomm,jeon2026lragent}.
An integration record should retain supported versions, transformed representations, fallback behavior, and the outcome being checked.
Control mechanisms add further preconditions: ChainCaps requires policy metadata and proxy-visible effects, while RAC requires compensation operations and valid mappings of their inputs~\cite{jiang2026chaincaps,perera2026rac}.
Integration tests should exercise these assumptions and the behavior of the combined system, with optimization gains measured alongside their overhead and behavioral effects.

\paragraph{Testing: validate the assessment procedure as well as the Agent.}
The environment, case generator, execution record, and oracle each contribute to an assurance result.
ToolEmu checks whether emulated trajectories could represent realizable executions, while ST-WebAgentBench evaluates policy-compliant task completion~\cite{ruan2024toolemu,levy2026stwebagentbench}.
An assessment should report the tested version, exposure conditions, observed property, attempt budget, aggregation rule, and judgment procedure.
Repair and release decisions additionally require identifying which version was assessed: AEVAL separates first-attempt results from subsequent self-correction~\cite{anand2026aeval}.
When formal models provide evidence, the implementation must be checked against the properties actually verified, as illustrated by TraceFix's distinction between checked protocol constraints and runtime enforcement~\cite{xia2026tracefix}.

\paragraph{Maintenance: identify the changed object and its acceptance conditions.}
An intervention can alter a current execution, a persistent configuration, executable code, stored state, or a distributed version.
The Maintenance comparison shows that these changes require different checks.
SelfHeal can return a candidate that its critic rejected, SkillForge commits revised skill documents, and HLTM evaluates an incremental index update~\cite{islam2026selfheal,liu2026skillforge,xu2026hierarchicalmemory}.
A change record should distinguish the proposed modification, the checks performed, and the version admitted into use.
Nubank's versioned prompts, offline evaluation, progressive rollout, and model-specific baselines illustrate a connected deployment procedure~\cite{gupta2026customersupport}.
Further decisions concern how to monitor the change, detect regressions, and restore a suitable state; incidents can also trigger reconsideration of earlier requirements or design decisions.

\subsection{Evidence-Derived Research Directions}

The open problems identified within the five stages motivate several kinds of further investigation.
Requirements research needs evidence about specifications as stakeholder needs change; architecture and integration research need comparisons under shared assumptions and changing dependencies; testing research needs valid judgments and evidence of operational relevance; maintenance research needs evaluation across persistent changes and versions.
The cross-stage findings add questions about the consequences of transferring evidence or revising upstream decisions.
Table~\ref{tab:research-directions} connects six research directions to concrete source limitations and candidate validation designs.
These are proposed studies, with their comparisons and outcomes intended to make the questions empirically testable.

\begingroup
  \small
  \renewcommand{\arraystretch}{1.14}
  \setlength{\LTcapwidth}{\textwidth}
  \begin{longtable}{L{0.29\textwidth}L{0.23\textwidth}L{0.38\textwidth}}
    \caption{Evidence-grounded research directions. Each row links a source-specific evidence boundary to an open question and a proposed future validation design. Source cases motivate the questions within their reported settings.}
    \label{tab:research-directions} \\
    \toprule
    Evidence and boundary & Open engineering question & Candidate comparison and outcomes \\
    \midrule
    \endfirsthead
    \caption[]{Evidence-grounded research directions (continued).} \\
    \toprule
    Evidence and boundary & Open engineering question & Candidate comparison and outcomes \\
    \midrule
    \endhead
    \midrule
    \multicolumn{3}{r}{Continued on next page} \\
    \endfoot
    \bottomrule
    \endlastfoot
    \textbf{D1: Requirements continuity.} Four of 26 RE contributions are E3; none is E4. The industrial trust case connects elicitation to prototype evaluation, not sustained requirements management~\cite{axetorn2026trustrequirements}. & Do explicit authority and acceptance artifacts remain useful as stakeholder needs and system versions change? & Compare task-prompt-only and traceable specification workflows across matched changes. Measure stakeholder disagreement, stale requirements, downstream defects, and update effort. \\
    \textbf{D2: Configuration compatibility.} Agent Spec enables cross-framework execution, but the same configuration yields runtime-dependent behavior~\cite{amini2025agentspec}. & Which design assumptions and interface obligations must be rechecked after a runtime, model, or tool upgrade? & Hold tasks and configurations fixed while varying versioned dependencies. Compare task-only regression with contract-aware checks; report missed incompatibilities, false alarms, and migration effort. \\
    \textbf{D3: Assurance transfer.} Testing has 230 E2 and 10 E3/E4 contributions. Policy-aware completion differs from task completion; ST-WebAgentBench's expert validation shares one organizational context~\cite{levy2026stwebagentbench}. & Do benchmark cases and oracles identify failures encountered in other operating contexts? & Compare task-only and policy-aware release judgments against independently reviewed field incidents. Report missed failures, false positives, repeated-run variability, and oracle disagreement. \\
    \textbf{D4: Upstream feedback.} No implemented/evaluated MAINT--REQ or MAINT--ARCH link is recorded. Process observability demonstrates local configuration repair in a calculator example~\cite{fournier2025observability}. & When does an incident require revising requirements or design rather than only patching execution? & Compare local repair with incident-to-decision traceability on matched change histories. Measure missed upstream impacts, regression escape, decision disagreements, and review effort. \\
    \textbf{D5: Persistent evolution.} SelfHeal's repair is single-file; SkillForge changes skill documents using historical-ticket replay~\cite{islam2026selfheal,liu2026skillforge}. & Do accepted changes remain valid across interdependent artifacts and successive updates? & Extend repair tests to multi-file and skill/configuration change sequences. Compare one-shot acceptance with repeated regression and rollback checks; measure durability, cumulative failures, and recovery cost. \\
    \textbf{D6: Accountable oversight.} Confirmation scheduling optimizes time; AgentClick illustrates editable review rather than a controlled user study~\cite{zhou2026confirmation,zhuang2026agentclick}. & How should review timing, mandatory authority checks, and interaction design work together under consequential errors? & Compare time-optimal, risk-constrained, and fixed review policies in safe simulations and authorized field studies. Measure missed harmful actions, intervention delay, workload, and inappropriate reliance. \\
  \end{longtable}
\endgroup

The appropriate evaluation depends on the engineering question.
Requirements methods can be compared through stakeholder agreement, specification fidelity, and update effort; architecture and integration methods through property preservation, compatibility, and development cost; testing methods through failure detection and oracle validity; maintenance methods through repair acceptance, regression escape, and behavior across releases.
Where a proposed benefit depends on several activities working together, a study should also trace the changed artifacts and the decisions that consume them.
Matched change requests, incident histories, or successive releases can support such comparisons, with task performance, resource budgets, risk, and change rate recorded alongside engineering outcomes.

The research agenda thus concerns both the depth of support within each stage and the use of that support in development.
Existing methods establish useful local properties, while recurring limitations concern requirements continuity, composition and compatibility, assurance under operational conditions, and reliable evolution.
Cross-stage feedback is one part of this agenda.
Its value, like that of an individual method, should be assessed through the engineering outcome it is intended to improve.

%% file: sections/threats.tex
\section{Threats to Validity}
\label{sec:threats}

The findings describe an author-reviewed collection of research on engineering Agent software.
Their interpretation depends on the scope boundary, coverage of the search and publication records, consistency of the content analysis, and the settings represented in the source studies.

\paragraph{Scope and construct validity.}
The distinction between engineering an Agent system and using an Agent for an external task requires judgment.
Terminology also varies across communities, particularly for architecture, evaluation, and lifecycle activities.
We use the engineered object, author-stated contribution, lifecycle purpose, and source-located evidence to support scope and stage decisions.
The five stages organize engineering responsibilities while allowing iterative development and contributions to several stages.
Boundary cases remain possible where implementation controls, runtime assurance, and maintenance interventions overlap.
The exclusion of a conversational-privacy study after source checking illustrates how object identification can affect membership (Appendix~\ref{sec:source-scope-correction}).

\paragraph{Search and study-selection validity.}
Database indexing, query translation, year cutoffs, access restrictions, and incomplete citation coverage can omit relevant studies.
The current reconciled cohort excludes 432 records without accessible full text; those decisions concern access and can underrepresent less accessible sources.
Collection stopped at a dated snapshot after diminishing retrieval yield, and independently confirmed citation-search closure remains pending.
The first two authors independently reviewed scope decisions, with disagreements resolved by the corresponding author.
Shared interpretations can still produce correlated errors.
Recorded rationales and full-text evidence support inspection of those decisions; Appendix~\ref{sec:corpus-amendments} preserves the identity corrections and historical counts, with residual duplicate-version errors still possible.

\paragraph{Coding and synthesis validity.}
The authors checked the content records for the 355 main-analysis papers and the 150-paper supplement against locatable source passages.
The principal risks concern contribution granularity, stage assignment, and whether an evaluation directly supports the contribution to which it is attached.
We compare the engineering operation, artifact, assessed property, and conditions in the source studies, retaining amendments where an internal component, experiment, or runtime transfer had been overinterpreted.
These checks improve traceability, while their targeted correction counts do not estimate a population error rate.
Artifact names also differ in granularity, so their roles are compared through source-level interpretations rather than raw lexical frequency.
The 172 five-stage handoffs are within-paper deduplicated reporting units: several can share a study, and an implemented transfer can end at inspection or measurement without a validated software revision.
The source records retained in the replication materials distinguish these endpoints.

\paragraph{Publication and version validity.}
The main collection uses publication evidence captured at the selection snapshot.
Of its 355 papers, 154 rely on public acceptance claims, while targeted checks identified 25 officially published or accepted studies outside that selection.
The omissions can affect stage coverage and the apparent extent of research on particular problems.
Because those checks followed publication clues, their results do not estimate the overall rate of publication-status error.
Version differences introduce an additional concern: later preprints can contain results beyond the accepted text.
The cited ST-WebAgentBench evidence was checked against its conference-hosted PDF~\cite{levy2026stwebagentbench}, whereas DoomArena's inspected evidence remains tied to the author version.
Appendix~\ref{sec:corpus-amendments} records status and identity corrections; comprehensive publication reconciliation and accepted-version alignment remain necessary.

\paragraph{Supplemental sensitivity.}
The 150-paper supplement examines stage coverage and method boundaries outside the fixed main selection.
Its sampling frame and weights incorporate recorded identity corrections; residual duplicate identities and exchangeability remain assumptions.
Appendix~\ref{sec:supplement-stage-sensitivity} varies the coding of a conservative set of 90 papers carrying 64.3\% of the expansion weight.
Testing is largest under the current sample codes, but its ordering over Requirements and Maintenance is not secured under that exercise.
The resulting ranges are conditional recoding bounds, with search omissions and shared errors outside the selected set beyond their coverage.
All sampled records have undergone author review; that confirmation does not establish equal reporting completeness for fine-grained methods, artifacts, or handoffs across the main and supplemental collections.

\paragraph{External and temporal validity.}
The collection is concentrated in recent work and includes only part of the 2026 observation year.
Reporting delays, changing software versions, and unpublished industrial practices limit generalization.
Evaluation settings also differ in methodological rigor, exposure, and duration: several contributions can share a short deployment, while controlled experiments can use realistic artifacts and professional participants.
Appendix~\ref{sec:evidence-setting-cases} and Table~\ref{tab:field-evidence-conditions} retain missing durations, conflicting source periods, and non-equivalent outcome measures.
The stage and relationship frequencies therefore describe the selected literature.
Establishing the prevalence or effectiveness of these methods in practice requires representative development studies, outcome-specific comparisons, and observation across relevant changes.

%% file: sections/conclusion.tex
\section{Conclusion}

This survey examines the development of LLM-based Agents as software systems through Requirements Engineering, Architecture and Design, Implementation and Integration, Testing and Assurance, and Maintenance and Evolution.
The 355-paper main analysis identifies 570 lifecycle contributions and compares their engineering problems, methods and practices, artifacts, and evaluation evidence.
The literature provides methods for eliciting and specifying needs, organizing coordination and control, constructing executable integrations, assessing behavior and defects, and observing or changing existing systems.
The lifecycle organization makes this knowledge accessible through the development responsibility each contribution addresses.

The synthesis shows both substantial method development and uneven support across engineering activities.
Architecture and Testing account for much of the collected research, while Requirements and Maintenance receive less coverage.
Direct evaluation is common, predominantly in bounded empirical or formal settings; fewer studies observe actual development or continuing operation.
Method-level comparisons expose distinct questions about specification fidelity, design assumptions, integration compatibility, test validity, and the acceptance of persistent changes.
Explicit artifact and decision transfers provide an additional view of how development activities connect, with operational feedback to validated upstream revisions sparsely documented in the selected studies.

Further research should deepen the evidence for the problems within each stage and investigate how methods can be used together during development and change.
Priorities include requirements continuity, architectural and configuration compatibility, operationally relevant assurance, reliable evolution, and effective human oversight.
These directions extend task-oriented Agent construction with the questions needed to specify, design, integrate, assess, and maintain the resulting software.
The collection's access, publication, sampling, and reporting limitations constrain generalization; the source-linked comparisons provide a basis for examining those limits and extending the available engineering knowledge.

%% file: sections/appendix.tex
\section{Search Details and Replication Materials}
\label{sec:replication-snapshot}

\subsection{Reference Sets and Known-Paper Coverage}
\label{sec:known-paper-coverage}

The search developed through an early Agent-method reading list and subsequent lifecycle-focused pilots.
Table~\ref{tab:search-reference-sets} distinguishes the recorded reading and query-check sets from the later citation-expansion inputs.
The ten-paper reading list comprises MRKL Systems~\cite{karpas2022mrkl}, ReAct~\cite{yao2022react}, Toolformer~\cite{schick2023toolformer}, HuggingGPT~\cite{shen2023hugginggpt}, Reflexion~\cite{shinn2023reflexion}, Generative Agents~\cite{park2023generative}, Voyager~\cite{wang2023voyager}, AutoGen~\cite{wu2023autogen}, MetaGPT~\cite{hong2023metagpt}, and LATS~\cite{zhou2023lats}.
It supported early vocabulary exploration under a broader Agent-method scope.
The lifecycle pilots subsequently supplied a 45-paper query-check set drawn from full-text reading of an initial sample and a contrast sample.
The saved query-check records preserve the seed roles assigned provisionally at that historical snapshot.
The complete lists, their source snapshots, and per-paper keyword matches are included in \nolinkurl{anc/search/} with the manuscript source.

\begin{table}[H]
  \centering
  \small
  \renewcommand{\arraystretch}{1.1}
  \caption{Reference sets used during search development and citation expansion. Counts describe separate, overlapping historical sets; they are not added to obtain the final corpus size.}
  \label{tab:search-reference-sets}
  \begin{tabular}{L{0.25\textwidth}rL{0.58\textwidth}}
    \toprule
    Recorded set & Papers & Role and interpretation \\
    \midrule
    Early reading list & 10 & Agent-method vocabulary exploration; broader scope than the present primary-study criteria \\
    Lifecycle calibration set & 27 & Keyword-expansion and scope-boundary checks; includes relevant and off-scope examples \\
    Lifecycle query-check set & 45 & Saved v2 query checks; drawn from pilot and contrast studies, with provisional seed eligibility \\
    Working primary pool & 271 & Prior-pool coverage check and primary-study citation expansion; membership reflects the August 28 working snapshot \\
    Secondary source pool & 34 & Supplementary citation discovery from related secondary studies \\
    \bottomrule
  \end{tabular}
\end{table}

Table~\ref{tab:known-paper-coverage} reports two saved checks with their own denominators.
For the 45-paper set, the August 25 v2 searches retrieved 29 papers in the union of arXiv, DBLP, and Semantic Scholar.
For the 271-paper working pool, the August 28 four-source search retrieved 157 papers; the remaining 114 were explicitly carried forward from the existing pool.
Only a recorded keyword-search match counts as retrieval in these checks.
The sets were used in search development or assembled from earlier retrieval, so these results describe known-paper coverage at the recorded snapshots rather than an independent estimate of recall over all in-scope literature.

\begin{table}[H]
  \centering
  \small
  \renewcommand{\arraystretch}{1.1}
  \caption{Known-paper coverage in saved query runs. The 45-paper set uses the three-source v2 snapshot of August 25, 2026; the 271-paper pool uses the four-source v2 snapshot of August 28, 2026. Source counts overlap. Manual carry-forward records are excluded from keyword matches.}
  \label{tab:known-paper-coverage}
  \begin{tabular}{L{0.24\textwidth}L{0.43\textwidth}rr}
    \toprule
    Reference set & Keyword-search sources & Matched & Total \\
    \midrule
    Query-check set & arXiv & 26 & 45 \\
     & DBLP & 6 & 45 \\
     & Semantic Scholar & 24 & 45 \\
     & Three-source union & 29 & 45 \\
    \midrule
    Working primary pool & arXiv, DBLP, Semantic Scholar, OpenAlex & 157 & 271 \\
    \bottomrule
  \end{tabular}
\end{table}

The supplementary index records paper titles, historical set roles, source snapshots, query-family matches, and available retrieval positions.
Unmatched entries retain their recorded status; causes are left unassigned where the saved runs provide no diagnosis.
Earlier ten-paper and 27-paper pilot results remain in the search archive.
The 271 primary and 34 secondary studies served as the two citation-expansion inputs after the working pools had been assembled.
Their role differs from the earlier reading material used to develop search vocabulary.

\subsection{Data and Code Availability}

Table~\ref{tab:replication-snapshot} gives the entry points needed to inspect the current findings.
The main evidence ledger links each contribution to its paper, stage, source locator, artifacts, and direct evaluations; a separate ledger records handoffs.
Exploratory clustering catalogs and assignments remain archived as working history.
The supplemental layer preserves its own population, content, mapping, and sampling weights for separate analysis.
Manifests identify input, code, and output hashes. The companion archive preserves prior snapshots and per-record corrections; current denominators are defined by the active analysis manifest.

A portable statistical slice rebuilds aggregate and association tables from selected coded fields without model services or local PDF caches.
Its frozen outputs include historical classification displays that are no longer used in the manuscript; the source tables for retained stage, evaluation, and handoff statistics remain unchanged.
Relocation tests reproduced the frozen SVG hashes and claim values before this presentation revision.
This check establishes reproducibility of the coded statistics; paper selection, source interpretation, and supplemental sensitivity analysis require separate validation inputs and procedures.
Replaying source-text and evidence-packet checks additionally requires access to the locally archived PDFs.

\begin{table}[H]
  \centering
  \scriptsize
  \renewcommand{\arraystretch}{1.12}
  \caption{Current replication entry points. Directory names are relative to \nolinkurl{data/analysis/} in the companion repository. The archive contains the full predecessor index and change history.}
  \label{tab:replication-snapshot}
  \begin{tabular}{L{0.13\textwidth}L{0.48\textwidth}L{0.29\textwidth}}
    \toprule
    Layer & Artifact directory & Use and boundary \\
    \midrule
    Main analysis & \nolinkurl{agentware_arch_boundary_v71_20260912} & 355 papers, 570 lifecycle contributions, 172 handoffs; source tables unchanged \\
    Statistical replay & \nolinkurl{agentware_arch_statistics_v71_20260912} & Portable coded-field inputs and scripts; includes retired displays, not a full-text or screening replay \\
    Historical catalog & \nolinkurl{agentware_arch_f11_catalog_v74_20260912} & Exploratory classifications and source corrections; retained for audit, not a reported taxonomy \\
    Supplement content & \nolinkurl{agentware_leni_source_v75_20260912} & Current selected-content index in \nolinkurl{selection/}; Leni annotations and artifact replay; 150 papers, 364 units, 484 evaluations; stage statistics remain in v60 \\
    Historical mapping & \nolinkurl{agentware_arch_boundary_v71_20260912} & Exploratory supplemental assignments and review queues; not method-category findings \\
    Sample design & \nolinkurl{agentware_eu_sensitivity_sample_v3_identity_20260911} & 1,158-record frame; fixed allocation and seed, revised stratum weights \\
    Publication audit & \nolinkurl{agentware_eu_publication_overlay_v5_contextual_20260912} & 30 targeted checks; retains prior observation dates; not a population error-rate estimate or automatic P promotion \\
    Claim traceability & \nolinkurl{agentware_arch_claims_v71_20260912} & Frozen claim mappings and population IDs; includes the retired classification claim and display labels \\
    Screening history & \nolinkurl{agentware_final_primary_corpus_v3_20260909} & Original 5,357-record ledger and 1,517-record scope snapshot behind the historical diagnostic \\
    Revision archive & \nolinkurl{agentware_manuscript_narrative_v48_20260912} & Complete prior artifact index and explanation of amendments; not another corpus freeze \\
    \bottomrule
  \end{tabular}
\end{table}
\FloatBarrier

The screening and coding manifests retain protocol and schema versions, input and output hashes, source-text locators, validation results, and record counts.
Source-specific queries, pagination logs, and version-family identity edges preserve retrieval and reconciliation decisions.
The claim-traceability records connect reported findings to source tables, figures, target sections, and supporting papers.
Historical automated preparation and preliminary assessment records remain in the archive, separately from the author-confirmed screening and coding described in Section~\ref{sec:methodology}.

The current coding snapshot contains 355 papers, 571 contributions (570 lifecycle contributions and one audit-only contribution), 690 evaluation records, and 232 retained stage/interface relations.
It has no unresolved content record and passes identifier, stage, coverage, assignment, packet, and hash checks.
Earlier records and their replacements or field amendments remain available for comparison.
The technical checks establish internal consistency and source traceability, while author review confirms the content interpretations, including contribution boundaries, E3/E4 settings, and sparse upstream feedback links.
Alignment with formally accepted paper versions remains a separate bibliographic task.

\subsection{Analysis Definitions}
\label{sec:analysis-definitions}

The following definitions specify the evaluation and relationship records used in the analysis described in Section~\ref{sec:evidence-coding}.
The statistics distinguish papers, engineering contributions, evaluations, and cross-stage relations.
Evaluation evidence is coded by study design and by the setting supporting each contribution.
The setting codes are:
\begin{itemize}
  \item \textbf{E0}: a concept, proposal, or mechanism without direct evaluation;
  \item \textbf{E1}: a worked example, toy case, or anecdotal demonstration;
  \item \textbf{E2}: a laboratory benchmark, controlled experiment, simulation, retrospective data, bounded prototype or expert assessment, or formal verification under an explicit model;
  \item \textbf{E3}: authentic engineering work on an actual system, an industrial case, or a field-like practitioner study; and
  \item \textbf{E4}: production field deployment, longitudinal observation, or independent replication.
\end{itemize}
These evaluation codes are separate from the publication stratum E.
Controlled laboratory work remains E2 when it uses professional participants or realistic tasks; authentic project work can support E3 before deployment.
An evaluation record links a contribution to a distinct evaluation setting, so multiple records can refer to one underlying experiment.
Contribution maturity is the highest level among its direct evaluations, or E0 when none is reported.
Independent replication requires a separate replication study.
Artifact availability is coded as Yes, Partial, or No, and configuration completeness as High, Medium, or Low, with missing reporting retained separately.
Study-design labels can be multiple-valued.

Cross-stage relations are recorded when the paper explicitly reports that an artifact, decision, result, or feedback item is consumed by another development activity.
Each record identifies the source, target, transferred item, direction, and status: Mentioned, Specified, Implemented, or Evaluated.
Only relations between the five development stages enter the handoff matrix; runtime and external-interface transfers are retained separately.
Unreported relations remain missing evidence of a connection, with recurring source-reported difficulties providing additional grounds for research directions.

\input{sections/source_amendments}

\input{sections/supplement_stage_sensitivity}

\input{sections/evidence_setting_cases}

%% file: sections/source_amendments.tex
\section{Corpus Boundaries and Source-Review Coverage}
\label{sec:corpus-amendments}

This appendix specifies the populations and interpretation rules underlying the results.
The main synthesis contains 355 papers, 570 lifecycle contributions, and 172 five-stage handoffs.
The fixed supplement contains 150 papers, 364 contributions, and 484 evaluation records, analyzed under separate denominators.
Table~\ref{tab:replication-snapshot} identifies the current evidence layers. The companion archive retains the counts, per-record changes, and source decisions for each preceding snapshot.

\subsection{Historical Screening Landscape}
\label{sec:historical-scope-landscape}

\begin{figure*}[t]
  \centering
  \includegraphics[width=0.98\textwidth]{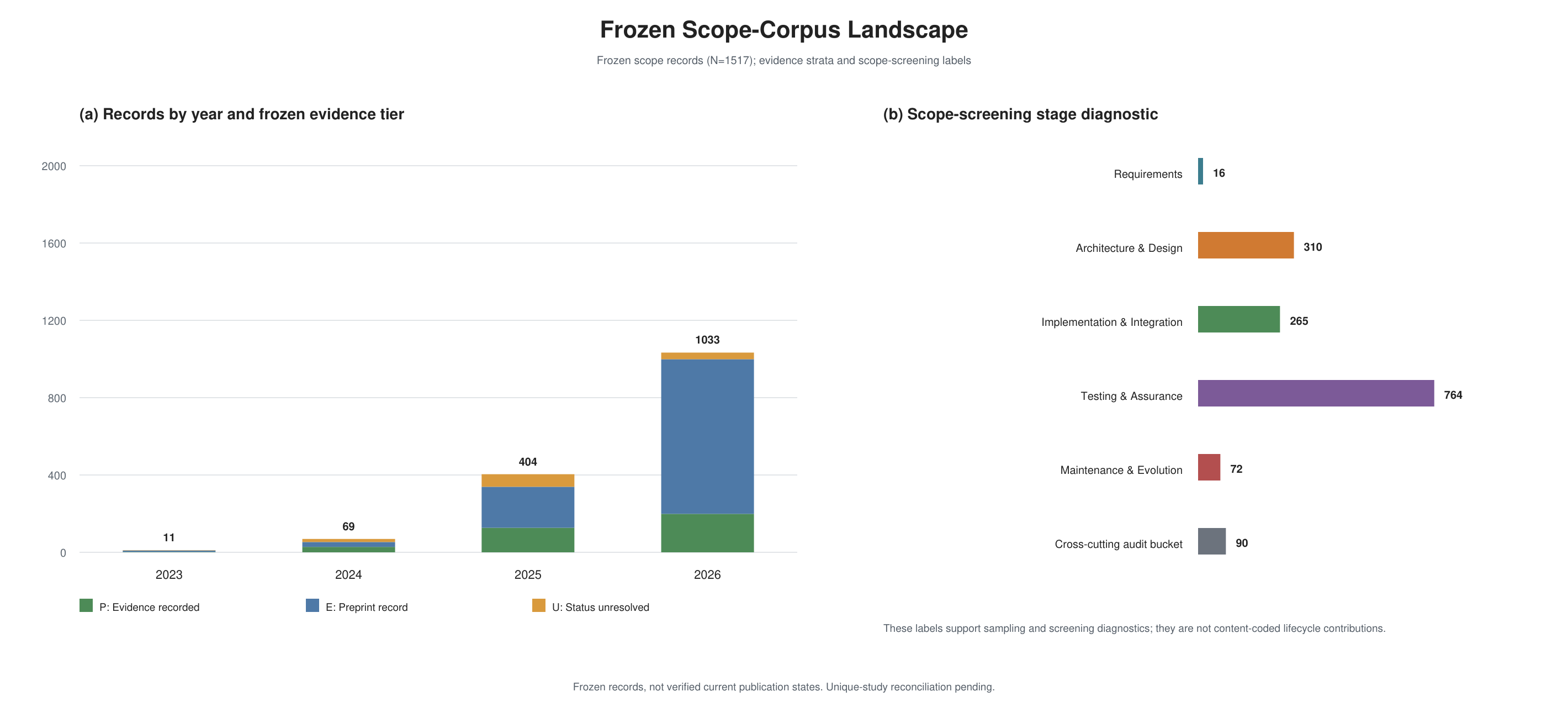}
  \caption{Historical landscape of the 1,517-record scope freeze, before identity and scope reconciliation. Panel (a) uses publication-evidence strata recorded at the freeze. Panel (b) reports scope-screening labels used to route records. Current publication status and contribution-level stages are assessed separately. The 2026 bar covers a partial observation year.}
  \Description{Panel a shows 11 records in 2023, 69 in 2024, 404 in 2025, and 1,033 in partial 2026. Frozen publication-evidence strata total 356 P, 1,044 E, and 117 U. Panel b shows 16 Requirements, 310 Architecture, 265 Implementation, 764 Testing, 72 Maintenance, and 90 cross-cutting scope labels. These historical record counts are not current unique-study or contribution totals.}
  \label{fig:scope-corpus-landscape}
\end{figure*}

The historical freeze in Figure~\ref{fig:scope-corpus-landscape} contains 1,437 records dated 2025 or the partial 2026 window (94.7\% of 1,517).
This temporal description uses retrieved first-public-year records from the historical screening population and a partial final observation year.
P records have captured publication or acceptance evidence; E records have a preprint without qualifying evidence captured; U records have unresolved status.
The original split is 356 P, 1,044 E, and 117 U, reflecting the publication evidence captured at that freeze.
The figure is retained only as a screening diagnostic; current paper-level coverage is reported in Section~\ref{sec:collection-results}, and contribution-level comparisons appear in Section~\ref{sec:lifecycle}.

\subsection{Current Identity and Sampling Frame}
\label{sec:identity-amendments}

The current disposition ledger contains 5,354 canonical records and 1,514 scope-primary records, of which 355 form the main P analysis.
Relative to the 5,357-record screening ledger, identity reconciliation merges two duplicate study pairs and a malformed, access-excluded InjecAgent alias; the scope correction below removes one further record from primary synthesis, not from the disposition ledger.
The remaining 1,159 scope-primary records lie outside P. Excluding one withdrawn record yields the 1,158-record supplemental sampling frame.
Reapplying the original stratified hash ranking, seed, and allocation after identity correction retains 149 sample members and substitutes ResMAS for one alias; the 150-paper sample uses recomputed stratum weights.
This targeted reconciliation resolves the identified aliases; residual version-family errors may remain.

A targeted publication audit, extended through September 12, 2026, covers 30 sampled records: 22 formal publications, three official acceptances, three non-archival workshop acceptances, and two author-only claims.
The additional journal record is the contextual-integrity study~\cite{ghalebikesabi2024contextual}; the preceding 29 checks retain their original observation dates.
The 25 official publication or acceptance records outside P identify missed metadata capture in cases selected through publication clues; the population error rate remains unknown.
P membership remains frozen, and alignment between analyzed and accepted text requires separate version checks.
The supplement consequently probes coverage outside the fixed P selection, including subsequently verified published or accepted studies.

\subsection{Scope and Contribution Boundaries}
\label{sec:source-scope-correction}

Eligibility requires the Agent system itself to be the engineering object.
Zhang et al.'s conversational-privacy study examines ChatGPT disclosures, mental models, and privacy controls, but does not establish an Agent system whose task, execution, or tool-use control is being engineered~\cite{zhang2024fairgameprivacy}.
It remains background context and is excluded from primary synthesis under the engineered-object criterion.
LCO illustrates a different boundary: its task-conditioned constraint synthesis configures an execution-time safety optimizer, so the contribution belongs to Implementation rather than a development requirements baseline~\cite{wan2026lco}.
Its two independently studied mechanisms remain separate units, but their internal runtime transfer is not a lifecycle handoff.
The 355-paper analysis therefore measures author-stated engineering contributions interpreted against their source evidence.

\subsection{Evidence Settings and Handoff Status}
\label{sec:source-coding-amendments}

The evidence-setting review covers all initially E3/E4 contributions, field/longitudinal/replication tags, records affected by a provisional E2 cap, and initially E2 practitioner interviews or user studies not tagged as controlled experiments or benchmarks.
After the scope exclusion above, this coverage comprises 50 contributions in 34 papers and all 68 evaluations linked to those contributions.
The review distinguishes authentic engineering activity from production exposure and changes 20 setting labels relative to the initial coding; its verification scope is the source-reported evaluations linked to these selected contributions.
For example, DiLLS supplies formative interview evidence rather than an unspecified evaluation design~\cite{sheng2026dills}.
Current E2/E3/E4 counts are 464/14/23. E4 requires production \emph{or} longitudinal observation \emph{or} independent replication; authentic project work can supply E3 before deployment.

Handoff status describes the reported transfer; a completed development cycle additionally requires evidence of resulting changes and their validation.
LLM-Conductor's changing quotas and schedules are runtime interfaces, whereas AgentIssue-Bench implements a transfer from executable benchmark tasks to maintenance issue categories~\cite{xu2026conductor,rahardja2025agentissues}.
GraphFlow's per-request workflow generation and KV reconstruction likewise describe an internal serving flow, not a development handoff~\cite{li2026graphflow}.
Its former L1 is archived alongside LCO's runtime flow without removing either method contribution or any evaluation.
Together, these distinctions delimit the current 172 five-stage links from 139 papers.
The source review of sparse directions retained in the replication materials assigns status from the documented implementation, evaluated consequences, or proposed transfer.
Section~\ref{sec:methodology} defines the current selection and evidence-setting rules; this appendix and the replication materials retain the historical screening counts and the current 355/570/172 analysis denominators.

\subsection{Retained Source Comparisons}

The source reviews distinguish an engineering operation from its implementation, its evaluation, and a claimed downstream guarantee.
The paper-level comparisons remain in Section~\ref{sec:lifecycle}; Table~\ref{tab:impl-f10-boundaries} retains a detailed comparison of reuse, recovery, and execution controls without classification identifiers.
Earlier catalogs, reassignment logs, and supplemental boundary notes remain in the source and replication archive as exploratory working material.
The retained synthesis uses the same paper inclusion, contribution records, evidence settings, and handoff counts.

\input{tables/impl_f10_boundaries}
\FloatBarrier

\subsection{Screening History}
\label{sec:screening-history}

The following snapshots document candidate processing before final author confirmation; the archive retains the original preliminary assessments and their provenance.
The initial database searches and known-paper set established the seed pool; citation expansion supplied 5,113 candidates for high-recall screening.
That pass recorded 4,327 provisional inclusions, 247 unresolved cases, 111 readable provisional exclusions, and 428 records without accessible full text.
Thus, 4,685 candidates had readable text: 4,574 positive or unresolved records entered stricter scope confirmation, and all 111 readable negatives received a separate high-recall recheck.
Legacy seeds were routed through the same scope rules, and the overlapping streams were reconciled by version family.
Their first frozen union contained 1,670 preliminary Primary records.

A complete precision-oriented scope pass over those 1,670 records retained 1,347 Primary records, excluded 153, and left 170 for boundary review.
Final adjudication combined that queue with nine earlier final-gate uncertainties, yielding 179 decisions: 170 retained as Primary and nine removed.
The historical v3 ledger therefore contained 5,357 records: 1,517 Primary, 3,372 scope-excluded, 35 Secondary, and 433 access-excluded, with no unresolved scope decision.
The access exclusions comprised the 428 initial cases and five supplemental cases.

The identity and scope corrections described above reduced this ledger to 5,354 canonical records: 1,514 Primary, 3,373 scope-excluded, 35 Secondary, and 432 access-excluded.
The current Primary split is 355 P, 1,043 E, and 116 U.
These counts supersede the earlier snapshot for current corpus accounting and describe the screening decisions confirmed by the authors.

%% file: tables/impl_f10_boundaries.tex
\begin{table*}[t]
  \centering
  \caption{Source-level comparison of reuse, recovery, and execution controls. Thirteen papers supply 15 currently coded units; Agent-X supplies three. The bands group related engineering operations for comparison, with evaluation statements scoped to the listed mechanism.}
  \label{tab:impl-f10-boundaries}
  \small
  \renewcommand{\arraystretch}{1.12}
  \begin{tabular}{L{0.18\textwidth}L{0.31\textwidth}L{0.40\textwidth}}
    \toprule
    Study & Engineering operation & Evaluation and validity boundary \\
    \midrule
    \multicolumn{3}{l}{\emph{Resource reuse and inference execution}} \\
    KVCOMM~\cite{ye2025kvcomm} & Approximate cross-prefix KV offsets; fall back to prefill & Same checkpoint and RoPE assumptions; latency and accuracy sensitivity, not semantic equivalence \\
    LRAgent~\cite{jeon2026lragent} & Compute attention from shared base and low-rank state & Multi-LoRA kernels and controlled traces; shared-A accuracy setting is a further condition \\
    Nova~\cite{xu2025nova} & Stream encoder weights through reusable GPU layers & Memory and latency ablations; overlap depends on bandwidth and execution length \\
    Agent-X~\cite{chung2026agentx} & Cache reordered prefixes; verify example-derived drafts; integrate both & Component and combined timing on macOS/TinyAgent; integration-unit granularity remains open \\
    GraphFlow~\cite{li2026graphflow} & Store base KV plus prefix residuals for frequent paths & Memory/performance and pruning ablations; rare paths are recomputed, not proved unreachable \\
    \midrule
    \multicolumn{3}{l}{\emph{State restoration, persistence, and separation}} \\
    RAC~\cite{perera2026rac} & Log tool effects; retry, substitute, or compensate & Failure scenarios; valid compensators and mappings are necessary, and may involve an LLM \\
    AGDebugger~\cite{epperson2025debugging} & Checkpoint state before messages; fork a debugging session & Bounded debugging study; internal restoration cannot undo arbitrary external effects \\
    ClawVM~\cite{rafique2026clawvm} & Select typed pages and validate scoped writeback & Fault/replay tests and a bounded live-hook check; content truth remains outside the contract \\
    He et al.~\cite{he2025securityagents} & Separate session histories and correlate model requests & Proposed session mechanism; sandbox attacks evaluate a different defense \\
    \midrule
    \multicolumn{3}{l}{\emph{Coordination checks and participation control}} \\
    TraceFix~\cite{xia2026tracefix} & Check topology operations and resource access at runtime & Coordination experiments retain failures; checks do not enforce complete protocol ordering \\
    BlindGuard~\cite{miao2026blindguard} & Detect anomalous participants and prune message edges & Prompt/tool/memory attack simulations; detection and pruning are not universal containment \\
    BAD-ACTS Guardian~\cite{noether2026badacts} & Classify inter-agent messages; terminate unsafe episodes & Attack and benign-message tests; stopping differs from successful recovery \\
    Koala II~\cite{houde2025participation} & Expose participation thresholds and presentation controls & Group ideation study with model/prompt changes; settings are not an authorization boundary \\
    \bottomrule
  \end{tabular}
\end{table*}

%% file: sections/supplement_stage_sensitivity.tex
\section{Supplemental Stage-Coverage Sensitivity}
\label{sec:supplement-stage-sensitivity}

Table~\ref{tab:supplement-stage-sensitivity} reports current paper-level stage codes from the 150-paper supplement, separately from the 355-paper main analysis.
This diagnostic checks the stability of observed stage orderings under hypothetical recoding of a fixed, historically flagged subset of the author-reviewed supplement.
The archived content flags cover 70 papers, while a dated exploratory classification review queue covers 58; their union contains 90 papers.
That historical queue defines part of the fixed sensitivity set and is retained for its reproduction.
We conservatively allow arbitrary stage recoding of that entire union, including historical flags and definition-fit cases that need not themselves be stage errors.
This post-coding stress test retains the full queue while holding other current codes fixed; errors outside that queue remain beyond the diagnostic's coverage.
A stage is present when at least one extracted contribution is assigned to it; multiple contributions in the same paper do not increase the indicator.
The historical flags identify the sensitivity set, not a current queue awaiting author confirmation.

\begin{table}[H]
  \centering
  \caption{Stage coverage in the frozen main analysis and supplement. Counts are papers and stages overlap. Reweighting uses the frozen tier-by-search-stage sample, not current publication status. The range permits arbitrary recoding of the flagged-sensitive papers while holding other current codes fixed; it is not a confidence interval.}
  \label{tab:supplement-stage-sensitivity}
  \small
  \renewcommand{\arraystretch}{1.15}
  \begin{tabular}{lrrrr}
    \toprule
    Stage & Main ($n=355$) & Sample ($n=150$) & Reweighted (\%) & Range (\%) \\
    \midrule
    Requirements & 26 & 13 & 6.8 & 0.2--64.5 \\
    Architecture & 127 & 59 & 38.9 & 7.9--72.1 \\
    Implementation & 76 & 45 & 29.9 & 8.1--72.4 \\
    Testing & 215 & 92 & 68.2 & 23.4--87.7 \\
    Maintenance & 49 & 31 & 17.4 & 8.7--72.9 \\
    \bottomrule
  \end{tabular}
\end{table}

For stratum $h$, let $w_i=N_h/n_h$ and let $z_{is}$ indicate a recorded contribution in stage $s$.
With the fixed frame denominator $N=1158$, the descriptive reweighted fraction is $\sum_i w_i z_{is}/N$.
For the flagged-sensitive set $\mathcal{B}$, the lower bound is $\sum_{i\notin\mathcal{B}}w_i z_{is}/N$ and the upper bound adds $\sum_{i\in\mathcal{B}}w_i/N$.
The 90 sensitive records carry 64.3\% of the total expansion weight.
All records in the fixed diagnostic frame remain in the denominator, including in the zero-positive-code boundary case.

The Testing--Requirements difference spans -41.1 to 87.4 percentage points; its positive ordering is not preserved across the conditional recoding range.
The corresponding Testing--Maintenance difference spans -49.6 to 79.0 percentage points, so that ordering is not secured by this diagnostic.
These bounds do not cover shared coding errors in unflagged records, unsampled studies, or search omissions.
Population interpretation additionally requires defensible version-family identities and exchangeable within-stratum hash ranks.
The table compares frozen selections, with the E/U strata including subsequently verified published or accepted studies.
The selected fields support the stage-coverage comparison; fine-grained method and artifact comparisons require their own source-level analysis.
This table uses the source-corrected contents pinned by its diagnostic manifest; earlier diagnostics remain in the replication archive.
The current counts group internal recovery mechanisms within their parent methods rather than inferring a Maintenance method from every feedback step.
Likewise, Leni's training recipe supports its loop architecture, and the type-directed privilege-separation study's case implementations and experiments support one architectural contribution rather than two additional development methods.
COCO's configuration and correction submechanisms are grouped into its platform and recovery outputs without changing its paper-level stage coverage.
The cross-stage source review moves TRIAGE's persistent harness-revision method from Implementation to Maintenance; CSI's compact scaffold remains one construction contribution with internal refinement.
AgentBuild's internal acceptance and domain assembly are grouped with construction, removing a separate Testing code without removing its held-out evaluation records.
The archived supplemental source notes explain these boundaries. The affected papers were already in the fixed sensitivity set, so the conditional bounds remain unchanged. The dated mapping supplies that review membership, not a retained taxonomy.

%% file: sections/evidence_setting_cases.tex
\FloatBarrier
\section{Evidence-Setting Interpretation Cases}
\label{sec:evidence-setting-cases}

This appendix records the source-specific interpretation of evaluation settings used in the stage reviews in Section~\ref{sec:lifecycle}.
The cases explain how evaluation setting relates to study design, measured outcome, and causal support within the existing main-corpus records.
The cases and tables below document production, authentic-engineering, longitudinal, and controlled-study boundaries.
Review coverage and source amendments remain in Appendix~\ref{sec:source-coding-amendments}.

\subsection{Production Exposure and Outcome Interpretation}

Interpreting production evidence requires separating deployment exposure from the comparison that supports a particular claim.
For example, HLTM's two-week indexing comparison and its longer production history answer different questions: the former compares update strategies, while the latter describes operational use~\cite{xu2026hierarchicalmemory}.
Similarly, ADR's sensing, investigation and prevention contributions share one deployment and its observation conditions~\cite{li2026adr}.
Wink evaluates intervention in live traffic, but its recovery measure concerns the detected misbehavior rather than end-to-end task correctness~\cite{nanda2026wink}.
Persistent records and a deployed system provide a setting for studying evolution; sustained benefit and generalization require outcome evidence across time and operating contexts.

Production evidence and longitudinal exposure must be distinguished even within E4.
DLBP reports operational incidents and a representative session but does not state the total deployment duration in the inspected conference paper~\cite{bass2026blackboard}.
Enterprise MCP describes eighteen months of production deployment, yet its study-period table starts in January 2024, before the November 2024 MCP launch stated in the same paper~\cite{guruvelli2026enterprisemcp}.
The period may include the pre-MCP baseline, but its duration relative to MCP deployment remains unresolved in the source.
Both production cases qualify as E4 under the disjunctive rule, with longitudinal exposure unspecified or uncertain.
Likewise, the 15-day production comparisons in OpenHands SDK and Wink qualify through their deployment setting; their benefit estimates remain tied to those observation windows~\cite{wang2026openhands,nanda2026wink}.

\FloatBarrier
\subsection{Engineering Work and Longitudinal Observation}

\input{tables/field_evidence_conditions}

Table~\ref{tab:field-evidence-conditions} separates engineering work, longitudinal observation, and controlled use of realistic artifacts.
MAP gathers first-hand accounts of engineering practice; its interviews and questionnaires support empirical characterization, not the superiority of a proposed architecture~\cite{pan2026map}.
Botender observes participants configuring and revising bots in their own communities; AEGIS records actual prompt-specification defects and fixes in a production project~\cite{kuo2026botender,calboreanu2026iterativeaudit}.
These engineering cases describe configuration and specification work within their respective projects; independently replicated effectiveness remains unestablished.
AEVAL instead illustrates its protocol through deliberately degraded skills and controlled cross-runtime runs; production artifact provenance alone does not establish observed engineering work~\cite{anand2026aeval}.

Repeated observation over time provides a distinct basis for the longitudinal setting code.
GPTracker repeatedly checks the same public GPT identifiers and analyzes removal and re-creation, supporting a longitudinal setting for its measurement contributions without showing operator adoption~\cite{shen2025gptracker}.
AgentEval follows development through regression alerts and configuration revisions; its E4 interpretation rests on those repeated observations, not simply a four-month calendar interval~\cite{guo2026agentevaldag}.
Separate offline experiments retain their own setting descriptions, and each contribution's highest code summarizes its strongest direct evaluation setting.

Field results also require contribution-specific interpretation.
AgentEval reports 23 regression alerts, of which eight were genuine, 12 borderline and three false positives; its before/after investigation times use retrospective reports and system logs with different measurement boundaries~\cite{guo2026agentevaldag}.
Those observations document use in development; differing timing measurements confound estimation of a tool-attributable speedup.
Wink's live monitoring and historical sample are separate evidence sources; conflicting counts in the body and table caption leave the historical sample's denominator and exact prevalence unresolved~\cite{nanda2026wink}.

\FloatBarrier
\subsection{Practitioner Evidence and Controlled-Study Boundaries}

\input{tables/capped_setting_source_review}

Table~\ref{tab:capped-setting-source-review} distinguishes reports of actual engineering work from controlled tool evaluations.
DiLLS's formative interviews concern participants' recent multi-agent development and diagnosis, while its separate controlled tool experiment stays at E2~\cite{sheng2026dills}.
The oversight study reports instruction-file maintenance and Agent control configuration performed by developers~\cite{dhanorkar2026oversight}.
C4's industry report and AgentArcEval's stakeholder walkthrough likewise document architecture work in actual projects, without establishing comparative effectiveness~\cite{rausch2026agenticc4,lu2026agentarceval}.

Further practitioner studies clarify this distinction.
XAgen elicits diagnostic needs from 12 practitioners' actual multi-agent work, while Tangent interviews ten practitioners with recent Agent development or deployment experience~\cite{wang2026xagen,pan2026tangent}.
Their separate prototype experiment and repository-mining evaluation remain E2.
Axetorn et al. conduct employee interviews and an expert workshop within an industrial partner's Agent project; this supports an E3 requirements-design case even though the later prototype validation is controlled~\cite{axetorn2026trustrequirements}.
By contrast, repeated expert review of AI-ZTMM refines an assessment model rather than observing an operational Agent system: its six monthly review sessions do not by themselves establish longitudinal engineering evidence~\cite{mok2026actioncentric}.

GUI-agent consequence reports remain a boundary case~\cite{zhang2025guiuc}.
Their retrospective accounts combine failures, implemented user controls, and expectations for future systems; we conservatively retain E2 rather than infer an observed engineering process from all of these accounts.
If both contributions are instead treated as field-like configuration and elicitation evidence, E3/E4 becomes 39/570 (6.8\%) across 23 papers, with five Requirements E3 contributions.
Both interpretations concern the reported configuration and elicitation activities; the alternative is a sensitivity check, while the main analysis uses the author-confirmed E2 coding.

These cases and tables illustrate author-reviewed setting boundaries within the targeted source comparisons specified in Appendix~\ref{sec:source-coding-amendments}.
The scarcity finding concerns documented engineering and operational settings, with methodological rigor requiring separate appraisal.

%% file: tables/field_evidence_conditions.tex
\begin{table}[htbp]
  \caption{Engineering cases, longitudinal observations, and controlled boundary cases in the initial source correction. These eight studies supply 13 reviewed contributions: seven E3, three E4, and three E2. Table~\ref{tab:capped-setting-source-review} illustrates additional low-setting and human-study boundaries.}
  \label{tab:field-evidence-conditions}
  \centering
  \footnotesize
  \setlength{\tabcolsep}{4pt}
  \renewcommand{\arraystretch}{1.12}
  \begin{tabular}{@{}>{\raggedright\arraybackslash}p{0.16\textwidth}>{\raggedright\arraybackslash}p{0.27\textwidth}>{\raggedright\arraybackslash}p{0.21\textwidth}>{\raggedright\arraybackslash}p{\dimexpr0.36\textwidth-24pt\relax}@{}}
    \toprule
    Study / code, units & Observation and sample & Engineering evidence & Interpretation boundary \\
    \midrule
    Botender~\cite{kuo2026botender} / E3, 2 & Five days, six communities, 31 participants; 137 proposals and 800 cases & Collaborative configuration and case-driven iteration & Compensated participation with minimum activity; intention is not adoption \\
    \addlinespace
    MAP~\cite{pan2026map} / E3, 3 & 20 interview cases; 86 deployed/pilot responses among 306 questionnaires & Reported architecture, implementation and evaluation practices & First-hand accounts, not a production experiment; item denominators vary \\
    \addlinespace
    AEGIS audit~\cite{calboreanu2026iterativeaudit} / E3, 2 & Nine audit-and-fix rounds on an actual production specification & Prompt-contract inspection and defect characterization & Protocol co-evolves with fixes; synthetic model repetitions are not independent replication \\
    \addlinespace
    GPTracker~\cite{shen2025gptracker} / E4, 2 & 16 collection rounds; repeated GPT-status checks and removal/re-creation analysis & Longitudinal security measurement infrastructure and findings & Observational tracking, not operator adoption; removal times are interval-based \\
    \addlinespace
    AgentEval~\cite{guo2026agentevaldag} / E4, 1 & Four-month development pilot; 18 engineers; 14 recorded configuration changes & Repeated regression observation and revisions & Eight genuine regressions among 23 alerts; before/after timing instruments differ \\
    \addlinespace
    AEVAL~\cite{anand2026aeval} / E2, 1 & Production skill artifacts; induced defects and six cross-runtime runs & First-attempt grading and auditable CI outputs & Controlled demonstration, not observed use in ongoing engineering work \\
    \addlinespace
    Workflow Graphs~\cite{park2025workflowgraphs} / E2, 1 & Deployed shopping interface; anonymized paired-response tests & Controlled architecture assessment & Availability of a live service does not make the comparison an operational outcome study \\
    \addlinespace
    G2CP~\cite{benkhaled2026g2cp} / E2, 1 & Historical industrial cases; 21 retained after expert consensus & Validation under realistic missing or ambiguous inputs & Nine-month log span is not nine months of deployment observation \\
    \bottomrule
  \end{tabular}
\end{table}

%% file: tables/capped_setting_source_review.tex
\begin{table}[htbp]
  \caption{Selected engineering cases and human-study boundaries recovered by checking low initial setting codes. Codes describe the direct evaluation setting of each contribution; separate evaluations retain their own descriptions. The companion ledger contains the full targeted review.}
  \label{tab:capped-setting-source-review}
  \centering
  \footnotesize
  \setlength{\tabcolsep}{4pt}
  \renewcommand{\arraystretch}{1.12}
  \begin{tabular}{@{}>{\raggedright\arraybackslash}p{0.18\textwidth}>{\raggedright\arraybackslash}p{0.12\textwidth}>{\raggedright\arraybackslash}p{0.32\textwidth}>{\raggedright\arraybackslash}p{\dimexpr0.38\textwidth-24pt\relax}@{}}
    \toprule
    Study & Stage / setting & Source basis & Interpretation boundary \\
    \midrule
    Oversight practice~\cite{dhanorkar2026oversight} & REQ / E3 & 17 developers; accounts of instruction maintenance and Agent configuration & First-hand practice, not observed production effects; 12 participants from one organization \\
    \addlinespace
    DiLLS formative study~\cite{sheng2026dills} & REQ / E3 & Eight practitioners recount recent multi-agent development and diagnosis & Elicitation of diagnostic needs; separate 12-person tool experiment remains E2 \\
    \addlinespace
    XAgen formative study~\cite{wang2026xagen} & REQ / E3 & 12 practitioners describe actual multi-agent development and debugging & Engineering needs, not deployed effectiveness; separate tool experiment remains E2 \\
    \addlinespace
    Trust requirements~\cite{axetorn2026trustrequirements} & REQ / E3 & Six employee interviews and a five-expert workshop within a partner project & Project-based elicitation can precede deployment; later prototype evaluation remains E2 \\
    \addlinespace
    Tangent~\cite{pan2026tangent} & TEST / E3 & Ten practitioners with recent Agent development or deployment experience & One organization's testing accounts; separate repository mining remains E2 \\
    \addlinespace
    C4 industry report~\cite{rausch2026agenticc4} & ARCH / E3 & Architecture descriptions and lessons from three industrial projects & Authors' experience, not measured maintainability gains or controlled readability tests \\
    \addlinespace
    AgentArcEval~\cite{lu2026agentarceval} & ARCH / E3 & Luna stakeholders refine scenarios, priorities, risks, and design decisions & One architecture-review case; scenario targets are not achieved performance \\
    \addlinespace
    GUI consequences~\cite{zhang2025guiuc} & TEST, REQ / E2 & 221 posts and 14 interviews: failures, user controls, future expectations & Retrospective boundary retained provisionally; treating both units as E3 yields 39/570 rather than 37/570 \\
    \bottomrule
  \end{tabular}
\end{table}